\documentclass[11pt,a4paper]{article}
\usepackage{jcappub}
\pdfoutput=1

\usepackage{graphicx}
\usepackage{amsmath}
\usepackage{amssymb}
\usepackage[english]{babel}

\newcommand{\INFN}{INFN - Sezione di Napoli, Complesso Univ. Monte S. Angelo, I-80126 Napoli, Italy}
\newcommand{\UNI}{Dipartimento di Fisica "Ettore Pancini", Universit\'a degli studi di Napoli "Federico II", Complesso Univ. Monte S. Angelo, I-80126 Napoli, Italy}
\newcommand{\DESY}{DESY, Platanenallee 6, 15738 Zeuthen, Germany}

\begin{document}

\title{Unified thermal model for photohadronic neutrino production in astrophysical sources}

\author[a,b]{Damiano F.G. Fiorillo}
\author[c]{Arjen van Vliet}
\author[a,b]{Stefano Morisi}
\author[c]{Walter Winter}

\affiliation[a]{\UNI}
\affiliation[b]{\INFN}
\affiliation[c]{\DESY}

\emailAdd{dfgfiorillo@na.infn.it}
\emailAdd{arjen.van.vliet@desy.de}
\emailAdd{stefano.morisi@gmail.com}
\emailAdd{walter.winter@desy.de}
\date{\today}

\abstract{High-energy astrophysical neutrino fluxes are, for many applications, modeled as simple power laws as a function of energy. While this is reasonable in the case of neutrino production in hadronuclear $pp$ sources, it typically does not capture the behavior in photohadronic $p\gamma$ sources: in that case, the neutrino spectrum depends on the properties of the target photons the cosmic rays collide with and on possible magnetic-field effects on the secondary pions and muons. We show that the neutrino production from known photohadronic sources can be reproduced by a thermal (black-body) target-photon spectrum if one suitably adjusts the temperature,  thanks to multi-pion production processes. This allows discussing neutrino production from most known $p\gamma$ sources, such as gamma-ray bursts, active galactic nuclei and tidal disruption events, in terms of a few parameters. We apply this thermal model to study the sensitivity of different classes of neutrino telescopes to photohadronic sources: we classify the model parameter space according to which experiment is most suitable for detection of a specific source class and demonstrate that different experiment classes, such as dense arrays, conventional neutrino telescopes, or radio-detection experiments, cover different parts of the parameter space. Since the model can also reproduce the flavor and neutrino-antineutrino composition, we study the impact on the track-to-shower ratio and the Glashow resonance.}

\maketitle

\section{Introduction}

Finding the origin of the cosmic-rays arriving at Earth is still an open problem in astrophysics; promising information may come from secondary neutrinos produced in cosmic-ray interactions. Astrophysical sites that are able to accelerate cosmic-ray protons can produce pions and subsequently neutrinos, either via purely hadronic collisions with gas or via photohadronic collisions with the radiation field in the sources; for the neutrino production off nuclei, see Ref.~\cite{Morejon:2019pfu}. High-energy neutrinos are not deflected by interstellar magnetic fields and do not suffer from interactions during their path to Earth: this is an advantage that can help in identifying the sources in which they have been produced. Neutrino telescopes such as IceCube~\cite{Goldschmidt:2001qd,Ahrens:2002dv} and KM3NeT~\cite{Kappes:2007ci} (see also, e.g., Refs.~\cite{Aslanides:1999vq,Aartsen:2014oha,Barwick:2007vba,Meures:2012fka,Martineau-Huynh:2015hae, Schulz:2009zz,Blaufuss:2015muc,Collaboration:2011nsa,Belolaptikov:1997ry,Riccobene:2017fpr}), are the key instruments in the detection of these astrophysical neutrinos. A diffuse flux of astrophysical high-energy neutrinos has been observed by IceCube over the last 10 years~\cite{Aartsen:2013bka,Aartsen:2013jdh,Aartsen:2014gkd,Aartsen:2015rwa,Aartsen:2020aqd}. However, due to directional  uncertainties, positional analyses have not yet been able to uncover the sources~\cite{Bartos:2016wud,Aartsen:2016lir,Aartsen:2017wea,Mertsch:2016hcd,Aartsen:2018ywr,Murase:2013rfa,Murase:2015xka,Ando:2015bva,Silvestri:2009xb,Ahlers:2014ioa,Murase:2016gly,Dekker:2018cqu}. For this reason, there is still a fundamental role to be played by the other neutrino telescopes with different energy and directional resolutions. We are mostly be interested in the energy threshold and the sensitivity energy range, which depend on the array spacing and technology; we therefore consider three representative experiments from three classes of experiments: dense neutrino arrays, neutrino telescopes, and radio arrays.

Many different possibilities have been proposed for the astrophysical sources responsible for the production and acceleration of cosmic-rays, including Active Galactic Nuclei (AGN)~\cite{Stecker:1991vm}, Gamma-Ray Bursts (GRBs)~\cite{Piran:2004ba} and Tidal Disruption Events (TDEs)~\cite{Wang:2011ip}. For all these candidates the dominant mechanism of neutrino production is believed to be the collision of cosmic-ray protons either with the target photons inside the source in the case of $p\gamma$ interactions or with gas in the case of $pp$ interactions. Some indications for individual source-neutrino associations have been presented so far, the blazar TXS~0506+056~\cite{IceCube:2018cha,IceCube:2018dnn} and the TDE named AT2019dsg~\cite{Stein:2020xhk} being the most prominent ones. In these source classes the neutrinos are presumably of photohadronic origin, see e.g.~Refs.~\cite{Gao:2018mnu,Cerruti:2018tmc,Keivani:2018rnh,Rodrigues:2018tku} for the blazar and Refs.~\cite{Winter:2020ptf,Murase:2020lnu,Liu:2020isi} for the TDE (although alternative $pp$ models exist, see e.g.~Ref.~\cite{Liu:2018utd} for the blazar). We therefore focus on photohadronic interactions in this work.

The spectral shape of the photohadronic neutrino spectrum is typically {\bf not} a simple power law, but depends on the primary proton and target-photon spectral shapes; see Ref.~\cite{Winter:2012xq} for a review of the neutrino production. This more complicated spectral shape affects the detection prospects depending on the energy range the detector is sensitive to, see e.g.~Ref.~\cite{Winter:2011jr}. In addition, the spectral shape, the relative number of each neutrino flavor produced, and the relative number of neutrinos and anti-neutrinos produced depend on the process in which they are produced (e.g. from pion, muon, kaon or neutron decays), possible cooling effects of the charged secondaries in magnetic fields, and the contribution of pion production processes other than the $\Delta$-resonance, see e.g.~Refs.~\cite{Mucke:1999yb,Kashti:2005qa,Lipari:2007su,Hummer:2010ai,Hummer:2010vx,Baerwald:2010fk,Baerwald:2011ee,Winter:2012xq,Roulet:2020yye}. Note that the flavor composition can be used to probe physics beyond the Standard Model, see e.g.~Ref.~\cite{Rasmussen:2017ert} for an overview. 

Because of the dependence on the target-photon spectrum, neutrino fluxes for different photohadronic sources are usually obtained by dedicated modeling for each source class. However, this approach is not suited for systematic scans; we therefore provide here a description of the photohadronic neutrino spectrum in a unified model relying on a thermal target-photon spectrum that depends on very few parameters only. This means that we simplify the dependence on the target-photon spectrum while we maintain the complexity in terms of spectrum and flavor composition coming from the neutrino production chain. We will demonstrate how this model can be used for conventional sources with typical non-thermal target photons, such as AGN and GRBs, whereas it trivially applies to sources with thermal spectra, such as to thermal TDE models. Therefore, the parameterization of the photon target as a black-body spectrum is sufficiently flexible to reproduce the neutrino production even in sources which do not posses a black-body photon target. The purpose of the model presented here is to describe the neutrino flux and flavor composition as accurately as possible, while the connection to the possibly complicated electromagnetic spectrum is given up. That has the advantage that the possible parameter space of neutrino spectra can be rather completely modeled, while the disadvantage is the weak connection with multi-messenger observations.
Earlier works in this direction, such as Refs.~\cite{Hummer:2010ai,Winter:2011jr,Winter:2013cla}, rely on a somewhat different assumption: there the target-photon spectrum is generated by synchrotron radiation off co-accelerated electrons -- which also reduces the number of parameters, but is probably only a good approximation for specific source classes.

We focus on two applications: the complementarity of different experiment classes regarding the parameter-space coverage of the model, and the impact on flavor- and neutrino-antineutrino-sensitive observables. Possible further applications include generic parameter-space studies of astrophysical neutrinos of all kinds. For example, Ref.~\cite{Bustamante:2020bxp} proposes using the magnetic field effects on the secondaries as ``cosmic magnetometers'' to determine the magnetic field in the sources. However, it was pointed out already in Ref.~\cite{Winter:2013cla} that the neutrino data fit is driven by the spectrum and not the flavor composition, which means that a realistic description of the spectral shape must not be neglected in a flavor analysis; our model could be used for such a purpose as well.  Other examples are neutrino multiplet studies leading to constraints on the source density, stacking analyses (where one could treat the model parameters in certain ranges as systematics) and flavor or neutrino-antineutrino composition studies, such as for the purpose to constrain physics beyond the Standard Model.

The structure of this work is as follows: in section~\ref{sec:valtherm} we discuss the qualitative features of the neutrino production from a thermal target-photon field. Based on these results, we show that it is possible to simulate the main aspects of the neutrino fluxes from a generic photon spectrum using an effective black-body photon target with a suitably chosen temperature. We also discuss our methodology for quantifying the sensitivity of an experiment to a generic astrophysical flux. In section~\ref{sec:resul} we show and discuss our results on the energy dependence of the fluxes and their possible detection at Earth. Finally, in section~\ref{sec:flavorresults} we study the impact on the flavor composition and its connection with the methodologies of flavor discrimination available in the different energy ranges. 

\section{Thermal model}\label{sec:valtherm}

Astrophysical neutrino production is more complicated to model in $p\gamma$  than in $pp$ sources. In $pp$ sources, cosmic-ray protons collide on a target of nuclei at rest; further, magnetic fields are typically not strong enough to significantly alter the secondary muon and pion spectra. Therefore, it is typically assumed that neutrinos follow the power-law spectrum of the parent cosmic-rays with a slope of around two and a high-energy cutoff fixed by the maximal acceleration energy of primaries (see Ref.~\cite{Ambrosone:2020evo} for an example of $pp$ spectrum beyond the single power-law approximation). On the other hand, in photohadronic sources, the neutrino spectrum will depend both on the cosmic-ray spectrum and on the spectrum of the low-energy photons acting as a target for $p\gamma$ collisions. The neutrino production is normally discussed for each source class individually (e.g. Refs.~\cite{Winter:2020ptf,Rodrigues:2017fmu,Murase:2020lnu}), where other astrophysical motivations are at play (such as multi-messenger signatures or the prediction of the flux normalization). There are also more generic approaches focusing on spectral shape and flavor composition: for example, in Ref.~\cite{Hummer:2010ai} target photons are assumed to be self-consistently produced by co-accelerated electrons via synchrotron radiation, which may however not apply to all sources; in Ref.~\cite{Hummer:2010vx} the photon spectrum in AGN or GRBs is (locally) modeled as a broken power law; in Ref.~\cite{Guepin:2017dfi} a broken power-law spectrum is assumed for the target photons, assessing the detectability of various transient photohadronic sources. There are alternative models for these sources adopting different choices for the target photons: for example Refs.~\cite{Murase:2006mm,Murase:2008sp,Kashiyama:2012zn,Murase:2013ffa} discuss neutrino production in GRBs, choked GRB jets and low-luminosity GRBs including a thermal component for the target photon field. In TDE models, the target photons may be described by  a thermal spectrum, see, e.g., Ref.~\cite{Winter:2020ptf}, a  broken power law, see, e.g.,  Ref.~\cite{Lunardini:2016xwi}, or a combination of the two, see, e.g., Ref.~\cite{Senno:2016bso}, where both a thermal target and a non-thermal synchrotron target is considered. These examples already show the variety of options which can arise; in all these cases the target-photon spectrum can locally (in the energy range relevant for the neutrino production) be approximated by a power-law, a broken power-law or a thermal spectrum.

We emphasize that the use of broken power laws is generally adopted as a simplified way of modeling the real spectrum, which is, in general, smooth and has a curved structure. A typical example is synchrotron radiation, for which the smooth spectrum can be approximated by a broken power law: this approximation is not supposed to be valid in the energy region close to the break energy, of course. As another example relevant for the benchmark cases, we will discuss, in Sec.~\ref{sec:thermalexamples}, the spectrum of GRBs modeled empirically with the aid of the smooth Band function~\cite{Band:1993eg} (see Eq.~\ref{eq:bandfunction}) which has continuous derivatives at the break energy. As we will show there, the use of curved target photon spectra does not lead to significant corrections to the applicability of the thermal model.

The aim of this section, paired with App.~\ref{app:thermalmodel}, is to show that an effective thermal model with target photons with an effective temperature $T$ is sufficiently general to reproduce the neutrino spectrum and flavor composition in most $p\gamma$ sources, even if the real photon spectrum of the source is non-thermal.  We will also discuss the properties of the target-photon spectra that are not easily reproducible by the thermal model. 

\subsection{Temperature mapping}
\label{sec:recipe}

The key point of the thermal model is that we can assign to any target-photon spectrum an equivalent black-body spectrum with a definite effective temperature. We now discuss this mapping, leaving for App. \ref{app:thermalmodel} the details of its derivation. To fix the notation, we note that a typical feature of photohadronic sources is the highly relativistic motion of the regions in which particles are accelerated, e.g., relativistic shocks or jets in AGN or GRBs. Throughout this work the quantities in the rest frame of the ``accelerator'', i.e. of the shock or the jet, are primed; unprimed quantities are measured in the rest frame of the observer at the Earth. The energies in the two rest frames are assumed to be connected by the relation $E=\Gamma E'$, where $\Gamma$ is a (simplified notation) Doppler factor.\footnote{If the Lorentz factor of the expansion/movement of the shock or jet is $\gamma$, the actual  Doppler factor is given by $\delta=\left[\gamma(1-\beta\cos\theta)\right]^{-1}$, where $\theta$ is the angle between the jet and the line of sight. For $\theta\sim1/\gamma$ one has $\delta \simeq \gamma$; consequently, $\Gamma \simeq \gamma$ for small redshifts. If the energy is in addition redshifted by a factor $1+z$, the simplified notation Doppler factor is related to $\delta$ by $\Gamma \equiv \delta/(1+z) \simeq \gamma/(1+z)$ for $\theta=1/\gamma$.}

Let the number of target photons per unit volume per unit energy be $n(\varepsilon'_\gamma)$. The interaction rate of a proton with this photon target is then proportional to the cross section for photohadronic interaction and to the photon {\em number} density $\varepsilon'_\gamma n(\varepsilon'_\gamma)$.\footnote{In principle the photon number density should be defined by $\int \text{d}\varepsilon'_\gamma n(\varepsilon'_\gamma)$; we will nonetheless use the definition $\varepsilon'_\gamma n(\varepsilon'_\gamma)$ throughout this section (which correctly display the proportionalities for power law spectra). We leave for App.~\ref{app:thermalmodel} a detailed derivation of why this quantity appears in the interaction rate.} The  interaction rate vanishes if the photon energy is too small (below threshold), since the interaction becomes kinematically forbidden. In particular, for a proton with energy $E'$, the photons that are energetic enough to induce photohadronic production must have an energy $\varepsilon'_\gamma\gtrsim y_\Delta m_p/E'$, where $y_\Delta \simeq 0.2$~GeV corresponds to the reasonable pitch-angle averaged ``threshold'' for the interaction; see Fig.~4 in Ref.~\cite{Hummer:2010vx}, and $m_p$ is the proton mass. For a proton distribution with a maximal energy $E'_\text{p,max}$ (we discuss in the next subsection the values of this maximal energy), the photons participating in the photohadronic interaction must satisfy the condition
\begin{equation}\label{eq:kincond}
    \varepsilon'_\gamma>\frac{y_\Delta m_p}{E'_\text{p,max}}.
\end{equation}
In the range defined by Eq.~\ref{eq:kincond} we now determine an energy $\bar{\varepsilon}'_\gamma$ which contributes most to the photohadronic production. We can distinguish different behaviors of the number density $\varepsilon'_\gamma n(\varepsilon'_\gamma)$ in this region:
\begin{figure}[t!]
    \centering
    \includegraphics[width=0.5\textwidth]{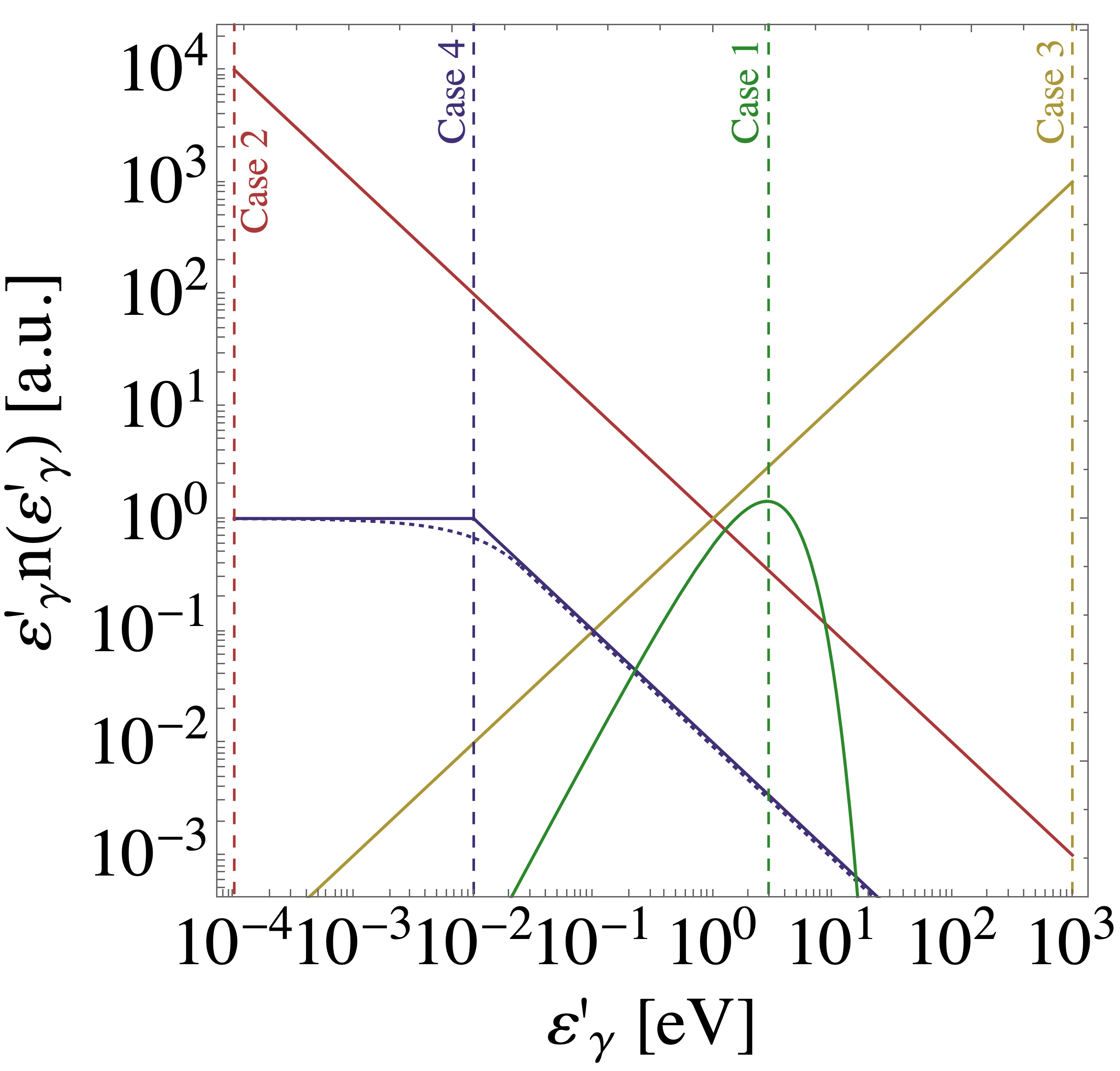}
    \caption{Target photon number density for the four cases mentioned in the text. The different cases are identified by different colors; the energy $\bar{\varepsilon}'_\gamma$ for each case is identified by a dashed line with the color corresponding to that case. For case 4, we also show the Band function interpolation to the broken power law as a dotted line (see text for more details).}
    \label{targetcases}
\end{figure}
\begin{itemize}

    \item \textbf{Case 1:} $\varepsilon'_\gamma n(\varepsilon'_\gamma)$ is maximum at a well-defined energy $\bar{\varepsilon}'_\gamma$ (e.g.~a black-body spectrum). Since the interaction rate is proportional to the number density, we expect photons of energy $\bar{\varepsilon}'_\gamma$ to contribute most to the photohadronic production;
    
    \item \textbf{Case 2:} $\varepsilon'_\gamma n(\varepsilon'_\gamma)$ is monotonically decreasing (e.g. a power law $n(\varepsilon'_\gamma)\propto\varepsilon^{'-\alpha}_\gamma$ with $\alpha>1$) for energies larger than $\varepsilon'_{\gamma,\text{min}}$. The maximum of the number density, subject to the condition in Eq.~\ref{eq:kincond}, is either $\varepsilon'_{\gamma,\text{min}}$ or $\frac{y_\Delta m_p}{E'_\text{p,max}}$, depending on which energy scale is larger. Therefore we define $\bar{\varepsilon}'_\gamma=\text{max}\left[\frac{y_\Delta m_p}{E'_\text{p,max}},\varepsilon'_{\gamma,\text{min}}\right]$;
    
    \item  \textbf{Case 3:} $\varepsilon'_\gamma n(\varepsilon'_\gamma)$ is monotonically increasing until a maximal energy $\varepsilon'_{\text{max}}$ (e.g. a power law $n(\varepsilon'_\gamma)\propto\varepsilon^{'-\alpha}_\gamma$ with $\alpha<1$). The maximum of the number density is at the upper limit of the spectrum, so we define $\bar{\varepsilon}'_\gamma=\varepsilon'_{\text{max}}$;
    
    \item  \textbf{Case 4:} $\varepsilon'_\gamma n(\varepsilon'_\gamma)$ is constant over an energy interval from $\varepsilon'_{\text{min}}$ to $\varepsilon'_{\text{b}}$ (e.g. a power law $n(\varepsilon'_\gamma)\propto\varepsilon^{'-\alpha}_\gamma$ with $\alpha=1$). There is no well-defined maximum, and thus there is no single energy contributing most to the neutrino production. We show in App.~\ref{app:thermalmodel} that here the best choice  is $\bar{\varepsilon}'_\gamma=\varepsilon'_{\text{b}}$.
    The reason is that in the (broken) power-law model, the $\Delta$-resonance contribution to the interaction is flat between $\varepsilon'_{\text{min}}$ and $\varepsilon'_{\text{b}}$ (and the multi-pion production only adds a correction on logarithmic scales), whereas in the thermal model, that flatness is reproduced by the multi-pion interaction rate being constant beyond threshold -- which is determined by the upper end of the photon energy range $\varepsilon'_{\text{b}}$. Note that for GRBs, magnetic field effects render any small remaining difference practically invisible -- as we will demonstrate later. 
\end{itemize}
These behaviors are identified in Fig.~\ref{targetcases}, where we show benchmark examples of the target photon number density for each case. The energy scales for the spectra shown in this figure are arbitrary and are chosen solely for graphical simplicity. For case 4, we report both a broken power-law spectrum (solid blue) and the parameterization given by the Band function (dotted blue) in Eq.~\ref{eq:bandfunction}: the break energy and the spectral indices $\alpha$ and $\beta$ are the same for both curves.

In the thermal model we now introduce a fictitious black-body number density peaked at the same energy $\bar{\varepsilon}'_\gamma$;  the black-body photon spectrum with an effective temperature $T'$ is given by
\begin{equation} \label{eq:thermalphoton}
    n^{\text{th}}_\gamma(\varepsilon'_\gamma) \text{d}\varepsilon'_\gamma=\frac{\varepsilon_\gamma^{'2}}{\pi^2} \frac{1}{e^{\varepsilon'_\gamma/T'}-1}\text{d}\varepsilon'_\gamma.
\end{equation}
The black-body photon number density $\varepsilon'_\gamma n^{\text{th}}_\gamma(\varepsilon'_\gamma)$ is peaked at an energy $2.8 \, T'$, as can be obtained by explicitly differentiating the function and equating the derivative to zero. The effective temperature of the black-body spectrum that simulates the target-photon spectrum $n(\varepsilon'_\gamma)$ is then chosen to be
\begin{equation} \label{eq:tempsource}
    T'=\frac{\bar{\varepsilon}'_\gamma}{2.8}.
\end{equation}
The agreement between the thermal and real neutrino production is of course especially good for target-photon spectra which are very peaked. However, as we will see in the examples below, the model also performs well in reproducing power-law and broken power-law spectra. The main limitation of the model can be found for  photon number densities  $n(\varepsilon'_\gamma)\propto \varepsilon^{'-\alpha}_\gamma$ with $\alpha$ close to one. In App. \ref{app:thermalmodel} we demonstrate that the thermal model reproduces the cases with $\alpha\lesssim 1$ quite well. On the other hand, it performs less accurately for the cases with $\alpha$ slightly larger than one for energies below the neutrino peak energy:  here the spectral index below the peak slightly too hard in the thermal case -- whereas for much larger values of $\alpha$ the decay kinematics of the secondary decays determines the spectral shape.

In our discussion, we focused mostly on examples that use a broken power-law shape for the target photons. However, the arguments given above only depend on identifying a single photon energy $\bar{\varepsilon}'_\gamma$ contributing most to the photohadronic production: therefore, they are valid also for photon spectra which are differentiable or curved. In Sec.~\ref{sec:thermalexamples} we will verify this with an explicit example for the GRB target photon spectrum parameterized as a Band function.

Finally, we comment on the role of the different pion production processes. At threshold,  the neutrino production from $p\gamma$ interactions is dominated by the $\Delta(1232)$-resonance 
\begin{equation}
	p + \gamma \rightarrow \Delta^+ \rightarrow \left\{\begin{array}{lc} n + \pi^+ & \frac{1}{3} \text{ of all cases} \\[0.2cm]  p + \pi^0 & \frac{2}{3} \text{ of all cases} \end{array} \right.  , \label{equ:Delta}
\end{equation}
whereas other processes contribute as well, such as higher resonances, $t$-channel production, multi-pion production, see Refs.~\cite{Mucke:1999yb,Hummer:2010vx} for details. For analytical considerations, it is sufficient to use the $\Delta$-resonance and the multi-pion production to capture the qualitative behavior, which are qualitatively distinguished by a peaked versus almost flat (as a function of center-of-mass energy) cross section, and the number of pions produced.  Especially in the cases close to $n(\varepsilon'_\gamma)\propto\varepsilon^{'-1}_\gamma$, the multi-pion production is the key feature allowing the reproduction of the neutrino flux with the thermal model. In fact, the $\Delta$-resonance in the thermal model always gives rise to neutrino spectra which are quite peaked. On the other hand, the multi-pion contribution, even for thermal or monochromatic targets, gives rise to neutrino fluxes which are not peaked but rather extended in energy, following the spectrum of the parent protons. This effect is frequently not taken into account in the literature, leading, e.g., to misleading estimates of the neutrino peak energy for thermal target-photon spectra; we refer the reader to App.~\ref{app:thermalmodel} for more details. The same effect allows for  a better reproduction of broken power-law target-photon spectra in effective thermal model than what one would naively expect, such as the cases with a flat (in energy) number density.

\subsection{Maximal proton energy model}
\label{sec:Emax}

Throughout this work we assume that protons are injected with a number density per unit energy $N_p(E')\propto E^{'-2} \exp\left[-E'/E'_{p,\text{max}}\right]$. However, the thermal model is valid also for more general choices of the proton spectrum, because it allows to reproduce the interaction rate of the protons (see App. \ref{app:thermalmodel}) which only depends on the photon spectrum.
The value of the maximal proton energy $E'_{p,\text{max}}$ depends on the magnetic field $B'$ and on the dimension $R'$ of the source. We assume that it is fixed by the balancing between synchrotron losses (assuming a cooling timescale $\tau'_{\text{ad}} \simeq R'/c$) and the acceleration timescale, see e.g.~\cite{Hummer:2010ai}.\footnote{Other processes, such as Bethe-Heitler pair production or photo-meson production can affect the maximal energy for high enough target densities as well; however, these can only be modeled using additional parameters.}
The acceleration timescale is taken as $\tau'_{\text{acc}}=E'/(\eta e B')$, where $e$ is the elementary charge, $B'$ is the magnetic field, $E'$ is the energy of the proton and $\eta$ is the acceleration efficiency, hereafter assumed to be $\eta=0.1$ as in Ref.~\cite{Hummer:2010ai}. With this choice the cosmic-rays are accelerated efficiently to high energies, whereas smaller values may be appropriate for certain acceleration mechanisms or sources. For example, for AGNs low values of the efficiency have been used~\cite{Palladino:2018lov} to explain the IceCube neutrinos, while high values could be used to explain the ultra-high energy cosmic-ray spectrum~\cite{Rodrigues:2020pli}.

For weak magnetic fields, the maximal proton energy can be approximated by the Hillas condition
\begin{equation}\label{eq:hillascond}
    E'_{p,\text{max}} \simeq 3\times 10^7\; \text{GeV} \frac{B'}{1\; \rm{G}} \frac{R'}{10^{10}\; \rm{km}}\frac{\eta}{0.1} \, ,
\end{equation}
which also corresponds to the confinement condition for $\eta=1$.
On the other hand, for strong magnetic fields, synchrotron losses are the main mechanism of energy loss: in this case, the maximal proton energy is set by the balance between the acceleration timescale and the timescale for synchrotron losses~\cite{Hummer:2010ai} and it can be approximated as
\begin{equation}\label{maxinmag}
    E'_{p,\text{max}} \simeq 5.9\times 10^{10}\; \text{GeV} \left(\frac{B'}{1\; \rm{G}}\right)^{-1/2}\left(\frac{\eta}{0.1}\right)^{1/2}.
\end{equation}

\begin{figure}[t!]
    \centering
    \includegraphics[width=0.5\textwidth]{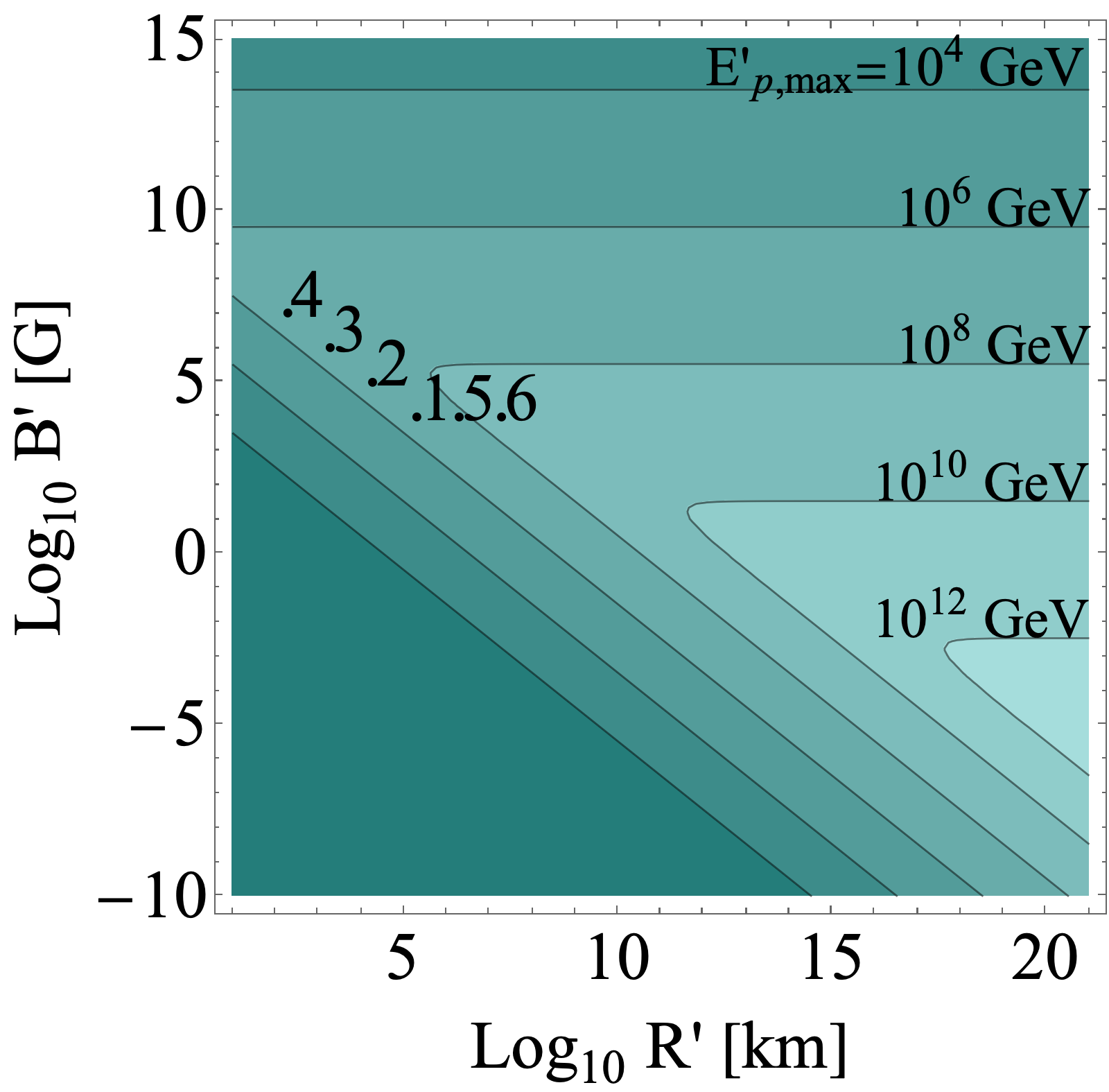}
    \caption{Maximal proton energy in the relativistically comoving frame as a function of $B'$ and $R'$ in our maximal proton energy model for $\eta=0.1$. Test points are identified by the numbers 1 to 6.}
    \label{hillasmax}
\end{figure}

We represent the contour plot of the maximal proton energy in Fig.~\ref{hillasmax} in the Hillas plane~\cite{Hillas:1985is}, namely in terms of the radius and the magnetic field of the source, which is generally used for classifying the cosmic accelerators. Since we show the maximal proton energy in the relativistically comoving frame, it does not depend on the Doppler factor $\Gamma$: it is also independent of the target photons, which are not involved in the acceleration of cosmic-rays. For weak magnetic fields the contours of fixed maximal proton energy are inclined along the directions of constant $B' R'$, as dictated by the Hillas condition in Eq.~\ref{eq:hillascond}. For higher magnetic fields, synchrotron losses limit the energies of cosmic-rays, and the contours of maximal proton energy become horizontal, according to Eq.~\ref{maxinmag}.

Since the maximal proton energy depends on the magnetic field and on the dimension of the source, we can now identify the free parameters of our model:  the Doppler factor $\Gamma$, the effective temperature of the photon spectrum $T'$, the magnetic field $B'$ and the source radius $R'$ (the latter two parameters determine univocally the maximum proton energy $E'_{p,\text{max}}$). Besides determining the maximal proton energy, the magnetic field and the source dimension play also a fundamental role in the neutrino production. After the secondary pions, kaons and muons from $p\gamma$ interactions are produced, they are subject to synchrotron energy losses changing their energy shape. These cooling effects can significantly modify both the energy spectrum and the flavor composition of the produced neutrinos (the effects on the flavor composition are discussed in more detail in App.~\ref{app:flavor}). In our treatment the cooling effects are automatically taken into account in the numerical computation. All the neutrino spectra throughout this work are computed using the software NeuCosmA~\cite{Hummer:2010vx,Hummer:2010ai,Baerwald:2011ee,Hummer:2011ms}. The computation requires the photohadronic cross section for pion production, as well as the decay functions of pions, kaons  and the helicity-dependent decay functions of muons. The neutrons produced in $p\gamma$ interactions are assumed to escape the source and decay while they propagate to Earth; note that here a simple re-distribution parameterization for the neutrinos is used~\cite{Lipari:2007su}, which leads to an inaccurate reproduction of the spectral shape and flavor composition below the neutron peak (where  however the thermal model does not accurately reproduce the flavor composition of the power-law model anyways, see below). The model for the photohadronic cross section is based on a parameterization of the SOPHIA software~\cite{Mucke:1999yb}, while the decay functions are taken from Ref.~\cite{Lipari:2007su}. With this information NeuCosmA proceeds to determine the injection spectrum of secondary pions, muons, kaons and neutrons from the injected proton spectrum; these injection spectra are subsequently used to determine the steady state number densities of secondaries and the injection spectrum of neutrinos.

\subsection{Benchmark examples}\label{sec:thermalexamples}
To exemplify the procedure we study the cases of AGN and GRBs. For both these sources typically a broken power law
\begin{align} \label{eq:brokenpower}
    n_\gamma(\varepsilon'_\gamma)\propto\begin{cases} (\varepsilon'_\gamma/\varepsilon'_\text{b})^{ -\alpha}, \hspace{0.2cm} & \varepsilon'_{\text{min}} \leq \varepsilon'_\gamma\leq\varepsilon'_{\text{b}} \\ (\varepsilon'_\gamma/\varepsilon'_\text{b})^{-\beta},\hspace{0.2 cm} & \varepsilon'_{\text{b}} < \varepsilon'_\gamma\leq \varepsilon'_{\text{max}} \end{cases}
\end{align}
is assumed,
where the minimal, break, and maximal energy, $\varepsilon'_{\rm min}$, $\varepsilon'_{\rm b}$ and $\varepsilon'_{\rm max}$, as well as the spectral indices $\alpha$ and $\beta$, are left as free parameters. The choice of these parameters for the two benchmark cases is shown in Table~\ref{tabpara}, as well as the numerical values of the effective temperatures used for the reproduction with the thermal model.

For the case of GRBs, we also provide a different parameterization for the spectrum using the Band function~\cite{Band:1993eg}, defined as
\begin{equation}\label{eq:bandfunction}
n_\gamma(\varepsilon'_\gamma)\propto\begin{cases} (\varepsilon'_\gamma/\varepsilon'_\text{b})^{ -\alpha} e^{-\varepsilon'_\gamma/\varepsilon'_\text{b}}, \hspace{0.2cm} & \varepsilon'_{\text{min}} \leq \varepsilon'_\gamma\leq\varepsilon'_{\text{b}}(\beta-\alpha) \\ (\varepsilon'_\gamma/\varepsilon'_\text{b})^{-\beta} (\beta-\alpha)^{\beta-\alpha} e^{\alpha-\beta},\hspace{0.2 cm} & \varepsilon'_{\text{b}}(\beta-\alpha) < \varepsilon'_\gamma\leq \varepsilon'_{\text{max}} \end{cases}
\end{equation}
in which we use the same values for the spectral indices reported in Table~\ref{tabpara} for the GRB case, and a break energy of $\varepsilon'_\text{b}=37$~keV. The use of the Band function allows us to simulate curved photon spectra with continuous derivatives as well, in place of the broken power laws which have discontinuous derivatives at the break energy.

\begin{figure}[t!]
    \centering
    \includegraphics[width=0.45\textwidth]{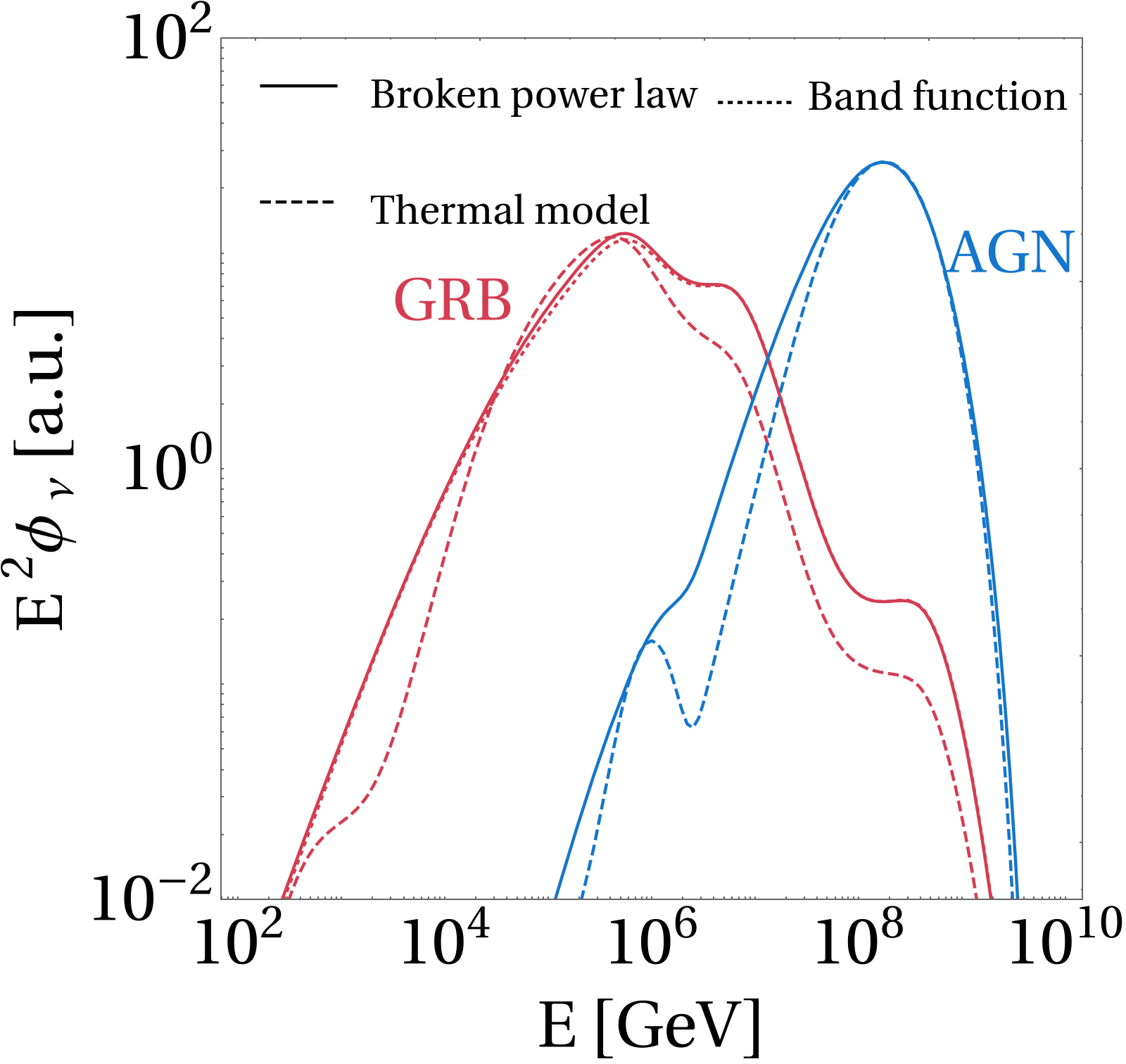}
    \includegraphics[width=0.445\textwidth]{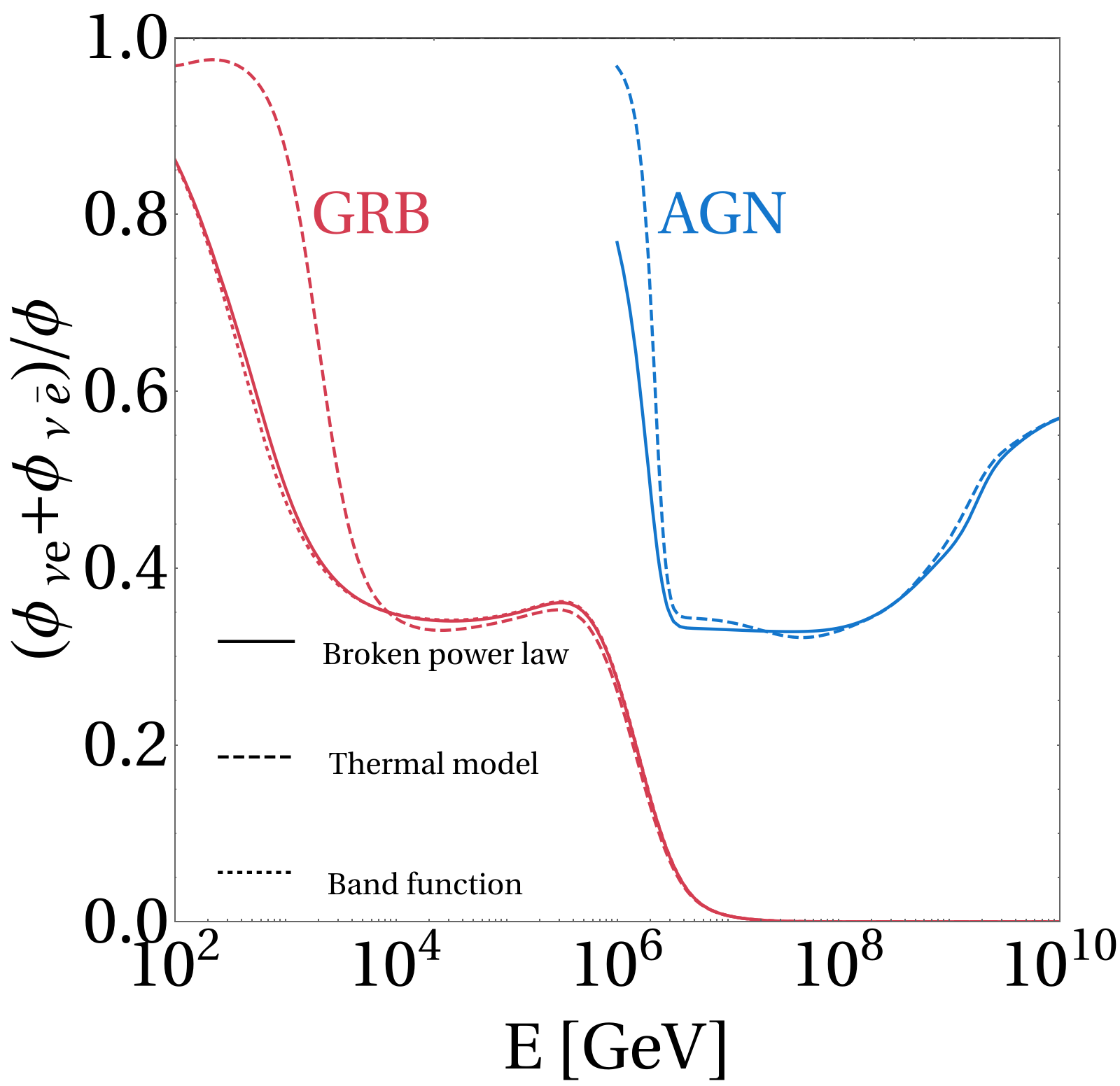}
    \caption{Comparison between the benchmark astrophysical neutrino fluxes, parameterized as broken power laws, and their reproduction using the thermal model. In the left panel we show the comparison for the neutrino fluxes produced in AGN and GRBs. We show the all-flavor neutrino flux $E^2 \phi$ in GeV~cm$^{-2}$ s$^{-1}$ (normalization in arbitrary units) as a function of the energy in the observer rest frame. In the right panel we show the fraction of electron neutrinos and antineutrinos in the differential flux at the source as a function of the energy in the observer rest frame, in the energy region in which the flux is at least $1/1000$ of its peak value. The solid curves are obtained from the astrophysical broken power-law source model, the dashed curves are obtained with the thermal model, the dotted curve is obtained from the Band function model for the GRB target photon spectrum. The parameters used in the generation of the astrophysical neutrino spectra are summarized in Table~\ref{tabpara}.}
    \label{thecomparison}
\end{figure}

The neutrino fluxes, both for the broken power-law model of the astrophysical source and for the thermal model, are shown in the left panel of Fig.~\ref{thecomparison}. We also show, as a dotted line, the Band function model of our benchmark GRB. We denote the neutrino flux observed at Earth as the number of particles per unit time per unit energy per unit surface $\phi_{\nu\alpha}=\frac{\text{d}N_{\nu\alpha}}{\text{d}t\text{d}E\text{d}S}$, where $\alpha=\left\{e,\mu,\tau,\overline{e},\overline{\mu},\overline{\tau}\right\}$ and $\phi_\nu=\sum_\alpha \phi_{\nu\alpha}$; since we are dealing with point sources, we do not consider the differential flux per unit solid angle. The main features are well captured by the thermal model near the peak of the spectrum in both cases. An interesting point to emphasize is that, in the case of GRBs, the accuracy of reproduction is strictly connected with the multi-pion production. In fact, if only the $\Delta$-resonance were taken into account, the thermal model would lead to a poor reproduction of the flux. We refer the reader to App.~\ref{app:thermalmodel} for more details on the multi-pion production and its role within the thermal model. The neutrino spectrum obtained using the Band function for the target photons is substantially identical to the spectrum obtained using the broken power law, and only very small differences can be observed near the peak of the spectrum due to the slight difference between the broken power law approximation and the curved, differentiable spectrum (this difference is evident between the solid and dotted blue lines in Fig.~\ref{targetcases}).

We also compare, in the right panel of Fig.~\ref{thecomparison}, the flavor composition obtained using the thermal model with the broken power-law flavor composition for the two benchmark examples for AGN and GRBs. In addition, we show the result for the Band function model of the benchmark GRB; the corresponding curve is practically superimposed on the solid line for the broken power-law model. As in the case of the all-flavor neutrino flux, the flavor composition also shows only very small changes in the region corresponding to the peak of the neutrino flux, which are related to the small changes in the target photon spectrum near the break energy. Since only electron and muon neutrinos are produced at the source, it is sufficient to represent the fraction of electron neutrinos at the source to capture the entire flavor structure\footnote{In this way we however lose the information on the ratio of neutrinos and antineutrinos.}. In both cases we find that the flavor composition is well reproduced in the high energy range. Only at low energies, far from the peak of the spectrum, the fraction of electron neutrinos and antineutrinos is not well reproduced by the thermal model.

\begin{table}[t]
    \centering
    \begin{tabular}{l|rrrrrrrrrr}
     \hline
       Source  & $\Gamma$ & B' [G] & R' [km] & $\varepsilon'_{\text{min}}$ [eV]& $\varepsilon'_{\text{b}}$ [keV]& $\varepsilon'_{\text{max}}$ [keV]& $\alpha$ & $\beta$ & T' [eV]\\
       \hline
       AGN & $10$ & $1$ & $10^{11}$ & - & $3\times10^{-5}$ & $1$ & $1$ & $2.64$ &  $0.2$\\
       GRB & $100$ & $3\times 10^5$ & $10^7$ & 0.3 & $14.8$ & $1000$ & $1$ & $2$ & $5300$ \\
       TDE & $10$ & $90$ &  $10^9$ & -& - & -& -& -& $840$ \\
       \hline
    \end{tabular}
    \caption{Summary of numerical values for the parameters used for the benchmark spectra shown in this work. For AGN we extract the parameters of the target-photon spectrum by a broken power-law fit to the low energy spectrum in Ref.~\cite{Gao:2016uld}; the astrophysical parameters $\Gamma$, $B'$ and $R'$ are fixed to typical values for AGN jets in Ref.~\cite{Hummer:2010ai}. For GRB the parameters of the target-photon spectrum and the magnetic field are taken from the reference values of Ref.~\cite{Baerwald:2011ee}, the Doppler factor is fixed to a typical value and the size of the accelerating region is taken to be the shell thickness $R'=\frac{\Gamma t_v}{1+z}$ with $t_v=0.3$~s being the variability timescale and $z\ll 1$. For TDE the parameters of the target-photon spectrum and the magnetic field are taken from Ref.~\cite{Winter:2020ptf}; the Doppler factor is fixed to a typical value and the size of the accelerating region is taken to be the shell thickness $R'=\frac{\Gamma t_v}{1+z}$ with $t_v=300$~s and $z\ll 1$.}
    \label{tabpara}
\end{table}

In order to understand the reason for the different behavior in the high and low energy regime of Fig.~\ref{thecomparison}, let us recall that in $p\gamma$ sources the flavor structure is characterised typically by transitions between different regimes: the \textit{pion-beam} regime (flavor composition (1:2:0) at the source), in which neutrinos are produced by the full decay chain of pions and muons; the \textit{muon-damped} regime (flavor composition (0:1:0) at the source), in which neutrinos are produced only by pion decays; the \textit{muon-beam} regime (flavor composition (1:1:0) at the source), in which neutrinos are only produced by muon decays; the \textit{neutron-beam} regime (flavor composition (1:0:0) at the source), in which neutrinos from neutron decays dominate at low energies. The properties of these different regimes is discussed in more detail in App.~\ref{app:flavor}. The transitions between pion-beam and muon-damped or muon-beam regime depend mainly on the magnetic field. Therefore these transitions are well reproduced by the thermal model, where only the shape of the photon spectrum is changed while the properties of the source do not change. On the other hand, the transition to the neutron-beam regime is, due to the presence of a bump of neutrinos from the neutron decay, well visible, for example, in the AGN case (first peak at low energy of the dashed curve) of the left panel of Fig.~\ref{thecomparison}. This bump (relative to the pion chain spectra) is the source of the mismatch between the flavor composition of the thermal model and the broken power-law model (a deeper discussion can be found in App.~\ref{app:flavor}).

In concluding this section, we note that the thermal model is useful to simulate neutrino production from generic astrophysical sources in an approximate way. There are in fact sources in which the target-photon spectrum is already by itself well represented by a thermal shape: an example of these are some TDEs (whereas some jetted TDEs exhibit non-thermal spectra). We will use these as benchmark astrophysical sources in Sec.~\ref{sec:resul}. The parameters of TDEs are also given in Table~\ref{tabpara}.

\section{Study of experimental sensitivity to astrophysical sources}
\label{sec:resul}

Within the thermal model introduced above, in conjecture with the maximal proton energy derivation used, the neutrino fluxes $\phi$ can be described by the parameters
\begin{equation}\label{eq:parameters}
    T',\; B',\; R',\; \Gamma,
\end{equation}
where $T'$ is the effective temperature of the target photons, $B'$ is the magnetic field, $R'$ is the source dimension and $\Gamma$ is the Doppler factor. Note that the acceleration efficiency $\eta$, which we fix to 0.1 in the following, is in principle another parameter of the model. 
As a first application of this model, in this section we study the dependence of the sensitivity of different neutrino experiments on these parameters. 

We can roughly divide the experiments potentially of interest for astrophysical neutrino detection according to the energy range in which they are most sensitive. For each class we identify an experiment representative studied as a benchmark for a certain energy range:
\begin{itemize} 
\item \textit{Dense neutrino arrays}, between $1$~GeV and $10^5$~GeV (e.g. PINGU, ORCA, DeepCore);
\item \textit{Neutrino telescopes}, between $10^5$~GeV and $10^7$~GeV (e.g. IceCube, KM3NeT, Antares); 
\item \textit{Neutrino radio arrays}, between $10^7$~GeV and $10^{12}$~GeV (e.g. ARIANNA, IceCube-Gen2 Radio Array, GRAND).
\end{itemize}
We choose as representatives DeepCore~\cite{Schulz:2009zz} for dense neutrino arrays, KM3NeT~\cite{Kappes:2007ci} for neutrino telescopes and IceCube-Gen2 Radio Array~\cite{Blaufuss:2015muc} for neutrino radio arrays, respectively. Our choices are only benchmark ones: we have tested other possibilities finding very similar results. In particular, we choose DeepCore since it has a better sensitivity above 100~GeV, and KM3NeT and IceCube-Gen2 Radio Array for their slightly larger effective areas compared to the others on the lists; see also Ref.~\cite{Song:2020nfh} for effective volumes of future neutrino telescopes.

Since the event rate is proportional to the exposure, it is useful to introduce in place of the flux $\phi_{\nu}$ the fluence $\mathcal{F}_\nu$, which is the flux times the observation time  $\mathcal{F}_{\nu\alpha}=\phi_{\nu\alpha} t$ for a steady flux (in units of GeV$^{-1}$~cm$^{-2}$).  In order to quantify the sensitivity for a given choice of the astrophysical parameters, we determine the normalization of the fluence  requiring that the expected number of events is 2.44; this number corresponds to the background-free Feldman-Cousins $90\%$ sensitivity limit \cite{Feldman:1997qc}. 
The computation of the number of events from the fluence is discussed in more detail in App.~\ref{app:methods}. We use the background-free limit for detection in the main text, whereas especially for the low energy experiments, such as the dense neutrino arrays, the presence of a background of atmospheric neutrinos can be significant -- see App.~\ref{app:methods} for the generalization to that case. Furthermore, note that for transients, the total fluence from individual sources (such as GRBs) with fluences  $\mathcal{F}_{\nu,i}$ is simply given by $\mathcal{F}_\nu=\sum_i \mathcal{F}_{\nu,i}$, whereas  the flux picture requires the concept of a quasi-diffuse flux -- such as in the GRB stacking analysis in Ref.~\cite{Abbasi:2009ig}.\footnote{The quasi-diffuse flux $\phi^{\mathrm{QD}}_\nu$ can be obtained from the total rate of transients as the product $\phi^{\mathrm{QD}}_\nu=\mathcal{F}_{\nu,i} \cdot \dot{N}^\mathrm{tot}$ if all transients are alike (and at the same redshift). For a more detailed computation, the redshift and luminosity distributions of the sources have to be taken into account. The number $\dot{N}^\mathrm{tot}$ can, in fact, also  be translated into the local (apparent) source density by a factor describing the redshift evolution, i.e., how representative the local density is for the whole sample, see Ref.~\cite{Baerwald:2014zga}. }

The question is now from the astrophysical perspective: how does that sensitivity translate into a physical constraint of the source energetics, such as the required cosmic-ray energy injected into the source? 
Since neutrinos originate mostly from pion decays, the emitted neutrino energy $E_\nu$ in the optically thin case is directly proportional to the injected (non-thermal) energy of protons $E_p$ and the fraction of the proton energy going into pion production, generally known as the ``pion-production efficiency'' $f_\pi$, which means that $E_\nu \propto f_\pi E_p$.  Whereas $E_p$ is often a free parameter connected with the X-ray or gamma-ray luminosity through a quantity called ``baryonic loading'', $f_\pi$ strongly depends on the size, geometry and Lorentz factor of the production region. 
For this reason, an interpretation in terms of the source parameters and of $E_p$ goes beyond the scope of this study. However, we can interpret the sensitivity in terms of the energy injected in neutrino $E_\nu$. 
We relate this to the (integrated) energy fluence detected in neutrinos, which we define as
\begin{equation} \label{eq:defxi}
    \xi=\int E \,  \mathcal{F}_\nu \text{d} E \, .
\end{equation}
For a point source at redshift $z$ one has (an isotropic-equivalent) $E_{\nu}=4\pi d_L^2(z)/(1+z) \, \xi$, where $d_L(z)$ is the luminosity distance.\footnote{The luminosity distance is given by $d_L(z)=(1+z)\int_0^z \frac{1}{H(t)} \text{d}t$ in a flat universe in natural units. Here $H(t)$ is the Hubble parameter, which we assume to vary with redshift as 
$H(z)=H_0\sqrt{\Omega_\Lambda+\Omega_m (1+z)^3}$
where $H_0=100$ $h$ km s$^{-1}$ Mpc$^{-1}$, $h=0.674$, $\Omega_\Lambda=0.685$ and $\Omega_m=0.315$ \cite{Zyla:2020zbs}.} Consequently, one roughly has $\xi\propto f_\pi E_p$ for a fixed redshift. Therefore, $\xi$ can be considered as a measure for the pion-production efficiency times the energy injected into protons into the source.

\begin{figure}[t!]
    \centering
    \includegraphics[width=0.45\textwidth]{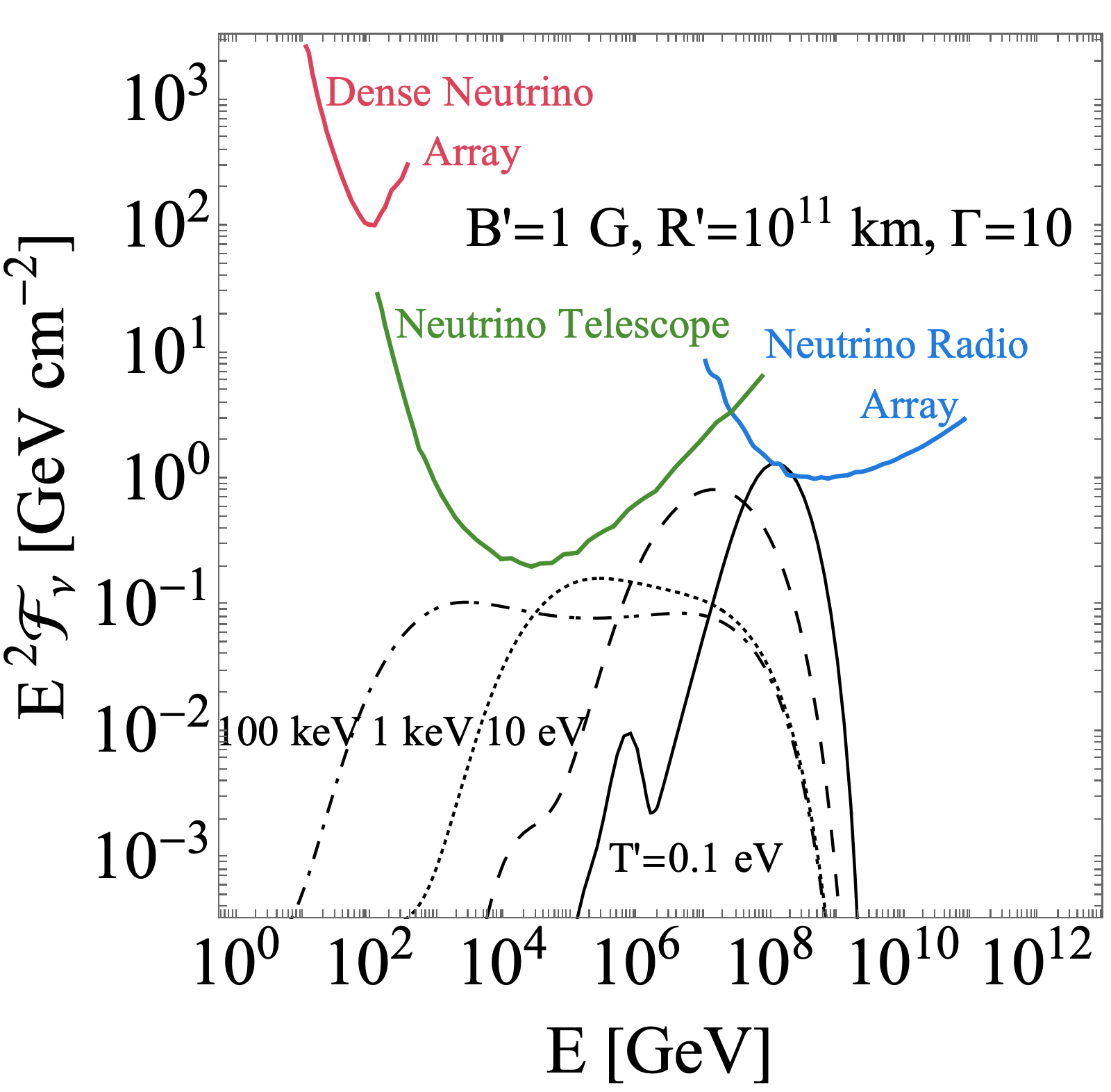}
    \includegraphics[width=0.45\textwidth]{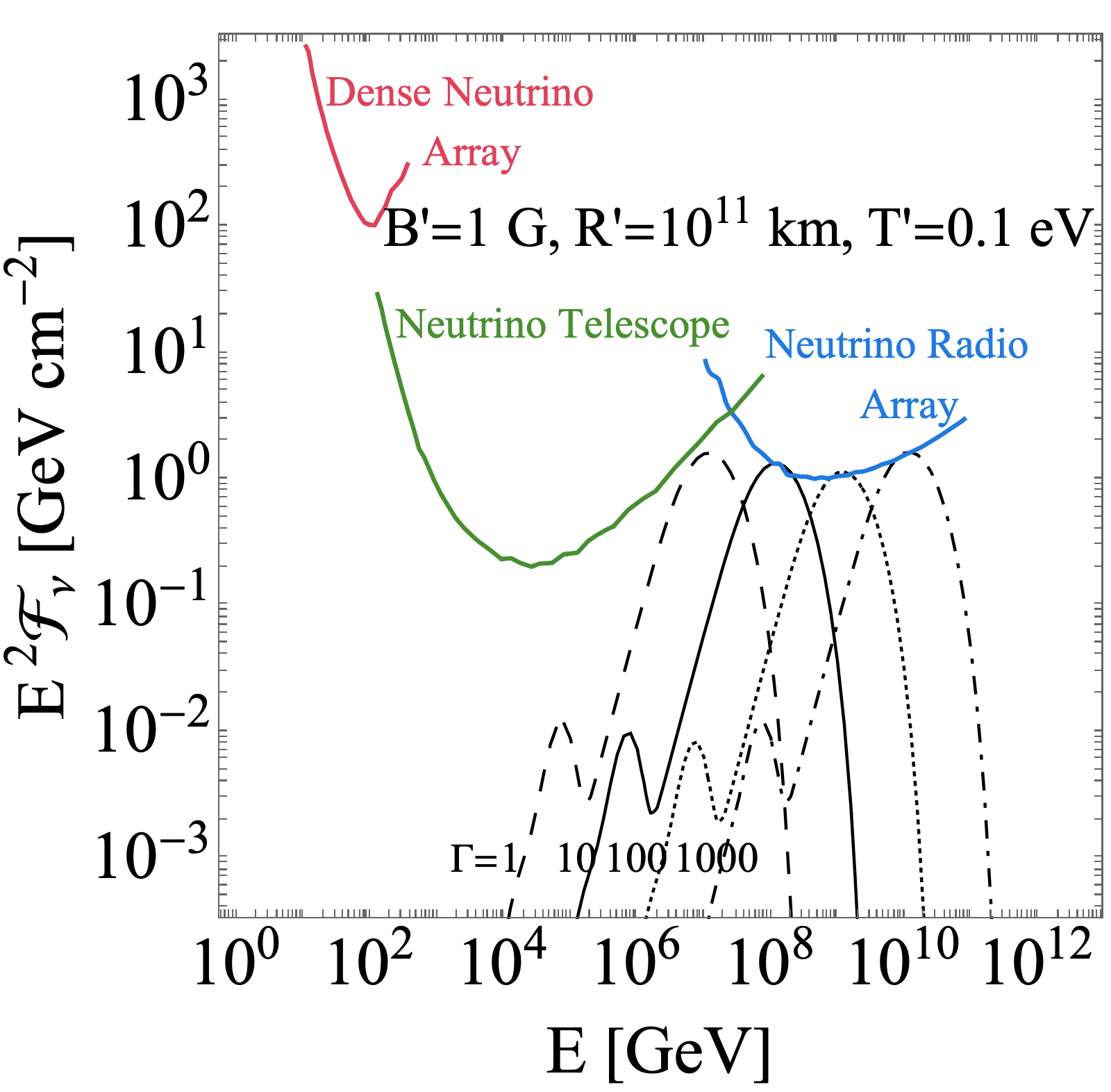}
    \includegraphics[width=0.45\textwidth]{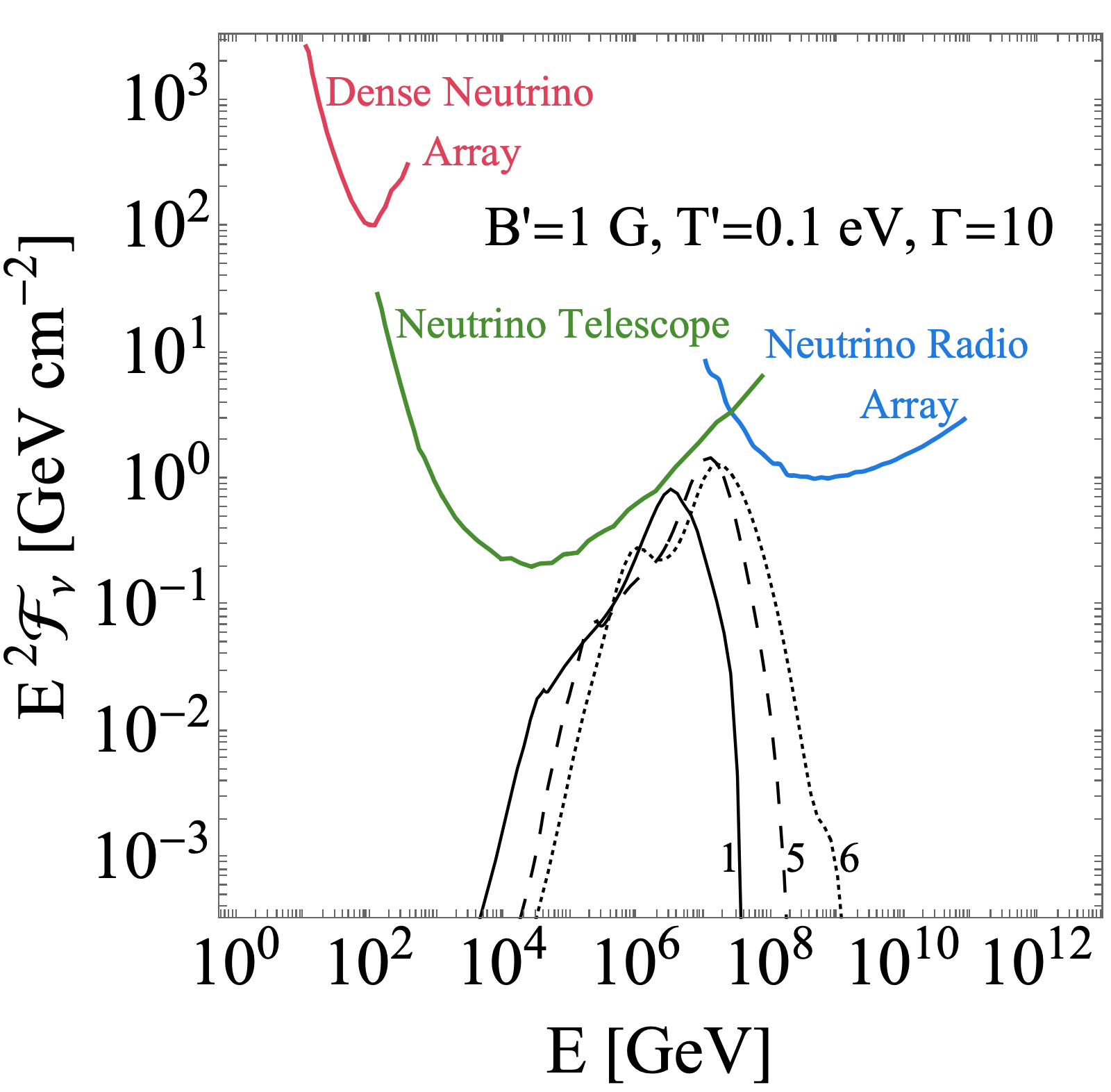}
    \includegraphics[width=0.45\textwidth]{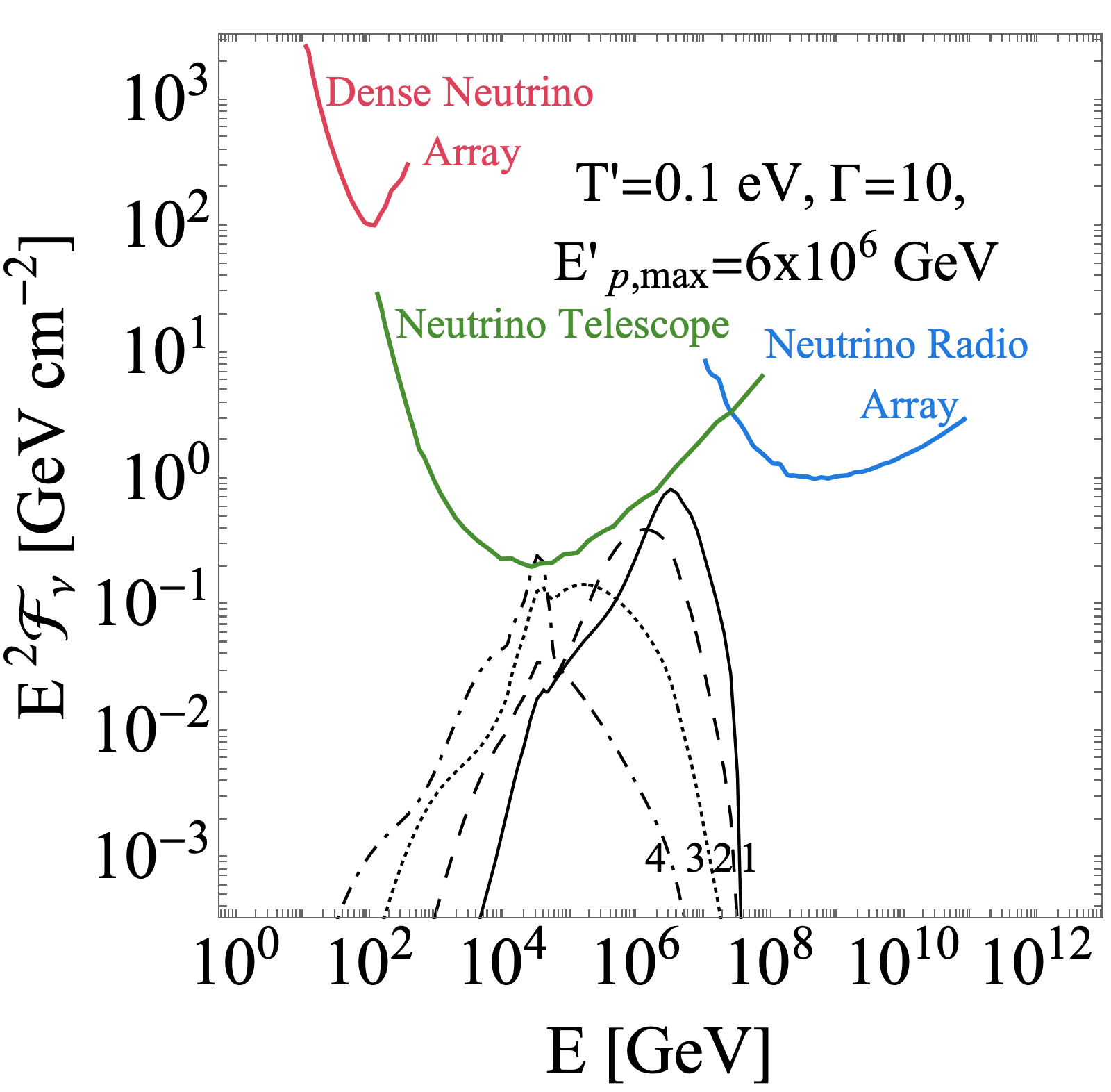}
    \caption{Experimental sensitivities for different model parameters. The figures show the differential limits (colored curves) and the sensitivities to certain benchmark fluences (black curves) as a function of (observed) neutrino energy, where the curves refer to all flavors. 
    In the upper left panel we show different choices of the effective temperature $T'$, in the upper right panel we show different choices of $\Gamma$. In the lower left and right panels, we show curves for different maximal proton energies and magnetic fields, respectively, referring to the test points identified in Fig.~\ref{hillasmax}. The fluences are normalized so that the sum of the expected events at all three experiments together is 2.44. }
    \label{varyagn}
\end{figure}

We show in Fig.~\ref{varyagn} the differential limits of different experiments (in  colors, see App.~\ref{app:methods} for the definition of the differential limits) in comparison to selected model sensitivities for different parameter values of $T'$ (upper left panel), $\Gamma$ (upper right panel), maximal proton energy (lower left panel), and magnetic field (lower right panel). Lowering the effective temperature in the top left panel of Fig.~\ref{varyagn} leads to spectra which are more peaked near their maximal energies and are more easily detected by neutrino radio arrays. At higher effective temperatures instead a nearly flat region, originating from the multi-pion contribution, can be identified; these neutrinos are more easily detected by neutrino telescopes.
Raising the Doppler factor $\Gamma$ in the top right panel of Fig.~\ref{varyagn} simply moves proportionally the spectrum to higher energies, so that high Doppler factors are associated with detection at neutrino radio arrays.

In order to show the effect of changing the maximal proton energy and the magnetic field separately, we have chosen a series of test points highlighted in Fig.~\ref{hillasmax}: the test points 1, 5 and 6 are chosen at a fixed magnetic field, so that the fluence only changes because of the change in $E'_{p,\text{max}}$, whereas the test points 1, 2, 3 and 4 are chosen along lines of constant maximal proton energy $E'_{p,\text{max}}$, so that the fluence only changes because of the varying magnetic field\footnote{The test points are conveniently chosen in the region of the Hillas plane in which most of the sources appear (see Ref.~\cite{Hummer:2010ai}) and with magnetic fields sufficiently large to exhibit non-trivial effects.}. The corresponding fluences are shown in the lower two panels of Fig.~\ref{varyagn}.

By moving to higher maximal proton energies, as in the lower left panel of Fig.~\ref{varyagn}, the upper cutoff of the neutrino spectrum correspondingly increases. On the other hand, if we change the magnetic field for a fixed constant proton energy, as in the lower right panel of Fig.~\ref{varyagn}, a steepening of the spectrum in the highest energy part of the spectrum appears, caused by synchrotron losses. The effects of the magnetic field on the flux are even more evident on the flavor composition as a function of the energy: we will discuss this point in the next section.

\begin{figure}[t!]
    \centering
    \includegraphics[width=0.45\textwidth]{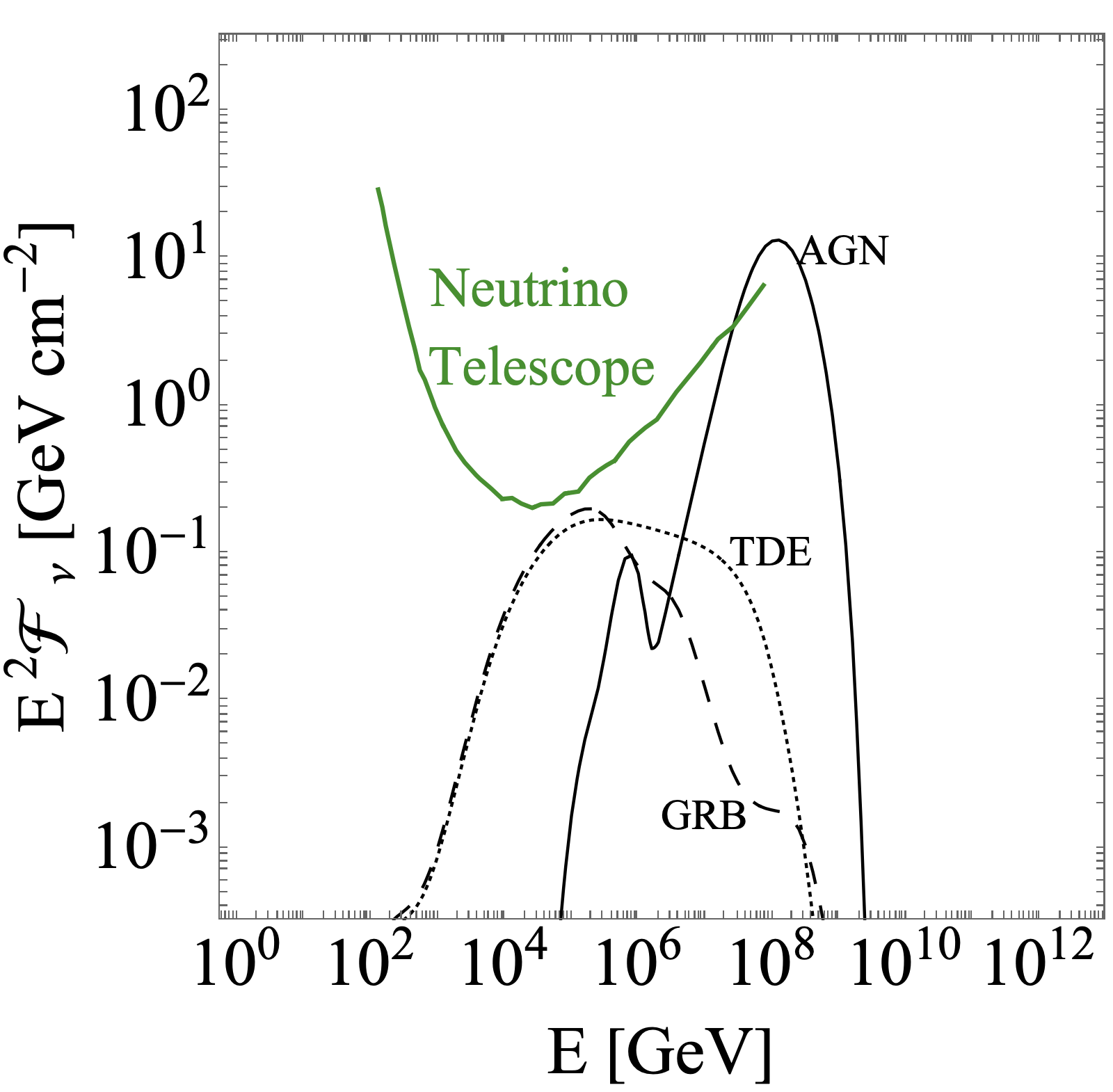}
    \includegraphics[width=0.45\textwidth]{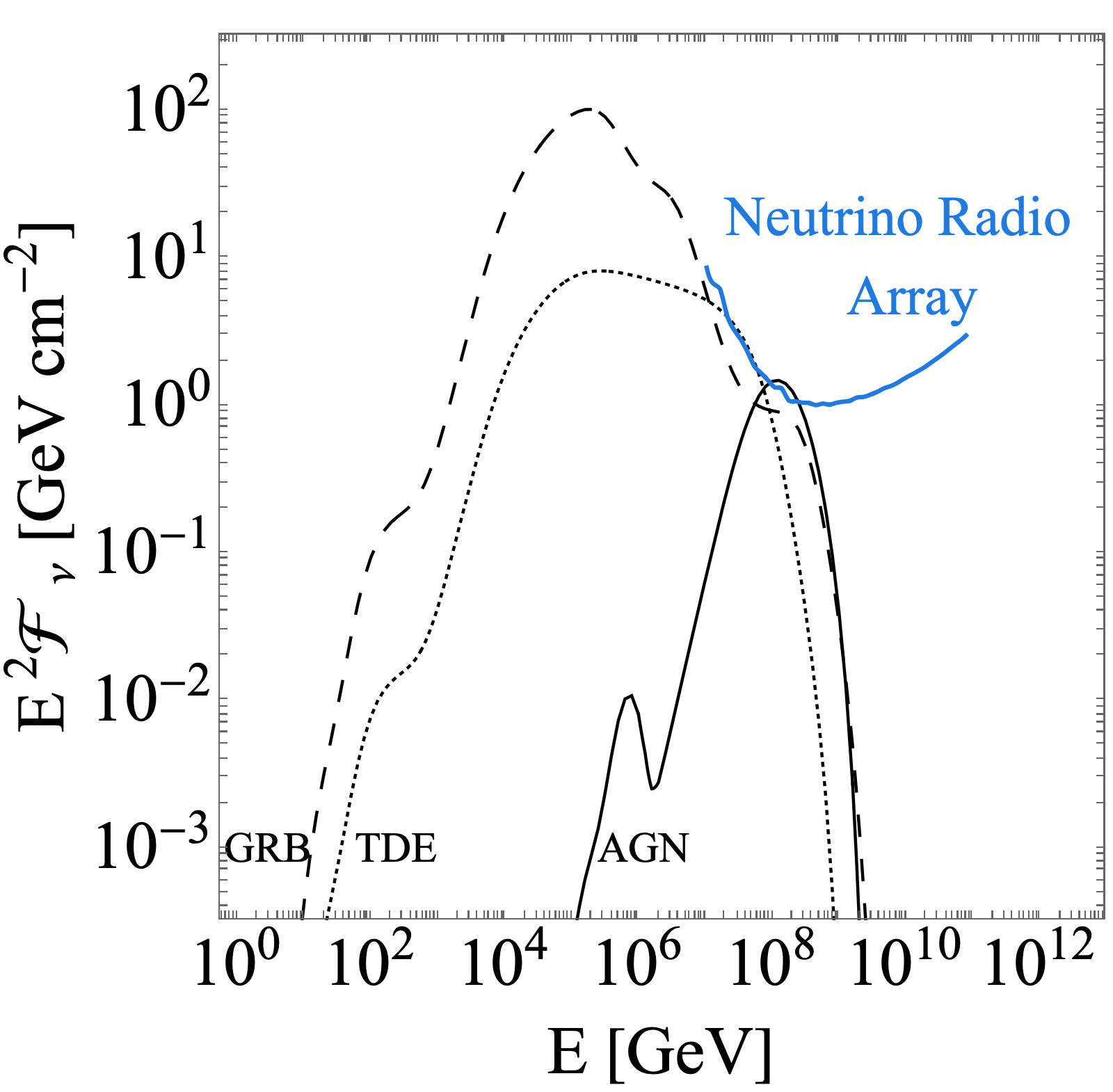}
    \caption{Experimental sensitivity to neutrino fluences from benchmark sources. In the left (right) panel we show the differential limit of our benchmark Neutrino Telescope (Neutrino Radio Array) in color, together with the sensitivities $E_\nu^2 \mathcal{F}_\nu$ in GeV~cm$^{-2}$ for sources with a thermal photon spectrum simulating GRBs, AGN and TDEs; all the spectra are normalized to the 90\% Feldman-Cousins limit for signal detection at the corresponding experiment. The numerical values for the parameters are reported in Table~\ref{tabpara}.}
    \label{comptest}
\end{figure}

We show the sensitivities of the experiments to the benchmark sources of Table~\ref{tabpara} as reproduced by our thermal model in Fig.~\ref{comptest}, where the sensitivity limits are represented together with the differential limits. 
We also give in Table~\ref{tabparaxi} the explicit values of $\xi$ needed for detection at neutrino telescopes and neutrino radio arrays\footnote{We do not compare these sources with dense neutrino arrays since the energy ranges are outside of the range of sensitivity of these experiments.}.

\begin{table}[t]
    \centering
    \begin{tabular}{l|rr}
     \hline
       Source  & $\xi$ at Neutrino Telescope $[\rm{erg\; cm^{-2}}]$ & $\xi$ at Neutrino Radio Array $[\rm{erg\; cm^{-2}}]$\\
       \hline
       AGN & $4.5\times 10^{-2}$ & $5.2\times 10^{-3}$\\
       GRB & $1.2\times 10^{-3}$ & $6.2\times 10^{-2}$ \\
       TDE & $1.7\times 10^{-3}$ & $8.3\times 10^{-2}$ \\
       \hline
    \end{tabular}
    \caption{Summary of the minimal energy fluences $\xi$ necessary for detection at different experiments for the benchmark spectra (reproduced in the thermal model) shown in this work.}
    \label{tabparaxi}
\end{table}
From the values of $\xi$, as well as from visual inspection of Fig.~\ref{comptest}, we can gather that the detection of AGN by neutrino radio arrays requires a smaller normalization, and therefore a smaller value of the energy fluence injected in neutrinos, compared to the detection by neutrino telescopes. We have the opposite situation in the case of GRB and TDE, which require a much higher normalization and energy fluence, by two or three orders of magnitude, for detection at neutrino radio arrays compared to neutrino telescopes.

\begin{figure}
    \centering
    \includegraphics[width=0.5\textwidth]{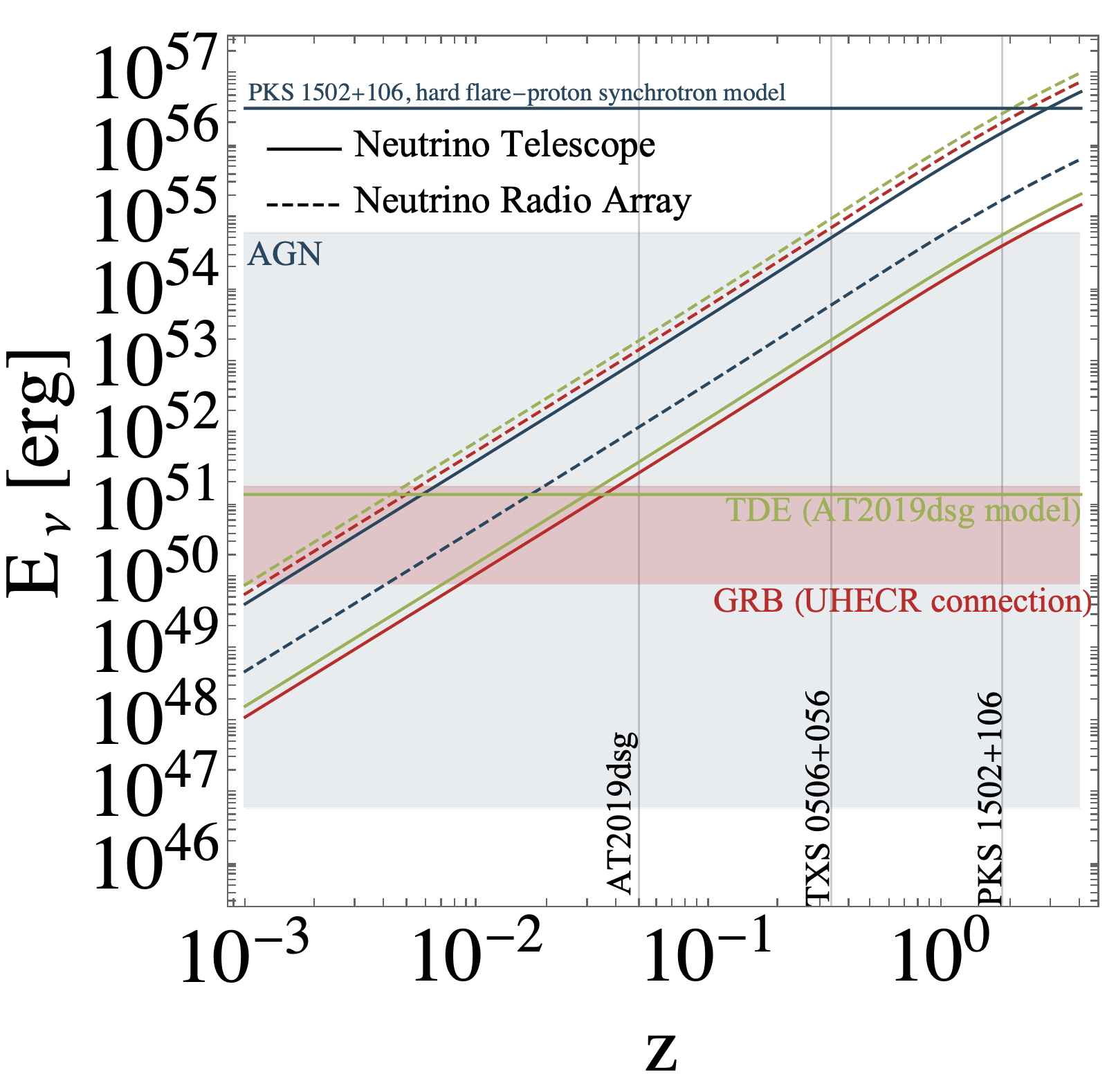}
    \caption{Isotropic-equivalent energy $E_\nu$ in neutrinos (all flavors)  required to observe {\bf one}  neutrino event as a function of redshift $z$. The different curves refer to neutrino telescopes (solid) and neutrino radio arrays (dashed), and the different benchmark sources are identified by different colors (AGN: blue, TDE: green, GRB: orange). We show  the redshifts of the observed sources AT2019dsg (see Ref.~\cite{Winter:2020ptf}), TXS~0506+056 (see Ref.~\cite{Gao:2018mnu}) and the blazar PKS~1502+106 (see Ref.~\cite{Rodrigues:2020fbu}) with vertical lines. Theoretically motivated  values for $E_\nu$ are identified with horizontal lines or shaded areas: for TDE we integrate the muon neutrino fluence from the AT2019dsg model in Fig.~2 of Ref.~\cite{Winter:2020ptf} to obtain the total energy output (we take into account a factor of three to obtain the all-flavor energy output); for GRB we show the range from the benchmark GRBs reported in Table~1 of Ref.~\cite{Heinze:2020zqb}, chosen to describe the UHECR origin as shaded area; for AGN we show
    the output energy expected in a 3.7~years hard flare of blazar PKS~1502+106 according to the proton synchrotron model in Fig.~2 of Ref.~\cite{Rodrigues:2020fbu}. We also show the range of the neutrino energy output per average year for the different benchmark AGN in the Normalisation~A model (with high maximal proton energy) in Table~6 of Ref.~\cite{Oikonomou:2019djc} as shaded area (multiplying by a factor of three to obtain the all-flavor energy output).}
    \label{fig:injectedenergyredshift}
\end{figure}

For an astrophysical discussion, it is of interest to compare the total energy $E_\nu$ that needs to be ejected in neutrinos for observation of one neutrino events at our benchmark experiments with theoretical expectations for these astrophysical sources.  As discussed earlier, that energy depends on redshift, which is why we show it as a function of redshift in Fig.~\ref{fig:injectedenergyredshift}.
In that figure,  the benchmark astrophysical spectra are identified by different colors, and the two benchmark experiments in Table~\ref{tabparaxi} are identified by different line styles. The total energy ejected in neutrinos increases with the redshift because of the geometrical factor $4\pi d_L^2(z)$, so that higher injected energies are necessary for observation at higher redshifts. Comparing the energy required for neutrino telescopes and neutrino radio arrays, we recover the result already stated above that for GRBs and TDEs neutrino telescopes require less energy, since the solid line is below the dashed line, whereas for AGN the opposite is true as our AGN spectra are chosen to have a peak at ultra-high energies. 

The total energies ejected into neutrinos are also compared with typical values for these astrophysical sources, identified by a horizontal line for TDE (assuming a model for  AT2019dsg~\cite{Winter:2020ptf}) and by horizontal bands for GRB and AGN with UHECR connection. If the required energy for detection exceeds these typical values, the source is not expected to be detected by the corresponding experiment. This means that for each source type at each benchmark experiment there will be a maximum redshift beyond which they will not be detected.  At the redshift of AT2019dsg, the required energy for TDE observation at neutrino telescopes is slightly larger than the value in Ref.~\cite{Winter:2020ptf}: this is consistent since there the number of expected neutrinos was 0.26. For AGN we also provide a comparison with the output energy obtained for PKS~1502+106 in Ref.~\cite{Rodrigues:2020fbu}.  
PKS~1502+106 is a particularly bright source and is very far away (redshift 1.84), and has a very large super-massive black hole (and consequently a large Eddington luminosity). The required energy output for detection at neutrino radio arrays (blue dashed line) is below the predicted energy output, suggesting that the source could be detected by experiments aimed at detection of ultra-high energy neutrinos. Another source potentially of interest for comparison with our model would of course be TXS~0506+056. However, such a comparison would not be straightforward because we are using an AGN model with high acceleration efficiency and ultra-high neutrino energies, whereas the TXS~0506+056 neutrinos have been detected in the IceCube energy range (see, however,  Suppl. Fig.~6 in Ref.~\cite{Gao:2018mnu} for a model with an ultra-high neutrino energy). For this reason, we do not show separately the typical energies for TXS-like AGN models in Fig.~\ref{fig:injectedenergyredshift}.

\begin{figure}[t]
    \centering
   \includegraphics[width=0.3\textwidth]{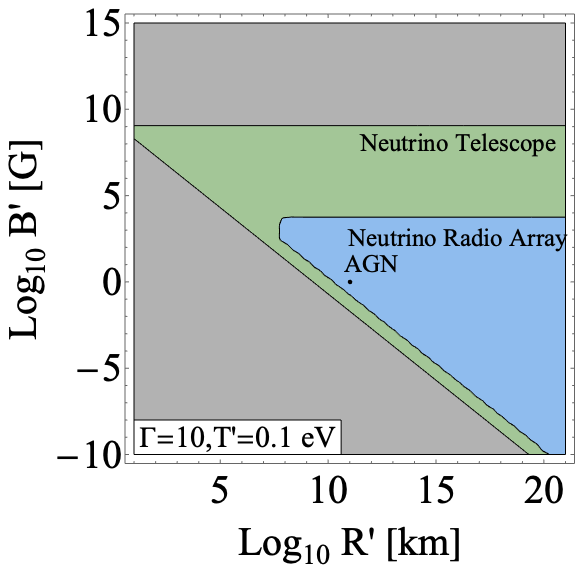}
   \includegraphics[width=0.3\textwidth]{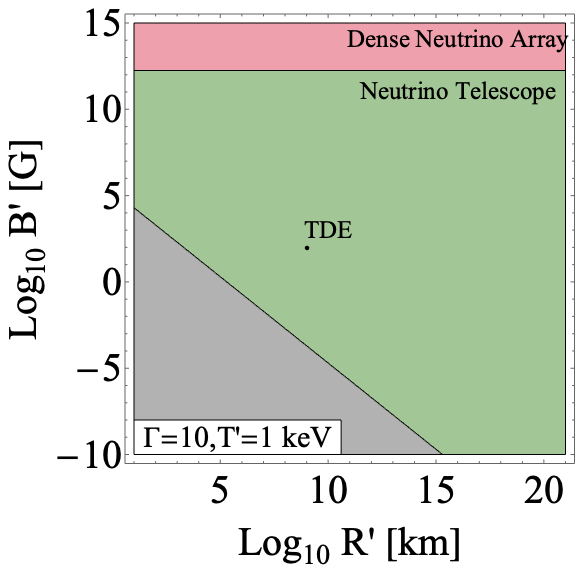}
   \includegraphics[width=0.3\textwidth]{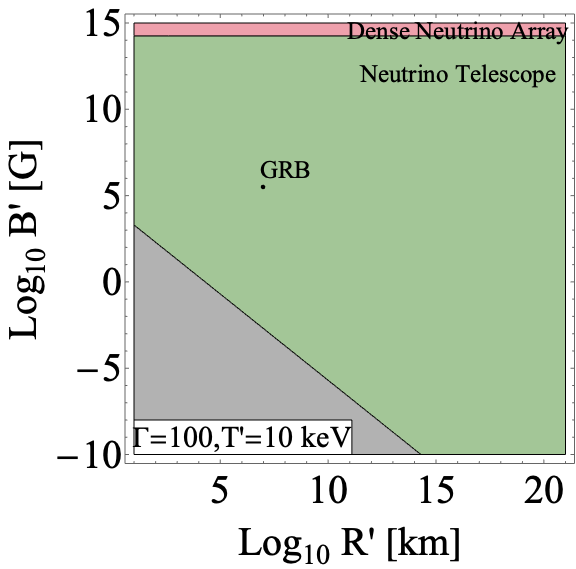}
    \caption{Determination of the most suitable experiment for the detection of astrophysical sources with parameters $\Gamma$, $T'$, $B'$ and $R'$. We show the Hillas plane divided according to which experiment is most suitable for detection: the red, green and blue regions correspond, respectively, to our benchmark Dense Neutrino Array, Neutrino Telescope and Neutrino Radio Array as most sensitive experiments. The three panels correspond to different values for the effective temperature and Doppler factor. Typical values of $B'$, $R'$, $T'$ and $\Gamma$ for AGN, TDE and GRB sources are identified (the sources are identified by the order of magnitude of the effective temperature rather than by the precise value). In the gray regions, pion production, and therefore neutrino production, is inefficient, because the maximal proton energy is below the threshold for pion photoproduction for all target photons.}
    \label{resulsensit}
\end{figure}

Finally, we perform a scan of the parameter space to show the complementarity of different experiments. The most sensitive experiment for a given point in  parameter space  is determined by the lowest value of $\xi$ needed for detection. We show this result in Fig.~\ref{resulsensit} in the Hillas plane, for three benchmark choices of effective temperature $T'$ and of $\Gamma$. The Hillas plane has been divided here in regions depending on the benchmark experiment most sensitive for detection.

Neutrino telescopes and neutrino radio arrays by themselves are already able to probe most of the parameter space with an energy fluence as low as $10^{-3}$~erg~cm$^{-2}$. 
Whereas neutrino radio arrays are most sensitive to sources with low effective temperatures and high neutrino energies, neutrino telescopes test most of the conventional parameter space best.
Dense neutrino arrays, on the other hand, require larger energy fluences, which typically reach values of $1$~erg~cm$^{-2}$, and are only competitive in the region of very large effective temperatures ($\gtrsim 100$~eV), very low maximal proton energies ($\lesssim 1$~TeV) and low Doppler factors ($\lesssim 10$). For sources with these properties DeepCore and, in general, the low energy neutrino telescopes might be the key instruments for detection.

\section{Flavor structure} 
\label{sec:flavorresults}

In this section we discuss the flavor composition of the astrophysical fluxes within the thermal model. The main complication in analyzing the flavor structure is its energy dependence~\cite{Hummer:2010ai}. In fact, the flavor ratio undergoes several transitions mentioned in section~\ref{sec:valtherm}. The different flavor regimes are typically classified as pion beam, muon-damped regime, muon beam, and neutron beam. The properties of these flavor regimes, as well as the energy range in which they appear for different benchmark sources, are summarized in Table \ref{tabregimes} (see also App.~\ref{app:flavor}), where we also show the flavor ratio at the source. After arriving at the Earth, the flavor composition changes due to mixing: in this work we take into account neutrino oscillations assuming the mixing parameters \cite{Zyla:2020zbs} $\sin^2\theta_{12}=0.307$, $\sin^2\theta_{23}=0.545$, $\sin^2\theta_{31}=2.18\times 10^{-2}$ and $\delta=1.36\pi$.

\begin{table}[t]
    \centering
    \begin{tabular}{l|rrrrr}
     \hline
       Composition  & Flavor ratio & Process & AGN  & GRB& TDE\\
        & at source & & E (GeV) & E (GeV) & E (GeV)\\
       \hline
       Pion beam & (1:2:0) & $\pi\to\mu+\bar{\nu}_\mu$,   & $10^6-10^9$ & $10^3-10^6$ & $10^4-10^8$\\
       &  &  $\mu\to e+\bar{\nu}_e+\nu_\mu$  &  &  & \\
       Muon-damped regime & (0:1:0) & $\pi\to\mu+\bar{\nu}_\mu$ & - & $>10^6$ & $-$ \\
       Muon beam & (1:1:0) & $\mu\to e+\bar{\nu}_e+\nu_\mu$ & $>10^9$ & - & $>10^8$ \\
       Neutron beam & (1:0:0) & $n\to p+e+\bar{\nu}_e$ & $<10^6$ & $<10^3$ & $<10^4$ \\
       \hline
    \end{tabular}
    \caption{Summary of the flavor compositions studied in the main text, as well as the typical ranges of the neutrino energy at which they appear in our benchmark sources. We do not show the charge-conjugated processes, which are, however, taken into account in all our computations. The muon-beam regime, for the benchmark examples in the table, only appears near the cutoff; however, this regime can also appear at low energies, due to a pile-up of muons after synchrotron losses: this behavior is visible for TP~G1 in Fig.~\ref{fig:trackshowerglas}.}
    \label{tabregimes}
\end{table}

As a first step, we will only take into account the flavor ratio at the peak of the fluence, where its detection is most probable. We will provide a general classification of the parameter space (in the Hillas plane) according to this observable. Next, to study the energy dependence we focus on the predicted track to shower ratio as a function of the energy for our benchmark Neutrino telescope. We also analyze the behavior near the Glashow resonance. The prospects for flavor discrimination at neutrino telescopes have also been discussed in Refs. \cite{Beacom:2003nh,Bustamante:2015waa,Bustamante:2019sdb,Bustamante:2020bxp,Song:2020nfh}: in this work we focus on the interplay between the flavor discrimination at neutrino telescopes and the source parameters within the thermal model.

\subsection{Flavor ratios of the high-energy astrophysical neutrino flux}

\begin{figure}[t!]
    \centering
    \includegraphics[width=0.3\textwidth]{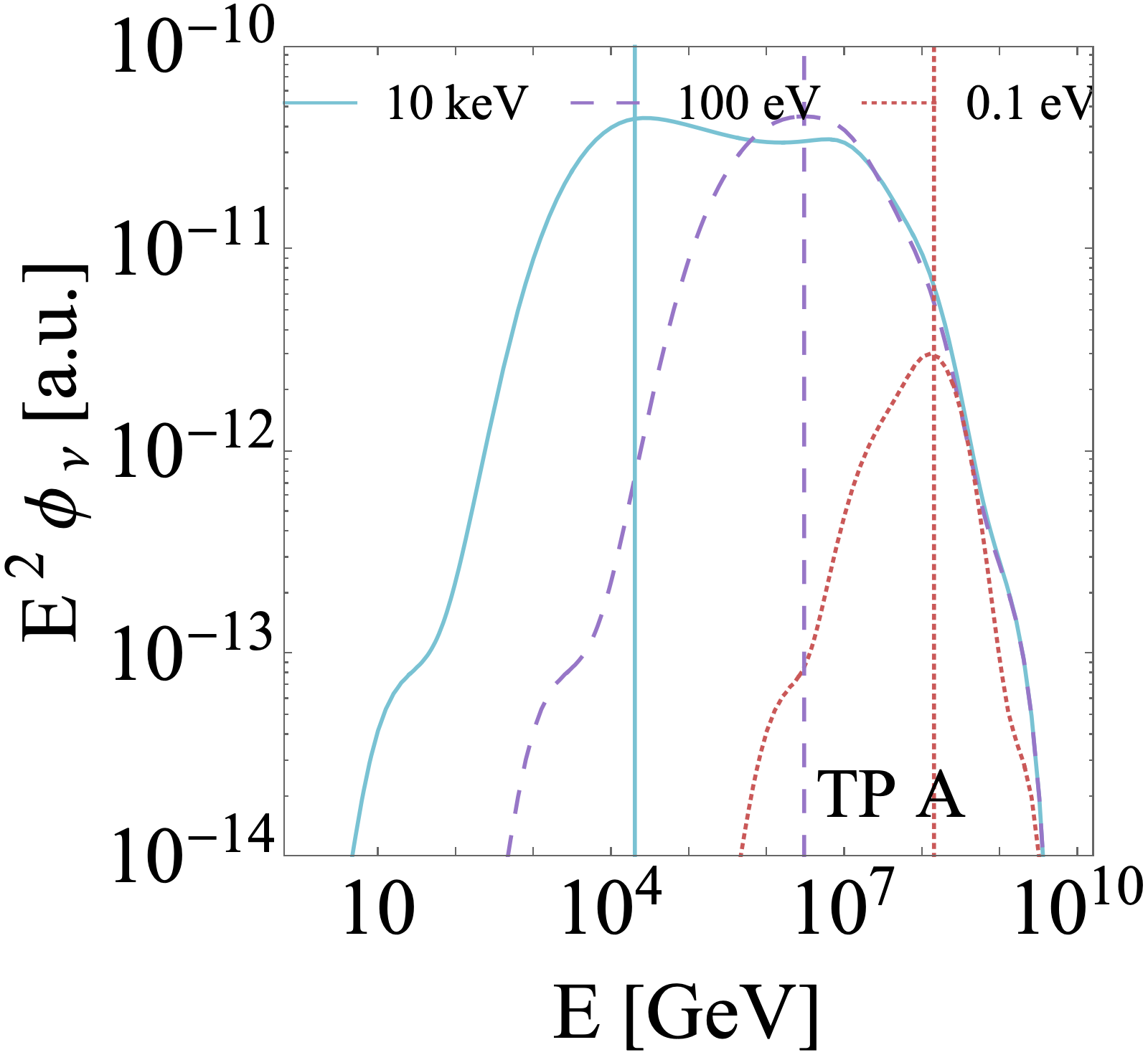}
    \includegraphics[width=0.3\textwidth]{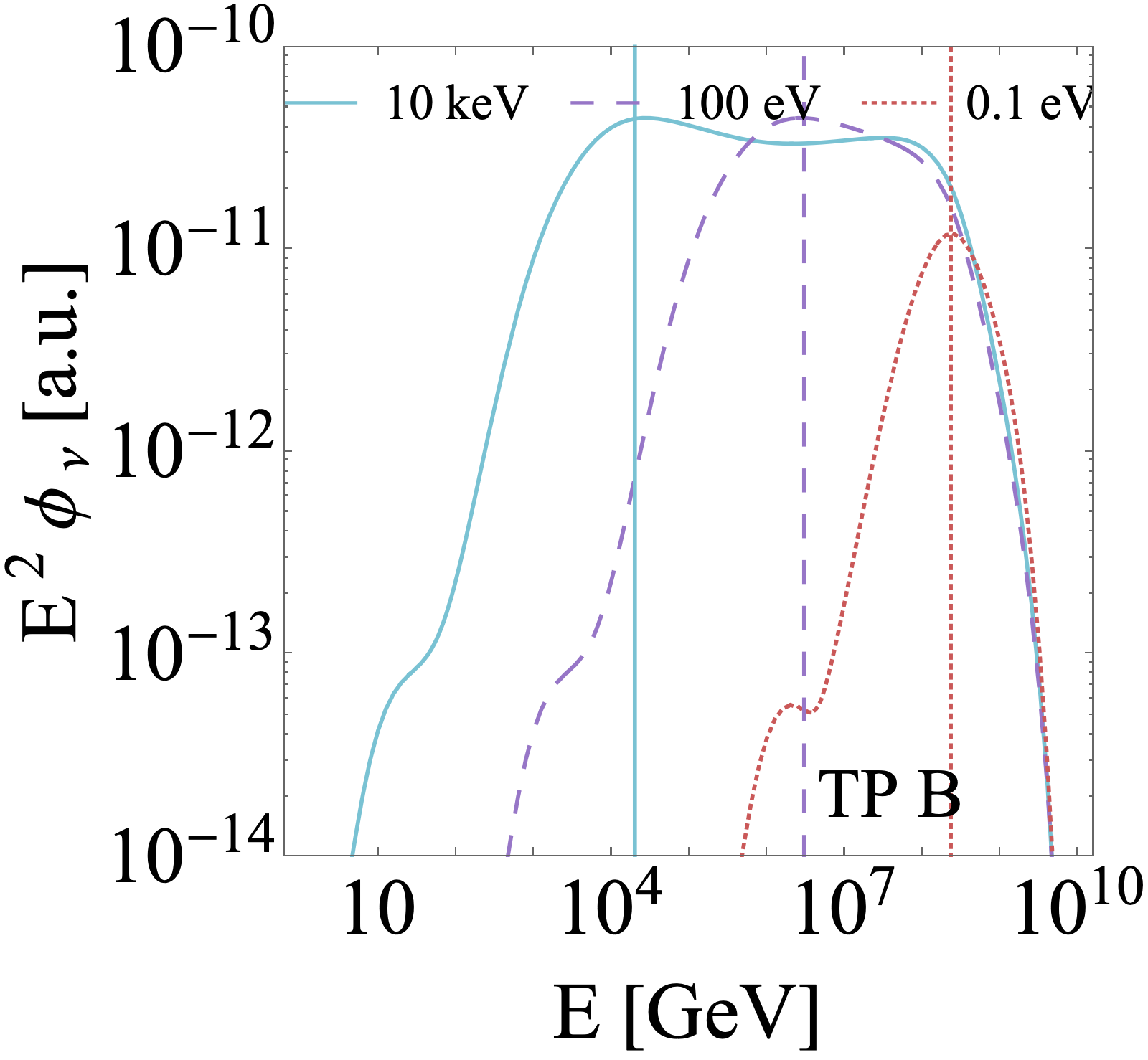}
    \includegraphics[width=0.3\textwidth]{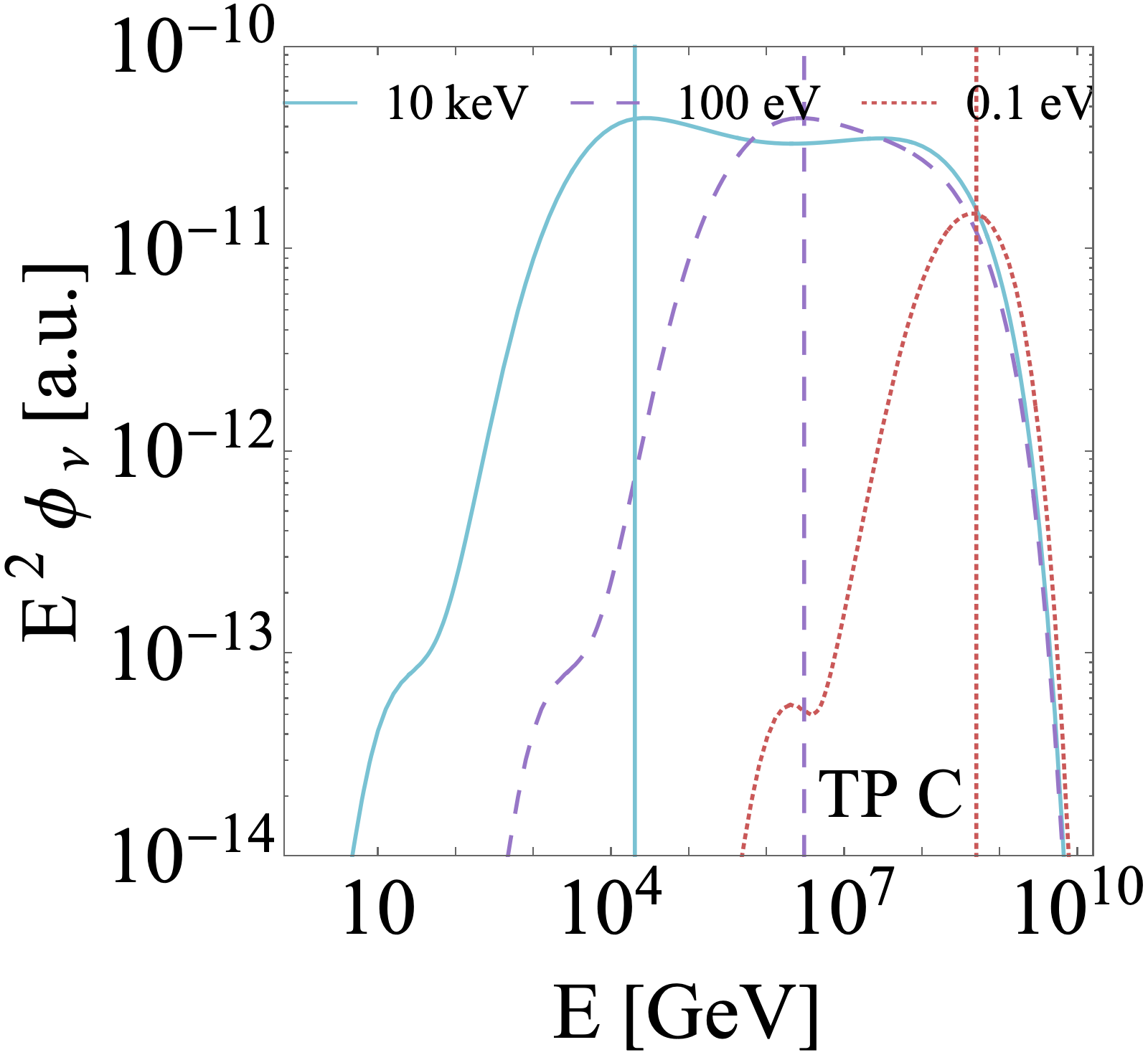}
    \includegraphics[width=0.3\textwidth]{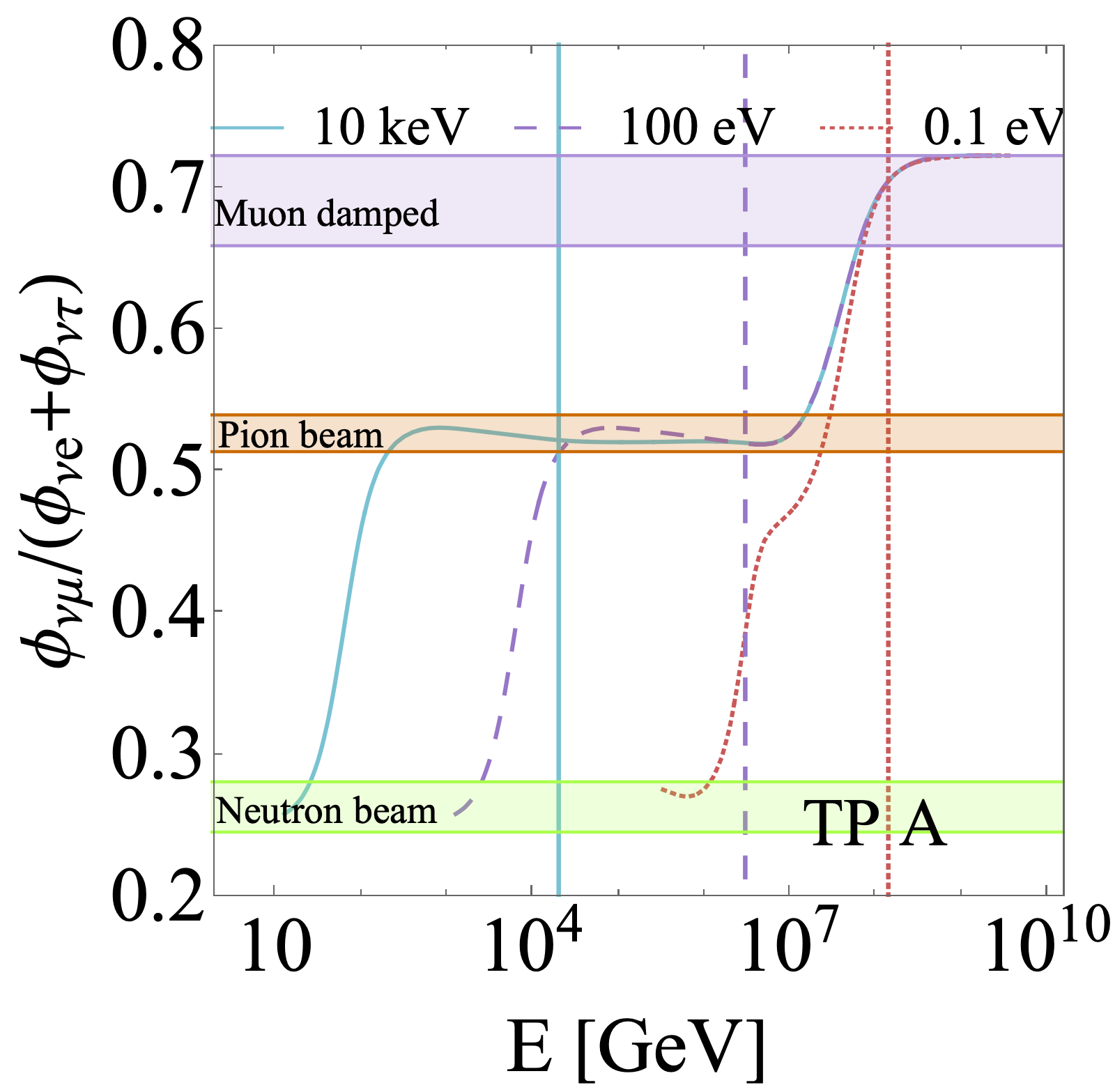}
    \includegraphics[width=0.3\textwidth]{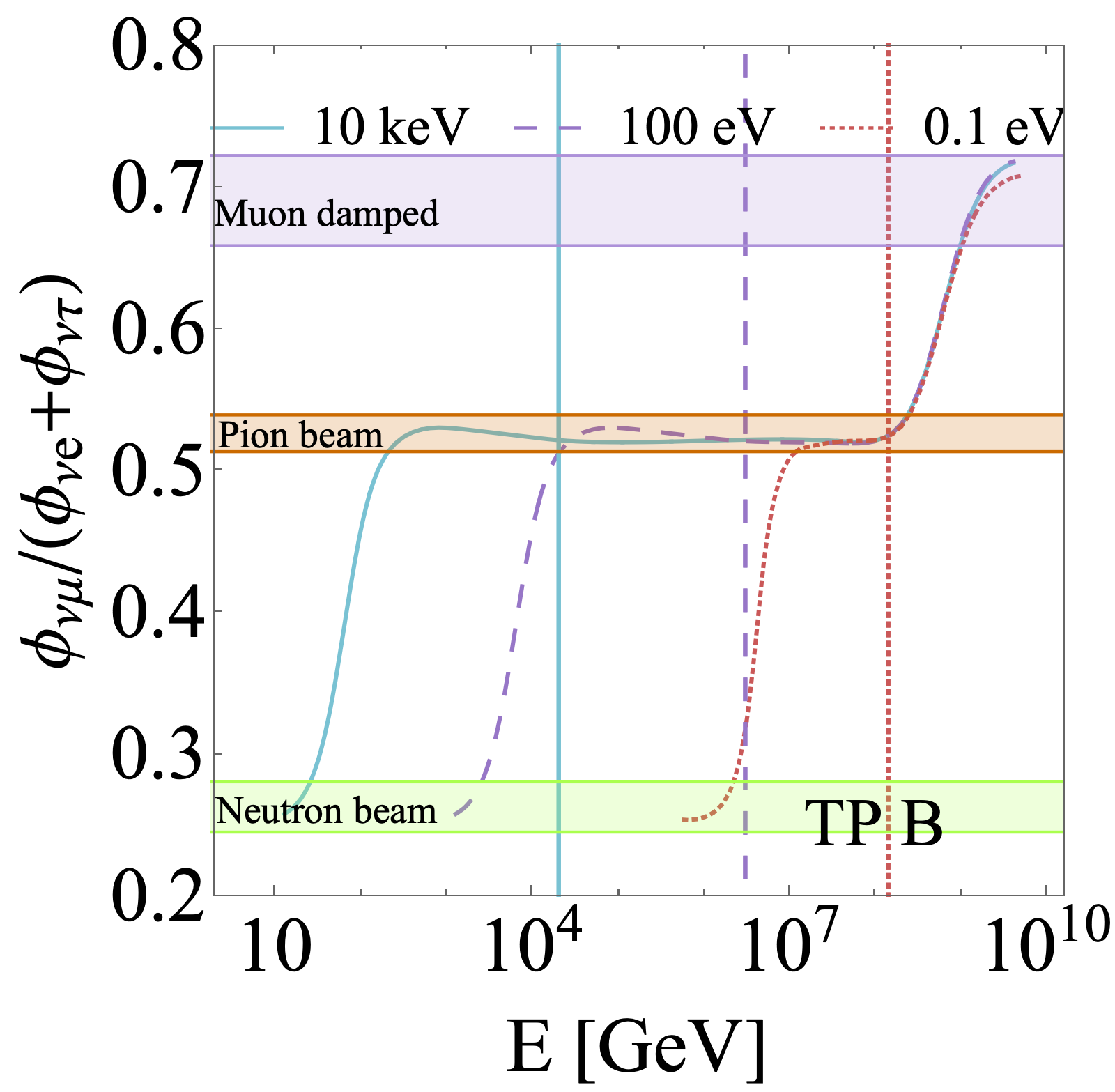}
    \includegraphics[width=0.3\textwidth]{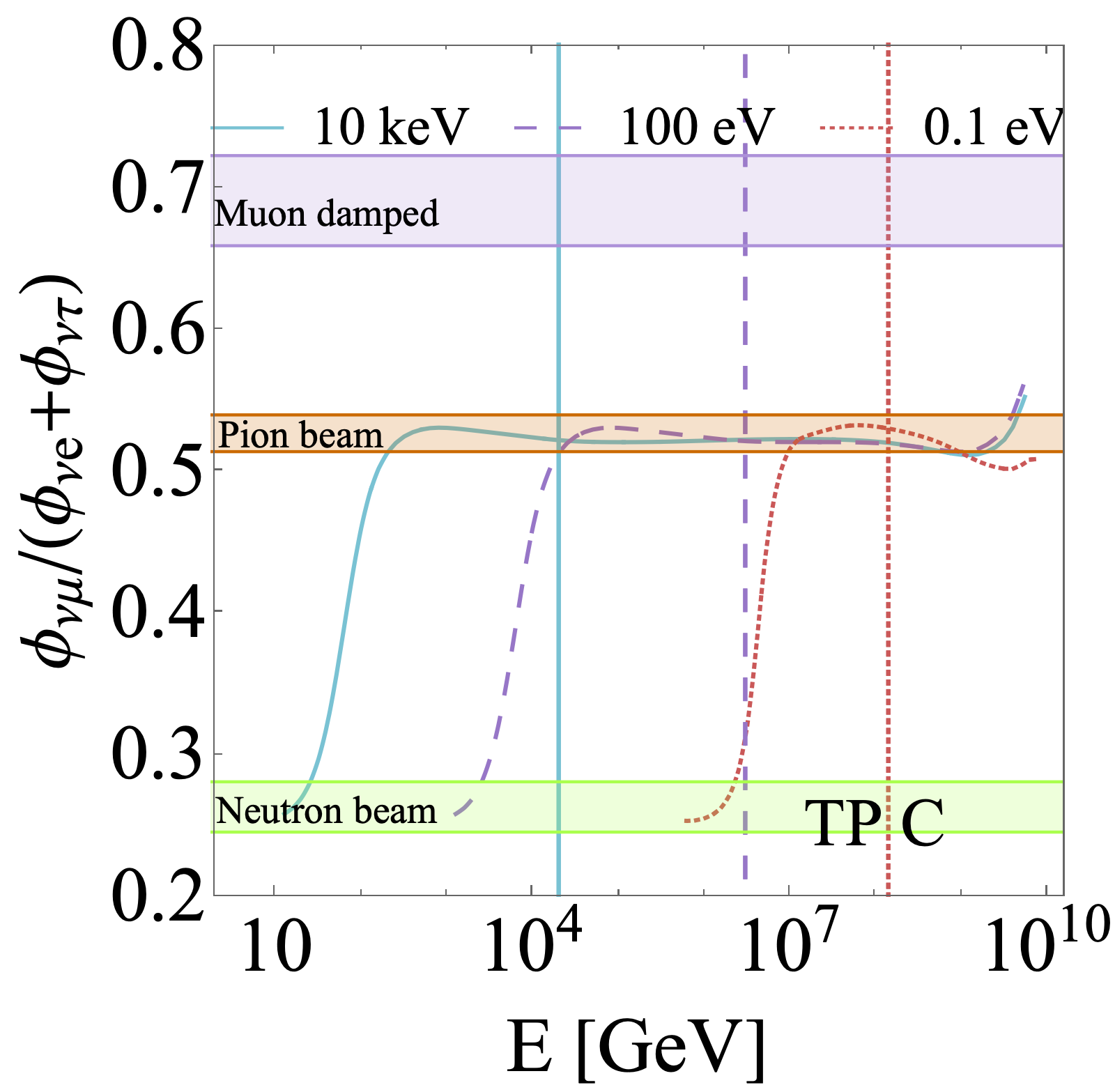}
    \caption{Neutrino fluxes and flavor ratios at Earth as a function of energy. In the top panels we show the all-flavor neutrino fluxes as a function of the energy for the three TPs in Fig.~\ref{flavorratios}, which are chosen to simulate lower magnetic fields from left to right. The three curves in each panel correspond to $T'=0.1$~eV, $T'=100$~eV and $T'=10$~keV; in all cases $\Gamma=10$. In the bottom panels we show the ratio between the muon and the sum of electron and tau neutrino and antineutrino differential flux as a function of the energy. The curves are represented only in the region in which the flux is at least $1/1000$ of its peak value. The horizontal bands identify the different flavor regime according to the quantitative criterion defined in the caption of Fig.~\ref{flavorratios}. For the neutrino mixing parameters the best-fit parameters of Ref.~\cite{Zyla:2020zbs} are chosen. Both in the top and the bottom panels we identify the peak energies of the neutrino fluxes with vertical lines: each color corresponds to the effective temperature according to the legend.} 
    \label{testpointse}
\end{figure}

\begin{figure}[t!]
    \centering
    \includegraphics[width=0.3\textwidth]{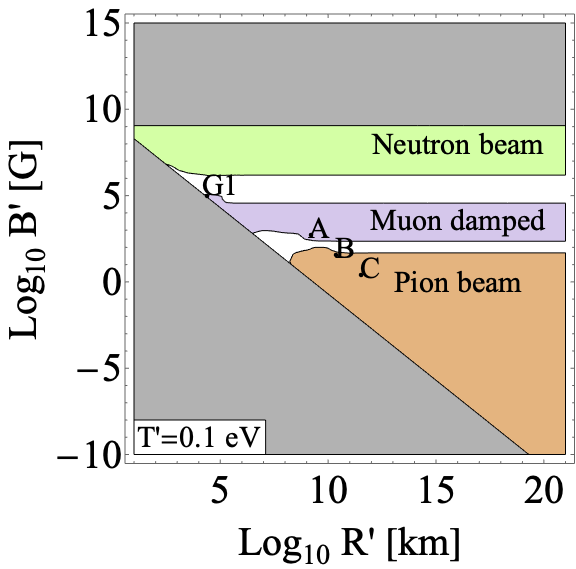}
    \includegraphics[width=0.3\textwidth]{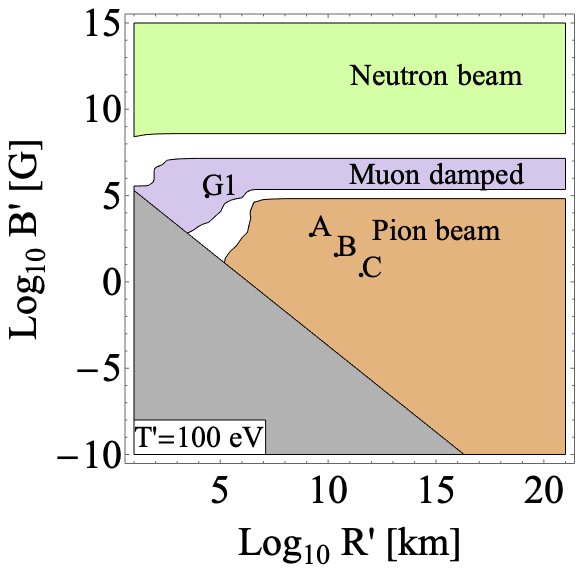}
 \includegraphics[width=0.3\textwidth]{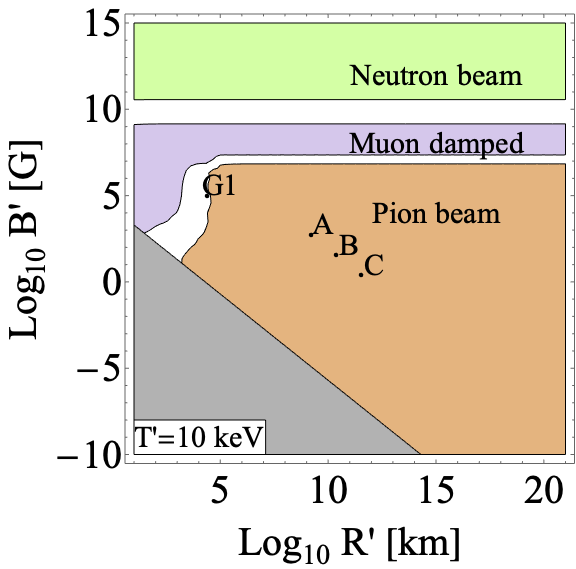}
    \caption{Flavor structure at the peak of the spectrum as a function of $R'$ and $B'$ (Hillas plot). We divide the Hillas plane according to the flavor composition at the peak of the spectrum: the neutron-beam region has a flavor ratio (at the source) between $(0.9:0.1:0)$ and $(1:0:0)$; the muon-damped region has a flavor ratio between $(0.1:0.9:0)$ and $(0:1:0)$; the pion-beam region has a flavor ratio between $(0.31:0.69:0)$ and $(0.36:0.64:0)$. In the white regions the flavor composition does not belong to any of the previous regimes. The three panels correspond to $T'=0.1$~eV, $T'=100$~eV and $T'=10$~keV. The gray regions correspond to inefficient pion production. Test points are indicated by A, B and C and by G1.}
    \label{flavorratios}
\end{figure}

The energy dependence of the flavor composition is most easily discussed with the aid of some benchmark examples. In Figure \ref{testpointse} we show the neutrino fluxes and the flavor composition as a function of the energy for three benchmark choices of magnetic field and source size (later identified by test points A, B and C in Fig.~\ref{flavorratios}). The test points are chosen for similar values of the  maximal proton energy. We let the effective temperature vary between $0.1$~eV and $10$~keV in order to highlight its effect. We quantify the flavor composition by the ratio between the muon neutrino flux and the electron and tau neutrino flux $\phi_{\nu\mu}/(\phi_{\nu e}+\phi_{\nu\tau})$. In order to collect the information into a single variable, we refer to the sum of the neutrino and antineutrino fluxes. While this choice does not capture the relative contribution of the processes involving $\pi^+$ and $\pi^-$, it can differentiate between the flavor regimes listed in Table \ref{tabregimes}. For all choices of the parameters, the ratio of the fluxes defined above exhibits marked transitions between different flavor regimes, which are identified by horizontal bands in Fig.~\ref{testpointse}; in particular, it passes from neutron beam to pion beam to muon-damped regime for increasing energies.

The critical energy at which the flux passes from pion beam to muon-damped regime decreases with increasing magnetic field (from TP~C to TP~A). In App.~\ref{app:flavor} we show that this decrease is approximately $E'_{\text{c},\mu}\propto B^{'-1}$. 
Decreasing the effective temperature, on the other hand, pushes the peaks of the spectra to higher neutrino energies, as shown by the upper panels of Fig.~\ref{testpointse}. Therefore the neutron-beam region moves to higher energies for lower effective temperatures. In addition, the peak of the spectrum moves progressively towards the region of muon damping. For this reason TP~A, for an effective temperature of 0.1~eV, is in the muon-damped regime at its peak. 

We emphasize that this succession of flavor compositions (neutron beam, pion beam and muon-damped regime) only refers to the specific combination of parameters that we have chosen. The spectrum may exhibit different flavor structures as a function of the energy for different values of the parameters. 
Such energy dependence is well captured by the thermal model, and the corresponding fluxes separated per flavor have been made available at Ref.~\cite{Datarelease}.
The energy dependence makes it difficult to provide a comprehensive study throughout the parameter space. For this reason, we limit a systematic study to the flavor composition at the peak of the spectrum, namely at the energy at which the spectrum is maximum. For each combination of parameters we determine the flavor composition at this energy and we classify it according to which flavor regime it is closer to. The result of this classification is shown in Fig.~\ref{flavorratios}. We find no extended region in the Hillas plane in which the flavor ratio at the peak of the spectrum is muon beam.

Increasing the magnetic field, the flavor composition at the peak of the spectrum passes from pion-beam to muon-damped to neutron-beam regimes. For weak magnetic fields in fact there are no cooling effects and the full decay chain $\pi^+\to\mu^+ + \nu_\mu$, $\mu^+\to e^+ + \nu_e +\bar{\nu}_\mu$ is active. Higher magnetic fields inhibit the muon decay, leading to the muon-damped regime. Finally, when the magnetic field is very strong, all charged species lose energy very quickly due to synchrotron radiation, which implies that the  maximal proton energy is low and that the neutrinos from pion and muon decays are both suppressed. Only neutrinos from neutron decay survive, since their parents (as neutral particles) are not subject to synchrotron losses. 

On the other hand, lowering the effective temperature causes the appearance of muon-damped and neutron-beam regimes already at lower values of the magnetic field. In fact, decreasing the effective temperature pushes the peak of the spectrum at higher energies into the muon-damped regime, as mentioned in reference to Fig.~\ref{testpointse}. At the same time, lowering the effective temperature raises the energy of the bump of neutrinos from neutron decay, moving the transition to the neutron-beam regime to lower magnetic fields. 

\subsection{Track and shower discrimination}

Track events are typically produced by muon neutrinos, which, after the collision with nuclei, produce a muon, subsequently detected as a track. Shower events are instead typically produced by electron and tau neutrinos, as well as by muon neutrinos interacting via neutral currents. In these processes the neutrinos collide with nuclei and give rise to an electromagnetic or hadronic shower. In reality there is also a small contamination of track events originating from tau neutrinos due to tau decay. Furthermore, tau neutrinos can give rise to a different topology, namely the double bang events, in which a first shower from the tau neutrino interaction is followed by the propagation of a tau lepton decaying into a second shower. The observation of these events has been recently reported for the first time by the IceCube collaboration \cite{Abbasi:2020zmr}. 

In order to connect the flavor structure with observable quantities, we have studied the predictions for the track and the shower event rates $N_T$ and $N_S$. These are obtained by multiplying each flavor neutrino flux $\phi_\alpha$ by the effective areas of the benchmark Neutrino telescope. We assume that electron and tau neutrinos only produce shower events, while muon neutrinos produce tracks with a probability taken to be constant in energy and equal to $p_{T}=0.8$ \cite{Palladino:2015zua}. Consequently, with probability $1-p_{T}=0.2$ they produce showers via neutral current interactions. We take the effective areas for different flavors from Ref.~\cite{Adrian-Martinez:2016fdl}\footnote{Near the Glashow resonance the electron neutrino and antineutrino effective areas are different, while Ref.~\cite{Adrian-Martinez:2016fdl} only provides the average effective area for neutrinos and antineutrinos. In order to approximately obtain the two different effective areas, we associate the peak of the Glashow resonance in the average effective area entirely to the electron antineutrino effective area.}.

\begin{figure}[t!]
    \centering
    \includegraphics[width=0.3\textwidth]{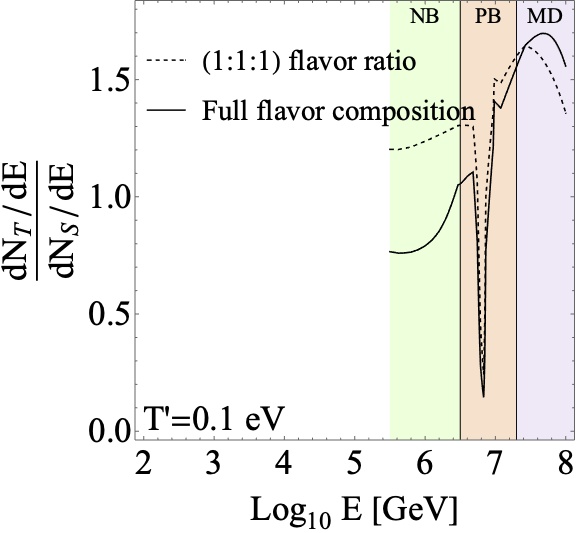}
    \includegraphics[width=0.3\textwidth]{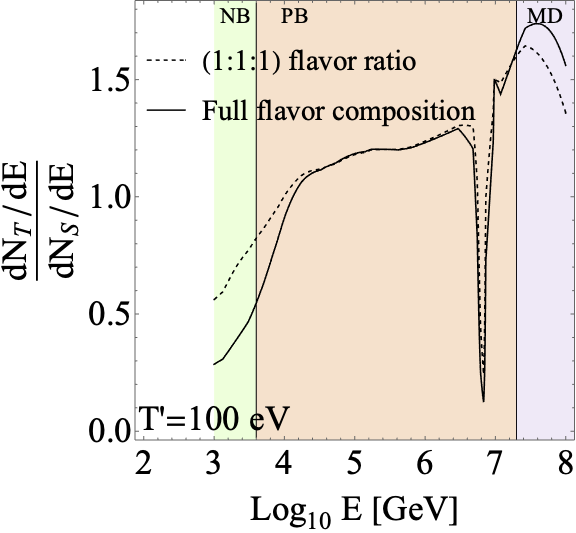}
    \includegraphics[width=0.3\textwidth]{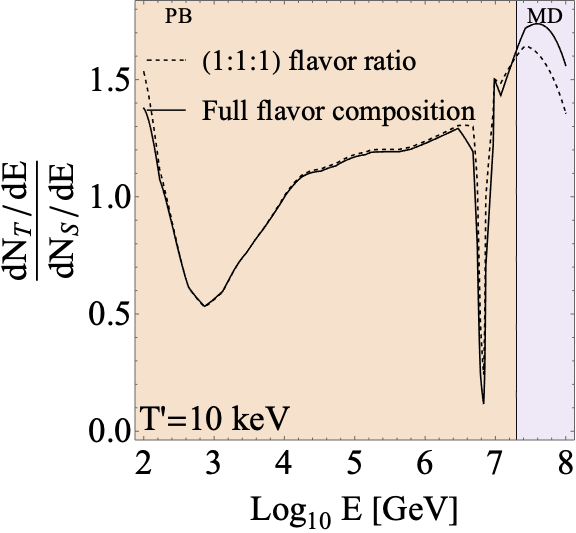}
    \caption{Track-to-shower ratio sensitivity to different flavor compositions for test point TP~A (see Fig.~\ref{flavorratios}) and $\Gamma=10$. We show the ratio between the expected differential number of track and shower events at our benchmark Neutrino Telescope as a function of the energy: dashed curves represent the prediction for a constant pion-beam flavor ratio (with no contamination from $\pi^-$, $(1:1:1)$ flavor ratio at Earth), while solid curves represent the full energy-dependent flavor composition as computed by NeuCosmA. The curves are represented only in the region in which the flux is at least 1/1000 of its peak value. The magnetic field and the source radius are indicated by TP~A in Fig.~\ref{flavorratios}, the three plots correspond to effective temperatures of $T'=0.1$~eV, $T'=100$~eV and $T'=10$~keV. For the full energy-dependent ratio, the flux passes between different flavor regimes highlighted with different colors, and denoted as neutron beam (NB, green), pion beam (PB, orange) and muon-damped regime (MD, purple).}
    \label{fig:trackshower}
\end{figure}

We study the ratio of the differential number\footnote{In principle it is also possible to study $N_T/N_S$ in different energy bins. This requires however the choice of a binning: the procedure we adopt in the text is binning-independent and reproduces the correct results for sufficiently small bins.} of tracks $\frac{\text{d}N_T}{\text{d}E}$ and showers $\frac{\text{d}N_S}{\text{d}E}$ per unit energy interval, which encloses the information on the energy dependence.
We show the corresponding result for TP~A in Fig.~\ref{fig:trackshower} as solid curves for different choices of the effective temperature in the three panels. We compare it with the result which would be expected in the case of a pure pion-beam source ($(1:1:1)$ flavor ratio at Earth), shown by dashed curves. 

The flavor composition undergoes a series of transitions between pion beam, neutron beam and muon-damped regimes, highlighted with different colors. The energy ranges in which these regimes appear depend on the choice of the parameters (for an analytical estimate see App.~\ref{app:flavor}). Raising the effective temperature, the neutron-beam region moves to lower energies. Compared with the predictions of the pure pion-beam flux, the neutron beam leads to a sizable decrease in the track-to-shower ratio, due to the larger presence of electron antineutrinos. On the other hand, in the muon-damped regime, a slighter increase in the track-to-shower ratio is found. This suggests that the identification of a neutron-beam regime among the data collected by neutrino telescopes such as IceCube and KM3NeT should be easier in the case of sources with sufficiently low effective temperatures. 

\subsection{Glashow resonance}

Another flavor-dependent process is the collision of electron antineutrinos with electrons via the so-called Glashow resonance \cite{Glashow:1960zz}; $\bar{\nu}_e+e^-\to W^-$ producing a shower. The resonant increase in the cross section for the detection of electron antineutrinos appears at a neutrino energy of $6.3$~PeV; this causes the dip in the track-to-shower ratio in Fig.~\ref{fig:trackshower}. 

If the flavor composition is muon-damped, only neutrinos, with a small contamination of antineutrinos, are expected, because neutrinos are mainly produced via the decay $\pi^+\to\mu^+ + \nu_\mu$ in the $\Delta$-resonance approximation. In the absence of antineutrinos, even after oscillations, no electron antineutrinos will reach the detector. There is therefore no enhancement in the event rate. The multi-pion contribution can actually give rise to a small contamination of antineutrinos from the decay of $\pi^-$, and therefore to a small decrease in the track-to-shower ratio near the Glashow resonance. This conclusion is also reached in Ref.~\cite{Biehl:2016psj}, where it is emphasized that the Glashow resonance could potentially provide a diagnostic to identify the muon-damped regime (see also Ref.~\cite{Hummer:2010ai}).

\begin{figure}[t!]
    \centering
    \includegraphics[width=0.3\textwidth]{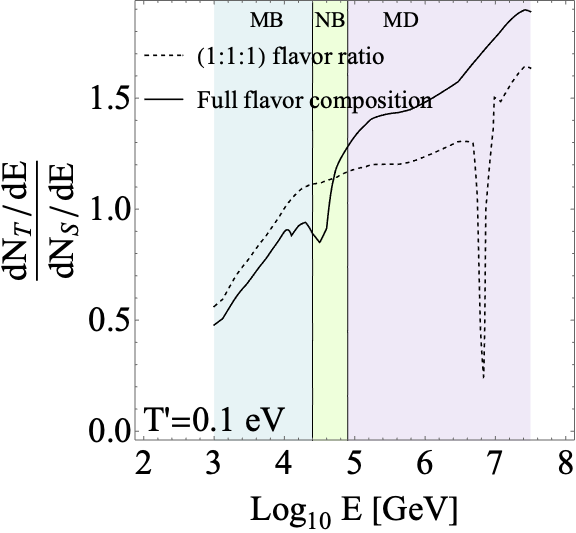}
    \includegraphics[width=0.3\textwidth]{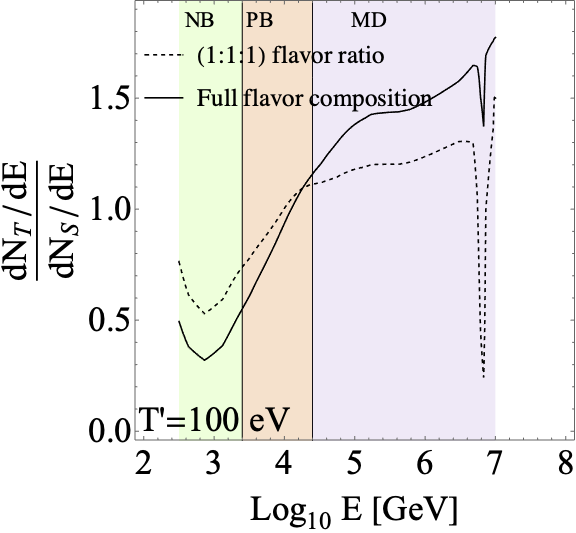}
    \includegraphics[width=0.3\textwidth]{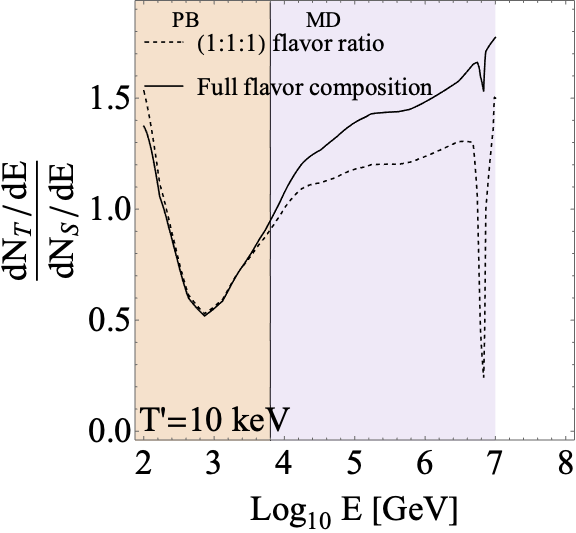}
    \caption{Track-to-shower ratio sensitivity to different flavor compositions for the test point TP~G1 (see Fig.~\ref{flavorratios}). We show the ratio between the expected differential number of track and shower events at our benchmark Neutrino Telescope as a function of the energy: dashed curves represent the prediction for a constant pion-beam flavor ratio (with no contamination from $\pi^-$, $(1:1:1)$ flavor ratio at Earth), while solid curves represent the full energy-dependent flavor composition as computed by NeuCosmA. The curves are represented only in the region in which the flux is at least 1/1000 of its peak value. The magnetic field and the source radius are indicated by TP~G1 in Fig.~\ref{flavorratios}, the three plots correspond to effective temperatures of $T'=0.1$~eV, $T'=100$~eV and $T'=10$~keV. For the full energy-dependent ratio, the flux passes between different flavor regimes highlighted with different colors, and denoted as neutron beam (NB, green), muon beam (MB, cyan), pion beam (PB, orange) and muon-damped regime (MD, purple).}
    \label{fig:trackshowerglas}
\end{figure}

To exemplify the differences that can arise in the track-to-shower ratio, we have chosen a test point, identified in Fig.~\ref{flavorratios} as G1. For this specific test point, the Glashow resonance energy falls in the muon-damped regime. In Fig.~\ref{fig:trackshowerglas} we represent the differential track-to-shower ratio $\frac{\text{d}N_T}{\text{d}E}/\frac{\text{d}N_S}{\text{d}E}$ as a function of the energy. As before, we refer to the benchmark Neutrino Telescope and we show the result for different effective temperatures. We compare this with the differential track-to-shower ratio in the case of a constant pion-beam flavor composition ($(1:1:1)$ flavor ratio at Earth).

For a pure pion-beam source (dashed in Fig.~\ref{fig:trackshowerglas}) the track-to-shower ratio exhibits a pronounced dip near the Glashow resonance, due to the increasing number of shower events from electron antineutrinos. On the other hand, for the muon-damped case (solid curves in Fig.~\ref{fig:trackshowerglas}) the track-to-shower ratio shows a small dip (middle and right panel) or no dip at all (left panel). This is consistent with our observation that only a few electron antineutrinos are present in this regime, coming from $\pi^-$ decays.
The track-to-shower ratio near the Glashow resonance can therefore differ significantly from the pion-beam prediction, providing a useful discrimination tool for the identification of the muon-damped regime. Such an identification could help to determine the intensity of the magnetic field, which causes the passage to the muon-damped regime.

This conclusion is particularly interesting in light of the recent discovery of an event due to the Glashow resonance made by IceCube \cite{IceCubeGlashow2021}. This observation suggests the presence of electron antineutrinos in the astrophysical flux. According to our discussion, this presence can be explained either by assuming that the flux is not in the muon-damped regime, or by assuming that it is in the muon-damped regime with a large enough contamination of antineutrinos from $\pi^-$ decays: these possibilities are both pointed out in Ref.~\cite{IceCubeGlashow2021}. Within the thermal model presented here, the first assumption requires sufficiently weak magnetic fields (we discuss in App.~\ref{app:flavor} how the energy range of the muon-damped regime depends on the magnetic field), while the second assumption requires that the temperature is not too low, as exemplified by the left panel of Fig.~\ref{fig:trackshowerglas}, where the absence of a dip in the track-to-shower ratio signals the absence of antineutrinos for the lowest temperature of $T'=0.1$~eV.

\section{Summary and conclusions}

Many classes of astrophysical sources have been proposed as candidates for the production of the astrophysical neutrinos observed by IceCube. In order to evaluate the detector response to these sources, one needs models of neutrino production which are both simple and flexible enough to capture the physical aspects of the problem. For $pp$ sources these two requirements are met by a power-law model. On the other hand, for photohadronic sources the power-law description is not very good, due to the dependence on the target-photon spectrum, the possible magnetic field effects, and pion production processes beyond the $\Delta$-resonance. For this reason, separate modeling is typically performed for each of these sources. This, however, prohibits systematical parameter space studies spanning different classes of sources at once.

To allow for systematic studies of $p\gamma$ sources, we have proposed a unified model that can reproduce the spectral shape and the flavor composition of the neutrino spectrum from photohadronic sources. We have approximated the target photons by a black-body spectrum parameterized by an effective temperature. The temperature has been chosen with a simple set of rules depending on the target-photon spectrum in the relevant energy range and on the maximal proton energy, see Sec.~\ref{sec:recipe}. We have determined the maximal proton energy from the confinement condition in the presence of synchrotron losses. In this way the variety of source classes can be reduced to a few parameters: the effective temperature of the photon spectrum, the Doppler factor, the magnetic field, the source size and the acceleration efficiency of the protons (which we have fixed to a relatively high value in the main text). 

We have demonstrated that the unified model reproduces the neutrino spectra and flavor compositions from different astrophysical sources, such as AGN, GRBs, and TDEs. The accuracy in the reproduction is sufficient to allow for systematic studies of $p\gamma$ sources. The physical reason that such a unified approach works is the dominance of multi-pion processes above the threshold for pion photoproduction. In fact, neutrinos produced in multi-pion processes approximately follow the parent proton spectrum even for a peaked photon black-body target spectrum, provided that the maximal proton energy is high enough. This feature is frequently ignored in simple estimates of the neutrino peak energy based on the $\Delta$-resonance only, and in our case it allows to reproduce also spectra with photon number densities flat in energy such as in GRBs. The model also reproduces cooling effects in magnetic fields, which are independent of the target-photon spectra.

We have applied this thermal model in two different directions. First of all, we have studied the parameter space of the model in terms of the sensitivity of neutrino experiments. We have focused on three experiments representative of different classes, namely dense neutrino arrays, neutrino telescopes and neutrino radio arrays, which are sensitive to neutrinos in different energy ranges. We have identified the regions of the parameter space where an experiment class has the best sensitivity, and we have demonstrated that neutrino radio arrays are most sensitive to sources with low effective temperatures and high maximal proton energies, such as AGN with high acceleration efficiencies. Moreover, we have found that dense neutrino arrays are best suited to test sources with extremely strong magnetic fields ($B'\gtrsim 10^{12} \, \mathrm{G}$), and neutrino telescopes are best suited to test the largest part of the parameter space, in which also GRBs and TDEs (and AGN with lower maximal proton energies) fall.  Overall, a complementary parameter space coverage among different instruments has been demonstrated.
As a second application, we have studied the flavor composition of the neutrino flux as a function of the energy, identifying the parameter ranges for which the source is a pion beam, muon-damped source, and a neutron beam. 
We have shown how the track-to-shower ratio, which is an observable in neutrino telescopes to leading order, can be used to identify the flavor composition. It can also potentially identify a muon-damped source at the Glashow resonance, which works especially well for low effective temperatures. 

Limitations of the model are encountered for photon number densities which are slightly decreasing with energy. Nevertheless, the peak can be well reproduced in this case as well. In addition, the flavor composition in the neutron-beam regime is somewhat inaccurate, because it depends on the shape of the photon spectrum. The thermal model is ideally suited for the description of neutrino production; it is, however, agnostic towards other messengers such as electromagnetic radiation across a wide energy range, since only the minimal ingredients needed to accurately reproduce neutrino spectrum and flavor composition are considered. A universal model including other signatures as well requires significantly more parameters, as it cannot be dominated by the energy range relevant for the neutrino peak only. Furthermore, the efficiency of the pion production, which depends on the source size and geometry, is not captured; we have, therefore, parameterized our results in terms of the energy fluence of the neutrinos, or, equivalently, the (isotropic-equivalent) energy ejected into neutrinos.

In conclusion, the thermal model presented here captures the main features of the photohadronic neutrino production in astrophysical sources in terms of the expected spectral shape and flavor composition.
Our unified model provides a way to systematically study the parameter space of cosmic neutrino emitters, including the characterization of the diffuse flux, multiplet studies and stacking analyses of the data. Furthermore, this framework could help in studying Beyond Standard Model physics, since it allows to represent the standard neutrino production in terms of a few parameters, which can be used to quantify the ``astrophysical systematics''.  

\acknowledgments{We would like to thank Mauricio Bustamante, Sarah Mechbal, and Andrea Palladino for useful discussions and comments. The work of A. van Vliet and W. Winter has been supported by the  European Research Council (ERC) under the European Unions Horizon 2020 research and innovation programme (Grant No. 646623). The work of D. F. G. Fiorillo and S. Morisi was supported by the research grant number 2017W4HA7S “NAT-NET: Neutrino and Astroparticle Theory Network” under the program PRIN 2017 funded by the Italian Ministero dell’Istruzione, dell’Universita` e della Ricerca (MIUR), and INFN Iniziativa Specifica TAsP.}

\appendix

\section{Construction of the thermal model}\label{app:thermalmodel}

In this appendix, we discuss the neutrino production in photohadronic sources in more detail. In doing so we justify the relationship in Eq.~\ref{eq:tempsource} which gives the temperature of the black-body spectrum in the thermal model. 
For clarity, throughout this appendix, we will denote the proton, pion and neutrino energies by $E'_p$, $E'_\pi$ and $E'_\nu$ respectively (in the main text the energy of each particle is denoted by $E'$ regardless of the species, while the symbols $E_p$ and $E_\nu$ are reserved for the total energy injected in protons and neutrinos) and we will use natural units with $c=1$.

First of all, we are going to derive approximate analytical expressions for the pion spectra produced by $p\gamma$ interactions. Similar results have been obtained using analytical approximations in other works, see, e.g., Refs.~\cite{Dermer:2012rg,Murase:2014foa}. The neutrino spectra will be then obtained from the pion ones under the assumption of weak magnetic fields: this assumption is only for convenience of analytical calculations and is not made in the text. Following Ref.~\cite{Hummer:2010vx}, the injection spectrum of pions (differential in energy, volume and time) can be approximated by
\begin{equation}\label{eq:pioninj}
    Q_\pi (E'_\pi)=\sum_{\text{IT}} \frac{N_p(E'_\pi/\chi^\text{IT}_\pi) M^\text{IT}_\pi}{\chi^\text{IT}_\pi} \Gamma^\text{IT}_{p\to\pi}\left(\frac{E'_\pi}{\chi^\text{IT}_\pi}\right),
\end{equation}
where IT denotes the ``Interaction Type'' (e.g.~resonant production, $t$-channel production, multi-pion production), $\chi^\text{IT}_\pi$ is the fraction of the proton energy carried by the pion for each interaction type, $M^\text{IT}_\pi$ is the multiplicity of pions in the final state (average number of pions of a certain type produced per interaction) and $\Gamma^\text{IT}_{p\to\pi}(E'_p)$ is the interaction rate for a proton with energy $E'_p$. The values of $\chi^\text{IT}_\pi$ and $M^\text{IT}_\pi$ for each interaction type are provided in Ref.~\cite{Hummer:2010vx}.  
Assuming that the proton spectrum is $N_p(E'_p)\propto E^{'-2}_p e^{-E'_p/E'_\text{p,max}}$, it follows that
\begin{equation}\label{eq:energydeppion}
    E^{'2}_\pi Q_\pi (E'_\pi)\propto \sum_{\text{IT}} \chi^\text{IT}_\pi M^\text{IT}_\pi e^{-\frac{E'_\pi}{\chi^\text{IT}_\pi E'_\text{p,max}}} \Gamma^\text{IT}_{p\to\pi}\left(\frac{E'_\pi}{\chi^\text{IT}_\pi}\right).
\end{equation}
Therefore, the main energy dependence (for energies lower than the pion cutoff energy $E^{'\text{IT}}_{\pi\text{,max}} = \chi^\text{IT}_\pi E'_\text{p,max}$) comes from the interaction rate.

The interaction rate depends upon the target-photon spectrum, and can be written as
\begin{equation} \label{eq:interactionrate}
    \Gamma^\text{IT}_{p\to\pi}\left(\frac{E'_\pi}{\chi^\text{IT}_\pi}\right)=\int \text{d}\varepsilon'_\gamma
 n(\varepsilon'_\gamma) f_\text{IT} \left(\frac{\varepsilon'_\gamma E'_\pi}{m_p \chi^\text{IT}_\pi}\right)=\int \frac{\text{d}y}{y} \left[\varepsilon'_\gamma n(\varepsilon'_\gamma)\right]_{\varepsilon'_\gamma=\frac{y m_p\chi^\text{IT}_\pi}{E'_\pi}} f_\text{IT}(y),
 \end{equation}
where we have introduced the convenient re-parameterization $y \equiv \varepsilon'_\gamma E'_p/m_p$. The functions $f_\text{IT} (y)$, which describe the pitch-angle-averaged cross sections for each interaction type, are again provided in Ref.~\cite{Hummer:2010vx} (see e.g.~Fig.~4 for the secondary yields). In both forms it is evident that the interaction rate is proportional to the photon number density, as is well known in particle physics.

To get a qualitative description, we only consider the contributions from two interaction types, namely the $\Delta$-resonance and the multi-pion production. Other production channels, such as $t$-channel production and higher resonances, do not qualitatively change the properties of the neutrino spectra.

For the $\Delta$-resonance, the pion carries a fraction of the energy of the parent proton equal to $\chi^{\Delta}_\pi=0.22$ in the Sim-B model of Ref.~\cite{Hummer:2010vx}; the interaction rate in Eq.~\ref{eq:interactionrate} can then be written as
    \begin{equation}\label{eq:lrpion}
        \Gamma^\Delta_{p\to\pi}\left(\frac{E'_\pi}{\chi^\Delta_\pi}\right)=\int_0^{+\infty} \frac{\text{d}y}{y} \left[\varepsilon'_\gamma n(\varepsilon'_\gamma)\right]_{\varepsilon'_\gamma=\frac{y m_p \chi^\Delta_\pi}{E'_\pi}} f_\Delta(y).
    \end{equation}
    The function $f_{\Delta}(y)$, as reported in Ref.~\cite{Hummer:2010vx}\footnote{In Ref.~\cite{Hummer:2010vx} the $\Delta$-resonance is referred to as the Lower Resonance (LR).}, is strongly peaked at values of $y\simeq y_\Delta=0.2$~GeV. The largest contribution to the integral in Eq.~\ref{eq:lrpion} comes, therefore, from the region of integration near this value. Assuming the function $\varepsilon'_\gamma n(\varepsilon'_\gamma)$ is not varying too rapidly, we can approximate Eq.~\ref{eq:lrpion} as
    \begin{equation}\label{eq:lrpion2}
        \Gamma^\Delta_{p\to\pi}\left(\frac{E'_\pi}{\chi^\Delta_\pi}\right) \simeq   \left[\varepsilon'_\gamma n(\varepsilon'_\gamma)\right]_{\varepsilon'_\gamma=\frac{y_\Delta m_p \chi^\Delta_\pi}{E'_\pi}} \int_0^{+\infty} \frac{\text{d}y}{y_\Delta}  f_\Delta(y).
    \end{equation}
     The integral in Eq.~\ref{eq:lrpion2} does not depend on $E'_\pi$ anymore. We can substitute Eq.~\ref{eq:lrpion2} in Eq.~\ref{eq:energydeppion} to obtain that the $\Delta$-contribution to the pion injection spectrum depends on the energy as
\begin{equation}
E^{'2}_\pi Q^{\Delta}_{\pi} (E'_\pi)\propto \left[\varepsilon'_\gamma n(\varepsilon'_\gamma)\right]_{\varepsilon'_\gamma=\frac{y_\Delta m_p \chi^{\Delta}_\pi} {E'_\pi}} e^{-\frac{E'_\pi}{\chi^{\Delta}_\pi E'_{\rm{p,max}}}}.
\end{equation}
Thus if the target-photon energy density $\varepsilon'_\gamma n(\varepsilon'_\gamma)$, for $\varepsilon'_\gamma\geq\frac{y_\Delta m_p}{E'_{p,\text{max}}}$, is maximum at an energy $\bar{\varepsilon}'_\gamma$, the peak of the pion spectrum will be at an energy of $\bar{E}'_\pi=\frac{y_\Delta m_p \chi^{\Delta}_\pi}{\bar{\varepsilon}'_\gamma}$. If we neglect cooling effects, we can also determine the neutrino spectra approximately by observing that, in the pion decay, the neutrino typically carries an energy $E'_\nu\simeq 0.25 E'_\pi$. Therefore, if we introduce an effective $\chi^\Delta_\nu=0.25 \cdot \chi^{\Delta}_\pi \simeq 0.05$, we can deduce\footnote{The approximation that the neutrino carries a fixed fraction of the pion energy does not hold in the lower part of the energy range. More realistically, the neutrino energy from pion decay does not depend on the pion energy, so that $E^{'2}_\nu Q(E'_\nu)\propto E^{'2}_\nu$. This can give the dominant contribution at small neutrino energies.}
\begin{equation}
E^{'2}_\nu Q^{\Delta}_{\nu} (E'_\nu)\propto \left[\varepsilon'_\gamma n(\varepsilon'_\gamma)\right]_{\varepsilon'_\gamma=\frac{y_\Delta m_p \chi^\Delta_\nu} {E'_\nu}} e^{-\frac{E'_\nu}{\chi^\Delta_\nu E'_{\rm{p,max}}}};
\end{equation}

For the multi-pion production we refer to the model Sim-C in Ref.~\cite{Hummer:2010vx}. In this case, the pion carries a fraction  $\chi^{\text{MP}}_\pi=0.2$ of the energy of the parent protons. The function $f_{\text{MP}}(y)$ is not peaked as in the case of the $\Delta$-resonance contribution, but is rather slowly varying over the energy range of interest. For this reason, we can roughly set it equal to a constant $\bar{f}_{\text{MP}}$ above a threshold value\footnote{We adopt the value at which the multi-pion contribution approximately peaks in Fig.~4 of Ref.~\cite{Hummer:2010vx}. This value does not influence the rule for mapping a generic photon spectrum to the effective black-body spectrum in the thermal model.} of $y_{\text{MP}} \simeq 0.5$~GeV 
    \begin{equation}
        \Gamma^\text{MP}_{p\to\pi}\left(\frac{E'_\pi}{\chi^\text{MP}_\pi}\right)=\bar{f}_\text{MP}\int_{\frac{y_\text{MP} m_p \chi^\text{MP}_\pi}{E'_\pi}}^{+\infty} \text{d}\varepsilon'_\gamma n(\varepsilon'_\gamma).
    \end{equation}
    We can rewrite this in terms of the logarithm of the photon energy as
    \begin{equation}\label{eq:multipion2}
        \Gamma^\text{MP}_{p\to\pi}\left(\frac{E'_\pi}{\chi^\text{MP}_\pi}\right) \simeq \bar{f}_\text{MP}\int_{\log_{10}\left[\frac{y_\text{MP} m_p \chi^\text{MP}_\pi}{E'_\pi}\right]}^{+\infty} \text{d}\left(\log_{10}\varepsilon'_\gamma\right)\log(10) \left[\varepsilon'_\gamma n(\varepsilon'_\gamma)\right].
    \end{equation}
    The dependence of the integral on the pion energy is non-trivial to discuss in full generality. We will treat the case that the photon number density $\varepsilon'_\gamma n(\varepsilon'_\gamma)$ has a maximum at $\bar{\varepsilon}'_\gamma$.
    
For small pion energies, such that $E'_\pi<\frac{y_{\text{MP}} m_p \chi^{\text{MP}}_\pi}{\bar{\varepsilon}'_\gamma}$, only photon energies $\varepsilon' > \bar{\varepsilon}'_\gamma$ contribute to the integral. In this region, the integrand function $\varepsilon'_\gamma n(\varepsilon'_\gamma)$ is monotonically decreasing with energy. Therefore, the largest contribution to the integral in Eq.~\ref{eq:multipion2} comes from the region near the lower limit of the integration, namely for $\log_{10}\varepsilon'_\gamma$ close to $\log_{10}\left[\frac{y_\text{MP} m_p \chi^\text{MP}_\pi}{E'_\pi}\right]$. An order-of-magnitude estimate of the integral is, therefore, given by $\left[\varepsilon'_\gamma n(\varepsilon'_\gamma)\right]_{\varepsilon'_\gamma=\frac{y_{\text{MP}} m_p \chi^{\text{MP}}_\pi}{E'_\pi}}$. In this pion-energy range, the photon-energy dependence of the multi-pion contribution is similar to that of the $\Delta$-contribution, even though the normalization may be different.
    
    For large pion energies, such that $E'_\pi>\frac{y_{\text{MP}} m_p \chi^{\text{MP}}_\pi}{\bar{\varepsilon}'_\gamma}$, the largest contribution to the integral is for $\varepsilon'_\gamma$ close to $\bar{\varepsilon}'_\gamma$, independent of the precise value of the lower limit; the integral is, therefore, only weakly dependent on the pion energy. From Eq.~\ref{eq:energydeppion} it then follows that, approximately, the spectrum $E^{'2}_\pi Q^{\text{MP}}_\pi (E'_\pi) \propto e^{-E'_\pi/\chi^{\text{MP}}_\pi E'_{\rm{p,max}}}$ for this energy range, so it is essentially flat before the cutoff energy.
    
    As a final step in the multi-pion production case, neglecting the cooling effects, we can introduce an effective $\chi^\text{MP}_\nu=0.25\chi^\text{MP}_\pi$ as the fraction of the proton energy carried by neutrinos. Since $\chi^\text{MP}_\pi\simeq\chi^\Delta_\pi$, we will use the approximation (also in App.~\ref{app:flavor}) $\chi^\text{MP}_\nu\simeq\chi^\Delta_\nu=\chi_\nu=0.05$. We then find
    \begin{equation}
        E^{'2}_\nu Q^\text{MP}_\nu (E'_\nu)\propto e^{-E'_\nu/\chi_\nu E'_\text{p,max}} \begin{cases}
         \left[\varepsilon'_\gamma n(\varepsilon'_\gamma)\right]_{\varepsilon'_\gamma=\frac{y_{\text{MP}} m_p \chi_\nu}{E'_\nu}} & E'_\nu<\frac{y_{\text{MP}} m_p \chi^{\text{MP}}_\pi }{\bar{\varepsilon}'_\gamma} \\
        \text{constant} & E'_\nu>\frac{y_{\text{MP}} m_p \chi^{\text{MP}}_\pi }{\bar{\varepsilon}'_\gamma}
        \end{cases}
    \end{equation}

For both the $\Delta$-resonance and the multi-pion production scenarios, the main parameter determining the pion spectrum is $\bar{\varepsilon}'_\gamma$. Therefore it is possible to reproduce the neutrino production from a generic photon target with density $n_\gamma(\varepsilon'_\gamma)$ using an effective black-body spectrum $n^{\text{th}}_\gamma(\varepsilon'_\gamma)$ provided that the values of $\bar{\varepsilon}'_\gamma$ are the same both for the effective black-body target and for the real target. This leads to the rule given in Eq.~\ref{eq:tempsource}.

In Fig.~\ref{thecomparison} we show two explicit examples of astrophysical neutrino fluxes from the thermal model. For the AGN benchmark, whose parameters are reported in Table~\ref{tabpara}, the target-photon number density in the kinematically allowed region\footnote{Since the maximal proton energy is $E'_\text{p,max}\simeq 3\times10^8$~GeV, the kinematically allowed range is $\varepsilon'_\gamma\geq 0.6$~eV.} decreases as $\varepsilon'_\gamma n(\varepsilon'_\gamma)\propto \varepsilon^{'-1.64}_\gamma$. Therefore, the energy at which the number density is maximum is the lower bound of the kinematically allowed region, namely $\bar{\varepsilon}'_\gamma=0.6$~eV. Correspondingly, the effective temperature used for the reproduction in the text is $T'=\bar{\varepsilon}'_\gamma/2.8=0.2$~eV. In this way, the peak of the spectrum is automatically reproduced at the correct energy. 

If the target-photon number density has a single well-defined maximum (in the kinematically allowed range $\varepsilon'_\gamma\geq y_\Delta m_p/E'_\text{p,max}$), it is somewhat intuitive that a black-body spectrum is apt to reproduce this situation since it is characterized by a single maximum as well. Spectra which are flat or very slowly varying with energy are more complicated to reproduce: this happens for the benchmark example of GRBs, reported in Table~\ref{tabpara}, where the photon number density $\varepsilon'_\gamma n(\varepsilon'_\gamma)$ is flat for energies between $\varepsilon'_\text{min}$ and $\varepsilon'_\text{b}$. Therefore, there is no well-defined maximum, and the $\Delta$-contribution to the interaction rate is essentially constant for energies $E'_\pi> y_\Delta m_p \chi^\Delta_\pi/\varepsilon'_\text{b}$ as can be seen from Eq.~\ref{eq:lrpion2}. The multi-pion contribution to the interaction rate, from Eq.~\ref{eq:multipion2}, is slightly more complicated. For energies $E'_\pi>y_\text{MP} m_p \chi^\text{MP}_\pi/\varepsilon'_\text{b}$ there is a contribution to the integral that slowly increases as $\log_{10} \left[\frac{\varepsilon'_\text{b} E'_\pi}{y_\text{MP} m_p \chi^\text{MP}_\pi}\right]$. Since this is only a logarithmic growth, we will neglect it, which is justified if the range between $\varepsilon'_\text{min}$ and $\varepsilon'_\text{b}$ is not too large.  Therefore, in this case, the full interaction rate is flat or slowly growing for $E'_\pi > y_\text{MP} m_p \chi^\text{MP}_\pi/\varepsilon'_\text{b}$. To simulate this case with the thermal model, we can choose $\bar{\varepsilon}'_\gamma=\varepsilon'_\text{b}$ and $T'=\varepsilon'_\text{b}/2.8$. With this choice the interaction rate for the black-body spectrum for $E'_\pi > y_\text{MP} m_p \chi^\text{MP}_\pi/\varepsilon'_\text{b}$ is flat because of the multi-pion contribution, leading to an accurate reproduction of the pion and neutrino production; the slight difference between $y_\Delta$ and $y_\text{MP}$ does not produce significant discrepancies. For the numerical values of the parameters of the GRB benchmark given in Table~\ref{tabpara}, $\varepsilon'_\text{b}=14.8$~keV and consequently $T'=5.3$~keV.

\begin{figure}
    \centering
    \includegraphics[width=0.49\textwidth]{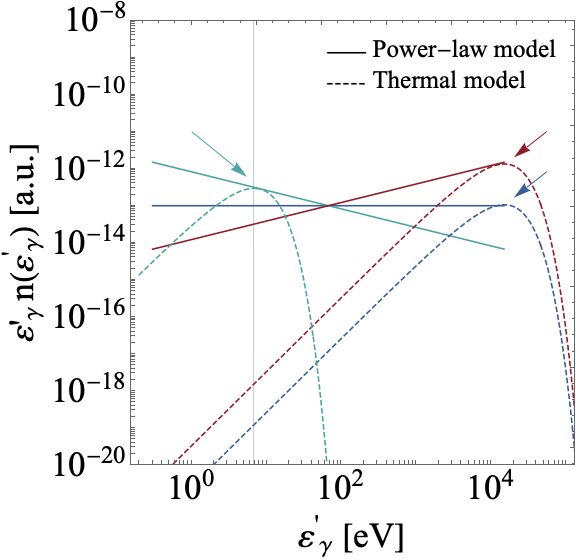}
    \includegraphics[width=0.49\textwidth]{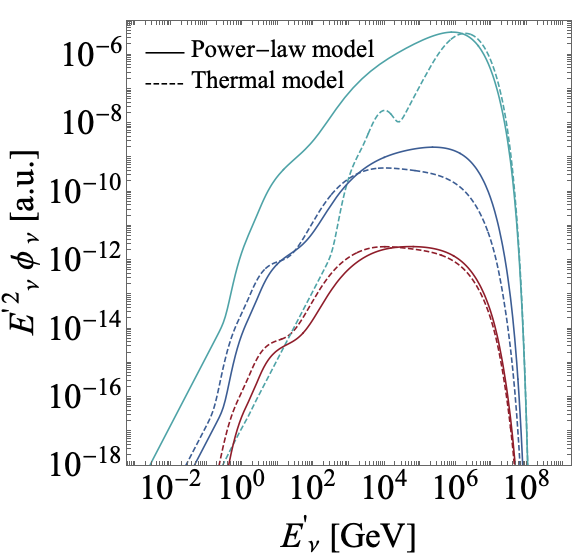}
    \caption{Target-photon and neutrino spectra in a power-law model and their reproduction in the thermal model, following the rules given in Sec.~\ref{sec:recipe}. In the left panel, we show the target-photon number density as a function of the energy for three different spectral indices of the power-law spectra (solid curves): green, blue and red lines correspond, respectively, to $\alpha=0.5$, $\alpha=1$ and $\alpha=1.5$. We also show the corresponding black-body spectra used within the thermal model (dashed curves), highlighting with arrows the thermal peak energies. The minimum photon energy which allows for photohadronic interactions is highlighted by a vertical line. In the right panel, we show the corresponding neutrino spectra: colors are the same as for the left panel. The normalizations both for the photon and for the neutrino spectra are arbitrary and are chosen only for graphical convenience.
    The parameters are fixed to the benchmark values $R'=10^{15}$~km, $B'=10^{-5}$~G (since we show all the quantities in the comoving frame there is no dependence on $\Gamma$).}
    \label{fig:limitationsthermal}
\end{figure}

In the case of GRBs, the performance of the thermal model is shown to be particularly good in Sec.~\ref{sec:thermalexamples}. This is connected with the presence of a strong magnetic field, which steepens the neutrino spectrum at high energies erasing the details from the target-photon spectrum. For this reason, we also discuss here the performance of the thermal model in the case of flat or slowly varying $\varepsilon'_\gamma n(\varepsilon'_\gamma)$ with very weak magnetic fields: these are expected to be the most complicated cases to reproduce with the thermal model since the photon spectra are far from the peaked structure of the black-body spectrum. We limit ourselves to a comparison between the thermal model and single-power-law models, $n(\varepsilon'_\gamma)\propto \varepsilon^{'-\alpha}_\gamma$, with $\alpha$ close to $1$: in particular, we choose as benchmark values $\alpha=0.5$, $\alpha=1$ (as in GRBs) and $\alpha=1.5$. We fix the other parameters to the benchmark values $R'=10^{15}$~km and $B'=10^{-5}$~G (small enough to avoid magnetic field effects on the secondaries). The corresponding photon number densities and neutrino fluxes are shown in Fig.~\ref{fig:limitationsthermal} -- following the rules given in Sec.~\ref{sec:recipe}.

For a slightly increasing photon number density (red curve), corresponding to $\alpha=0.5$, the value of $\bar{\varepsilon}'_\gamma$ is at the upper end of the power-law spectrum, $\varepsilon'_\text{b}=14.3$~keV. Correspondingly, the equivalent black-body photon spectrum is chosen to be peaked there. In this case, the reproduction with the thermal model is particularly good, due to the flat region in the $E^{'2}_\nu \phi_\nu$ neutrino spectrum coming from the multi-pion production.

For a constant photon number density (blue curve), corresponding to $\alpha=1$, the value of $\bar{\varepsilon}'_\gamma$ is again at $\varepsilon'_\text{b}=14.3$~keV, so the equivalent black-body spectrum is the same as in the previous case (we normalize it differently in Fig.~\ref{fig:limitationsthermal} for graphical clarity). The corresponding neutrino spectrum in the power-law model exhibits a slow (logarithmic) growth with energy. This feature is not captured by the thermal model, which instead predicts again a flat $E^{'2}_\nu \phi_\nu$ neutrino spectrum up to the maximal energy. We emphasize again that this also happens in the GRB scenario: however, the strong magnetic field, in that case, steepens the spectrum at high energies and erases this feature of spectral hardening, leading to a better performance of the thermal model.

The case of a slightly decreasing photon number density (green curve), corresponding to $\alpha=1.5$, is the hardest to reproduce with the thermal model. Here, the photon number density is larger at lower energies: in particular, the value of $\bar{\varepsilon}'_\gamma$ is determined by the minimum energy allowed by kinematics, namely $\bar{\varepsilon}'_\gamma=\frac{y_\Delta m_p}{E'_\text{p,max}}=6.7$~eV. The corresponding neutrino spectrum predicted by the thermal model is peaked near the maximal neutrino energy by construction, whereas the spectrum generated by the power-law target photons is softer below the neutrino spectral peak. 

A final point to consider is the dependence on the proton spectrum. Throughout the paper we have assumed the protons to be distributed as a power law in energy, $N(E'_\text{p})\propto E^{'-\alpha_\text{p}}_p$, with a fixed spectral index $\alpha_\text{p}=2$: we now discuss what happens when we relax this assumption and leave the spectral index free to vary. In this more general case, the results in Eqs.~\ref{eq:lrpion2} and \ref{eq:multipion2} are still valid since they give the interaction rates independently of the proton spectrum. The interaction rate for the $\Delta$-resonance follows the target-photon spectrum, and, therefore, for a thermal target, it is peaked in energy just as the black-body spectrum. The $\Delta$-resonance contribution will, therefore, be peaked regardless of the proton spectral index $\alpha_\text{p}$. On the other hand, the interaction rate for the multi-pion contribution becomes constant above the threshold energy for pion production. A constant interaction rate means that the neutrino fluxes will follow the parent proton spectrum. The multi-pion contribution to the neutrino spectrum will, therefore, be a power law with the same spectral index as the protons.

\begin{figure}
    \centering
    \includegraphics[width=0.5\textwidth]{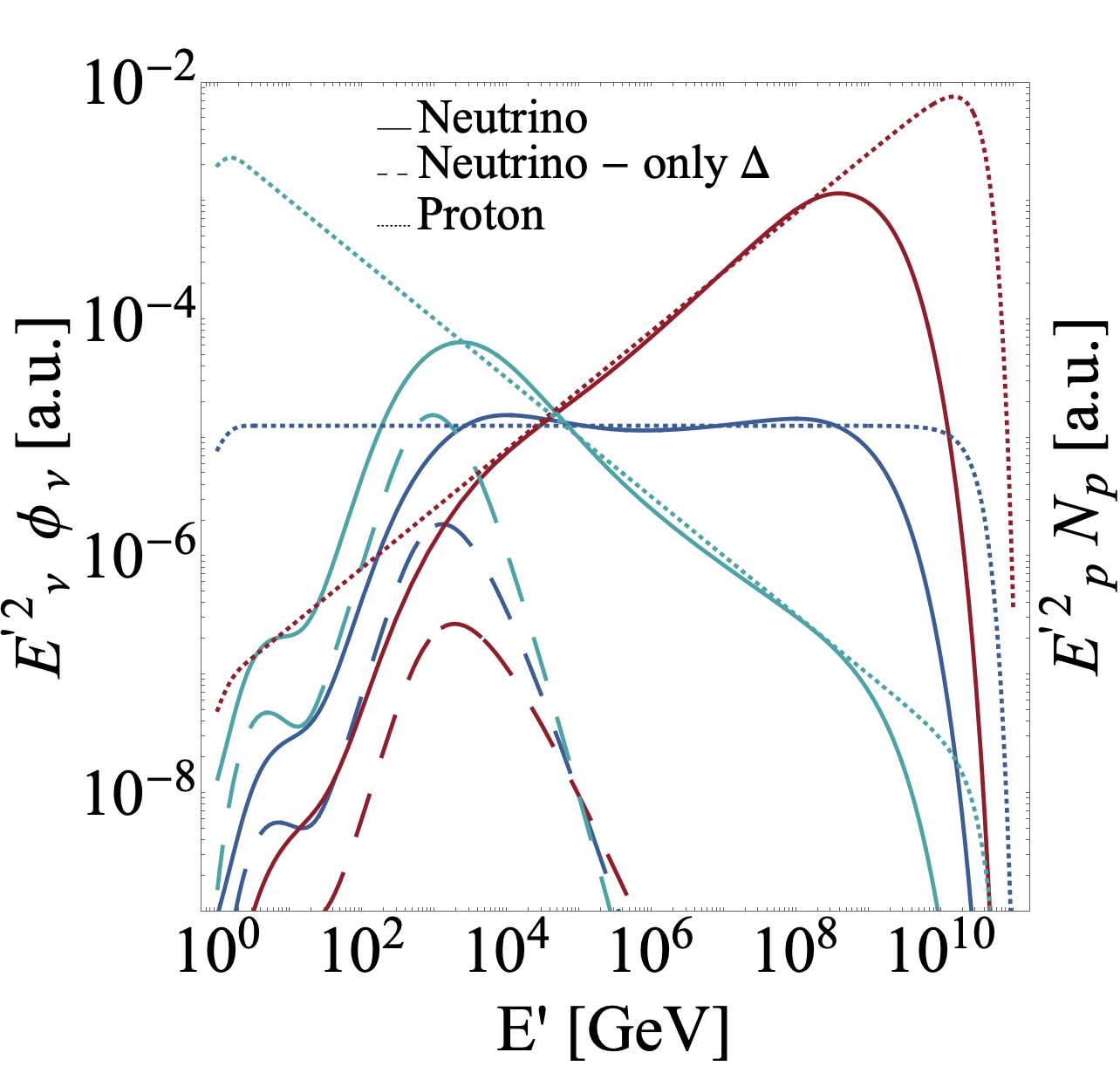}
    \caption{Neutrino fluxes for varying spectral indices of the proton spectrum. We show the neutrino fluxes (solid) as a function of the energy, as well as the $\Delta$-resonance contribution to the neutrino fluxes (dashed) and the parent proton spectra (dotted). The spectral index of the proton spectrum is chosen as $\alpha_\text{p}=2.5$ (green), $\alpha_\text{p}=2$ (blue) and $\alpha_\text{p}=1.5$ (red). All the other parameters are fixed to benchmark values, namely $T'=5.3$~keV, $B'=0.01$~G, $R'=10^{15}$~km (since we show all the quantities in the comoving frame there is no dependence on $\Gamma$). The normalization of the fluxes and the spectra are arbitrary.}
    \label{fig:spectralindexprotons}
\end{figure}

These conclusions are confirmed by Fig.~\ref{fig:spectralindexprotons}, where we show the shape of the neutrino spectra for varying spectral indices of the proton and large maximal proton energies. We also show the contribution from the $\Delta$-resonance separately; as expected, the $\Delta$-resonance contribution is peaked at the same energy independent of the proton spectral index. On the other hand, for all values of $\alpha_\text{p}$, the full neutrino spectrum behaves as a power law following the proton spectrum up to the maximal energy. 

\section{Analytical treatment of flavor regimes} \label{app:flavor}

In this appendix, we discuss the different flavor regimes and we estimate the energy range in which they can be expected. Our focus is on how these energy ranges depend on the astrophysical parameters. As in App.~\ref{app:thermalmodel}, we will denote the energies of different species with different subscripts.
\begin{itemize}
    \item \textbf{Neutron-beam regime:} in this regime neutrinos mainly come from neutron decay. A neutron produced in the reaction $p+\gamma\to n+\pi^+$ typically carries $4/5$ of the parent proton energy, and in turn the neutrino from the neutron decay typically carries a fraction $\chi=5.1\times 10^{-4}$ of the energy of the parent neutron~\cite{Lipari:2007su}. Therefore, for a proton energy $E'_{p}$, the energy for neutrinos from neutron decays will be $4\times 10^{-4} E'_{p}$. For a given effective temperature $T'$ of the photon spectrum, the protons which interact through the $\Delta$-resonance to produce neutrons typically have energies close to
\begin{equation}
    E'_p=\frac{y_\Delta m_p}{2.8 T'},
\end{equation}
where the factor of 2.8 takes into account that the number density of target photons is peaked at an energy $2.8 T'$.
Consequently, the neutron-beam regime will be expected for typical neutrino energies close to
\begin{equation}
    E'_\nu=4\times 10^{-4} \frac{y_\Delta m_p}{2.8 T'}.
\end{equation}
The transition to the neutron-beam regime is not always well-captured by the thermal model. This transition comes from the bump of neutrinos originating from neutron decay. The energy at which the passage to the neutron-beam regime happens is determined by the intensity of this bump compared with the neutrino flux from pion decays. As discussed in Sec.~\ref{sec:thermalexamples}, the neutrino flux from pion decays is not necessarily well captured by the thermal model at energies significantly lower than the peak of the flux. Therefore, when the transition to the neutron-beam regime happens at these energies, we do not expect it to be reproduced by the thermal model. Furthermore, the decay kinematics of the neutrons is not well reproduced in NeuCosmA by the chosen decay model, which means that the spectral shape below the neutron peak may be too hard; since this is far below the overall peak energy, and the thermal model is not very accurate there, we did not follow up on this issue in greater detail.

\item \textbf{Pion-beam regime:} in the pion-beam regime neutrinos mainly come from pion and muon decays. At high energies there are two processes that can inhibit the pion-beam regime, depending on the magnetic field of the source. If the magnetic field is sufficiently strong (see below for an estimate), the flux will transition to a muon-damped regime (eventually passing through an intermediate region of muon-beam regime coming from pile-up neutrinos); if the magnetic field is very weak, the flux will instead transition to a muon-beam regime.

\item \textbf{Muon-damped regime:} in the muon-damped regime, muon decays are inhibited by the synchrotron losses of the muons and neutrinos are predominantly produced by pion decays. By imposing the equality between the synchrotron energy-loss lifetime and the decay lifetime of muons~\cite{Kashti:2005qa,Lipari:2007su,Baerwald:2011ee,Bustamante:2020bxp}, one can define a critical energy above which the muon spectrum will be steepened by the energy losses
\begin{equation}
    E^{'}_{c,\mu}=4.8\times 10^9\; \text{GeV} \left(\frac{B'}{1\; \text{G}}\right)^{-1}.
\end{equation}
Since neutrinos carry an average energy of about $0.3$ of their parent muon energy, the corresponding neutrino energy can be estimated as
\begin{equation}
    E^{'\mu \text{ damp}}_{\nu}=1.4\times 10^9\; \text{GeV} \left(\frac{B'}{1\; \text{G}}\right)^{-1}.
\end{equation}
If this energy is smaller than the cutoff energy, namely
\begin{equation}
    E^{'\mu \text{ damp}}_{\nu}< 0.05 E'_{p,\text{max}},
\end{equation}
the flux will exhibit a transition from the pion-beam to the muon-damped regime at that energy.

At sufficiently high energies, pion decay can be inhibited too by the large synchrotron losses. By imposing the equality between the synchrotron energy-loss and decay lifetimes, we find that the critical energy for pions is
\begin{equation}
    E^{'}_{c,\pi}=8.9\times 10^{10}\; \text{GeV} \left(\frac{B'}{1\; \text{G}}\right)^{-1},
\end{equation}
and, since neutrinos carry a fraction 0.2 of the parent pion energy, the corresponding neutrino energy is
\begin{equation}
    E^{'\mu \text{ cut}}_\nu=1.8\times 10^{10}\; \text{GeV} \left(\frac{B'}{1\; \text{G}}\right)^{-1}.
\end{equation}

\item \textbf{Kaon-beam regime:} at energies higher than $E^{'\mu \text{ cut}}_\nu$ the dominant source of neutrinos will be kaon decays via the channel $K^+\to \mu^++\nu_\mu$. Kaons, due to their higher critical energy, are not suppressed in this range. 
Note that in this work we only consider the leading kaon production mode $p+\gamma\to K^++\Lambda/\Sigma$, as in Ref.~\cite{Baerwald:2011ee}.

\item \textbf{Muon-beam regime:} in the muon-beam regime neutrinos are produced mainly in muon decays. The muon-beam regime was originally proposed at very high energies from decays of heavy charmed mesons~\cite{Pakvasa:2010jj}. In Ref.~\cite{Hummer:2010ai} was shown that the muon-beam regime can also come from the pile up of neutrinos from muon decays which were depleted in the muon-damped regime and pile up at lower energies. We find that the muon-beam regime can also appear in a different context.

As we discussed above, if the condition
\begin{equation}
    E^{'\mu \text{ damp}}_{\nu}> 0.05 E'_{p,\text{max}}
\end{equation}
holds, the transition to a muon-damped regime will not appear as the spectrum will be cut off at an energy of $0.05 E'_{p,\text{max}}$. The flux will instead transition to a muon-beam regime: from the decay functions in Ref.~\cite{Lipari:2007su} can be shown that neutrinos from pion decay typically carry a slightly smaller fraction of the energy of the parent proton than neutrinos from muon decay because the decay function peaks at lower energies. While this is a small numerical difference, which does not produce noticeable effects for most of the energy range, it is instead significant near the cutoff. For neutrino energies larger than $0.05 E'_{p,\text{max}}$, neutrinos from pion decay will be cut off, while neutrinos from muon decay will not: the flavor regime will then be of the muon-beam type in a very narrow region near the end of the spectrum.
\end{itemize}

\section{Methodology of comparison with the experiments}\label{app:methods}

In this appendix, we discuss the methods used to compute the event rates at each of the three experiments in terms of the astrophysical flux. The differential limit in terms of fluence per flavor of a given experiment is given by
\begin{equation}\label{eq:sensi}
    S^{\alpha}(E)=\frac{2.44 E}{\log\left(10\right) A^{\alpha}_{\text{eff}}(E)},
\end{equation}
with $A^{\alpha}_{\text{eff}} (E)$ the effective area of the experiment (averaged over the angle) for neutrinos of flavor $\alpha$ as a function of the energy $E$.
$S^{\alpha}(E)$ represents the minimum fluence that produces $2.44$ expected events per decade of energy. With this definition, the number of events expected at the experiment is
\begin{equation}
    N=2.44 \sum_\alpha \int \frac{E^2\mathcal{F}_{\nu\alpha}}{S^\alpha (E)}\text{d}\left[\log_{10} E\right].
\end{equation}

In our analysis, for DeepCore and KM3NeT, we only study the sensitivity to muon neutrinos and antineutrinos, using the effective areas provided, respectively, in Refs.~\cite{Aartsen:2016zhm} and \cite{Adrian-Martinez:2016fdl}. The angular reconstruction of neutrino events is best effected for tracks originating from muon neutrinos and antineutrinos. On the other hand, for IceCube-Gen2 Radio Array we use the all-flavor differential limit given in Ref.~\cite{Blaufuss:2015muc}. 

Eq.~\ref{eq:sensi} refers to the background-free case, in which the sensitivities $S^{\alpha}(E)$ do not depend on the exposure time. The reason is that, for a given fluence of the source, the total number of events is the same independently of the exposure time. Correspondingly, the minimum energy fluence needed for detection $\xi$, as defined in Sec.~\ref{sec:resul}, is constant in time as well. Throughout Sec.~\ref{sec:resul} we stick to this background-free regime, so that all the experiments are treated in the same way. On the other hand, neglecting the background from atmospheric neutrinos can only be justified for the neutrino radio arrays and partially for the neutrino telescopes, because of the different energy range. For dense neutrino arrays, such as DeepCore, the results obtained in Sec.~\ref{sec:resul} should be taken as indicative, rather than precise. We now estimate the rate of background events in DeepCore and provide a simple analytical discussion of the dependence of $\xi$ on this rate.

We follow the methodology suggested in Ref.~\cite{Backstrom:2018plo} for the treatment of the background. In particular, using the atmospheric model proposed in Ref.~\cite{Honda:2015fha}, we compute the number of expected detected muon-neutrino events per unit time as
\begin{equation}
    \frac{\text{d}N}{\text{d}t}=\bar{\eta}\int A_{\text{eff}}(E) \frac{\text{d}\Phi_{\text{background}}}{\text{d}E\text{d}\Omega} \left[1-\cos(\theta_{\text{res}}(E))\right] \text{d}\Omega \text{d}E,
\end{equation}
where $\theta_{\text{res}}(E)$ is the angular resolution and $A_{\text{eff}}(E)$ is the effective area of DeepCore, both of which are provided in Ref.~\cite{Aartsen:2016zhm}; $\frac{\text{d}\Phi_{\text{background}}}{\text{d}E\text{d}\Omega}$ is the atmospheric flux. The factor $\bar{\eta}$ is the veto factor introduced to correct for the contamination of atmospheric muons; following Ref.~\cite{Backstrom:2018plo} we assume $\bar{\eta}=1.76$.

Using this estimate, we can obtain the number of background events expected at DeepCore per unit time
\begin{equation}\label{eq:bckratedeepcore}
    \frac{\text{d}N}{\text{d}t}=1.5\times 10^{-5}\; \text{s}^{-1}.
\end{equation}
For small exposure times, of the order of some hours, the number of background events is smaller than $1$ and can therefore be neglected. In this case, the results of Sec.~\ref{sec:resul} for background-free detection are valid. On the other hand, for larger exposures, the background will be non-negligible. The sensitivity is, therefore, necessarily decreased since a larger number of signal events are necessary to be distinguished over the background. More specifically, in the presence of a non-negligible number of background events, the 90$\%$ Feldman-Cousins limit for signal detection, namely the factor of 2.44 in Eq.~\ref{eq:sensi}, increases. For several specific realizations of the number of background events $b$ the corresponding 90$\%$ limit is tabulated, for example, in Ref.~\cite{Feldman:1997qc}. A simple formula that reproduces the value of the Feldman-Cousins limit $\bar{n}(b)$ reasonably well is (see Eq.~1 of Ref.~\cite{Kierans:2021kpg})
\begin{equation}\label{eq:feldman}
    \bar{n}(b)=\frac{n_\text{sig}^2+\sqrt{n_\text{sig}^4+4n_\text{sig}^2 b}}{2},
\end{equation}
with $n_\text{sig}=1.64$ is the number of standard deviations corresponding to $90\%$ confidence level; notice that for $b=0$ this formula gives $\bar{n}(0)=2.69$, which is close to the exact value of $2.44$, so it is reasonable to extend the validity of this formula down to arbitrarily small values of background events.

Using this relation, we see that the $90\%$ confidence level differential limit in Eq.~\ref{eq:sensi} is changed in the presence of a rate of background events per unit time $\frac{\text{d}N}{\text{d}t}$ (for DeepCore this is given in Eq.~\ref{eq:bckratedeepcore}) to
\begin{equation}\label{eq:generalformula}
    S^\alpha (E)=\frac{\bar{n}\left(\frac{\text{d}N}{\text{d}t}t\right) E}{\log(10) A^{\alpha}_{\text{eff}}(E)},
\end{equation}
where $t$ is the exposure time.

In the limit of a large number of background events, $b \gg 1$, we recover the expected asymptotic behavior $\bar{n}\left(\frac{\text{d}N}{\text{d}t}t\right)\propto \sqrt{\frac{\text{d}N}{\text{d}t} t}$ and $S^\alpha\propto\sqrt{\frac{\text{d}N}{\text{d}t}t}$. In particular, since the energy flux needed for detection, $\xi$, scales in the same way as the differential limit, it follows that $\xi \propto \sqrt{t}$ in the background-dominated regime. This is in agreement with our qualitative expectations: for larger exposure times larger fluences are needed to detect the signal over the background\footnote{However, the necessary flux, namely the ratio between the fluence and the exposure time, decreases as $1/\sqrt{t}.$}. From Eq.~\ref{eq:generalformula} the results in Sec.~\ref{sec:resul} can be generalized to arbitrary values of the exposure time.

\bibliographystyle{unsrt}
\bibliography{Biblio}

\begin{thebibliography}{10}

\bibitem{Morejon:2019pfu}
Leonel Morejon, Anatoli Fedynitch, Denise Boncioli, Daniel Biehl, and Walter
  Winter.
\newblock {Improved photomeson model for interactions of cosmic ray nuclei}.
\newblock {\em JCAP}, 11:007, 2019.

\bibitem{Goldschmidt:2001qd}
A.~Goldschmidt.
\newblock {The IceCube detector}.
\newblock In {\em {27th International Cosmic Ray Conference}}, pages
  1237--1240, 8 2001.

\bibitem{Ahrens:2002dv}
J.~Ahrens et~al.
\newblock {Icecube - the next generation neutrino telescope at the south pole}.
\newblock {\em Nucl. Phys. B Proc. Suppl.}, 118:388--395, 2003.

\bibitem{Kappes:2007ci}
A.~Kappes.
\newblock {KM3NeT: A Next Generation Neutrino Telescope in the Mediterranean
  Sea}.
\newblock 11 2007.

\bibitem{Aslanides:1999vq}
E.~Aslanides et~al.
\newblock {A deep sea telescope for high-energy neutrinos}.
\newblock 5 1999.

\bibitem{Aartsen:2014oha}
M.G. Aartsen et~al.
\newblock {Letter of Intent: The Precision IceCube Next Generation Upgrade
  (PINGU)}.
\newblock 1 2014.

\bibitem{Barwick:2007vba}
Steven~W. Barwick.
\newblock {ARIANNA: A New Concept for UHE Neutrino Detection}.
\newblock In {\em {30th International Cosmic Ray Conference}}, volume~5, pages
  1601--1604, 7 2007.

\bibitem{Meures:2012fka}
T.~Meures.
\newblock {The Askar'yan Radio Array, an UHE-neutrino detector at South Pole}.
\newblock {\em Proc. Int. Sch. Phys. Fermi}, 182:321--325, 2012.

\bibitem{Martineau-Huynh:2015hae}
Olivier Martineau-Huynh et~al.
\newblock {The Giant Radio Array for Neutrino Detection}.
\newblock {\em EPJ Web Conf.}, 116:03005, 2016.

\bibitem{Schulz:2009zz}
Olaf Schulz.
\newblock {The IceCube DeepCore}.
\newblock {\em AIP Conf. Proc.}, 1085(1):783--786, 2009.

\bibitem{Blaufuss:2015muc}
Erik Blaufuss, C.~Kopper, and C.~Haack.
\newblock {The IceCube-Gen2 High Energy Array}.
\newblock {\em PoS}, ICRC2015:1146, 2016.

\bibitem{Collaboration:2011nsa}
M.~Ageron et~al.
\newblock {ANTARES: the first undersea neutrino telescope}.
\newblock {\em Nucl. Instrum. Meth. A}, 656:11--38, 2011.

\bibitem{Belolaptikov:1997ry}
I.A. Belolaptikov et~al.
\newblock {The Baikal underwater neutrino telescope: Design, performance and
  first results}.
\newblock {\em Astropart. Phys.}, 7:263--282, 1997.

\bibitem{Riccobene:2017fpr}
Giorgio Riccobene.
\newblock {Status and development of KM3NeT/ARCA}.
\newblock {\em PoS}, NOW2016:054, 2017.

\bibitem{Aartsen:2013bka}
M.G. Aartsen et~al.
\newblock {First observation of PeV-energy neutrinos with IceCube}.
\newblock {\em Phys. Rev. Lett.}, 111:021103, 2013.

\bibitem{Aartsen:2013jdh}
M.G. Aartsen et~al.
\newblock {Evidence for High-Energy Extraterrestrial Neutrinos at the IceCube
  Detector}.
\newblock {\em Science}, 342:1242856, 2013.

\bibitem{Aartsen:2014gkd}
M.G. Aartsen et~al.
\newblock {Observation of High-Energy Astrophysical Neutrinos in Three Years of
  IceCube Data}.
\newblock {\em Phys. Rev. Lett.}, 113:101101, 2014.

\bibitem{Aartsen:2015rwa}
M.G. Aartsen et~al.
\newblock {Evidence for Astrophysical Muon Neutrinos from the Northern Sky with
  IceCube}.
\newblock {\em Phys. Rev. Lett.}, 115(8):081102, 2015.

\bibitem{Aartsen:2020aqd}
M.~G. Aartsen et~al.
\newblock {Characteristics of the diffuse astrophysical electron and tau
  neutrino flux with six years of IceCube high energy cascade data}.
\newblock {\em Phys. Rev. Lett.}, 125(12):121104, 2020.

\bibitem{Bartos:2016wud}
I.~Bartos, M.~Ahrens, C.~Finley, and S.~Marka.
\newblock {Prospects of Establishing the Origin of Cosmic Neutrinos using
  Source Catalogs}.
\newblock {\em Phys. Rev. D}, 96(2):023003, 2017.

\bibitem{Aartsen:2016lir}
M.G. Aartsen et~al.
\newblock {The contribution of Fermi-2LAC blazars to the diffuse TeV-PeV
  neutrino flux}.
\newblock {\em Astrophys. J.}, 835(1):45, 2017.

\bibitem{Aartsen:2017wea}
M.G. Aartsen et~al.
\newblock {Extending the search for muon neutrinos coincident with gamma-ray
  bursts in IceCube data}.
\newblock {\em Astrophys. J.}, 843(2):112, 2017.

\bibitem{Mertsch:2016hcd}
Philipp Mertsch, Mohamed Rameez, and Irene Tamborra.
\newblock {Detection prospects for high energy neutrino sources from the
  anisotropic matter distribution in the local universe}.
\newblock {\em JCAP}, 03:011, 2017.

\bibitem{Aartsen:2018ywr}
M.G. Aartsen et~al.
\newblock {Search for steady point-like sources in the astrophysical muon
  neutrino flux with 8 years of IceCube data}.
\newblock {\em Eur. Phys. J. C}, 79(3):234, 2019.

\bibitem{Murase:2013rfa}
Kohta Murase, Markus Ahlers, and Brian~C. Lacki.
\newblock {Testing the Hadronuclear Origin of PeV Neutrinos Observed with
  IceCube}.
\newblock {\em Phys. Rev. D}, 88(12):121301, 2013.

\bibitem{Murase:2015xka}
Kohta Murase, Dafne Guetta, and Markus Ahlers.
\newblock {Hidden Cosmic-Ray Accelerators as an Origin of TeV-PeV Cosmic
  Neutrinos}.
\newblock {\em Phys. Rev. Lett.}, 116(7):071101, 2016.

\bibitem{Ando:2015bva}
Shin'ichiro Ando, Irene Tamborra, and Fabio Zandanel.
\newblock {Tomographic Constraints on High-Energy Neutrinos of Hadronuclear
  Origin}.
\newblock {\em Phys. Rev. Lett.}, 115(22):221101, 2015.

\bibitem{Silvestri:2009xb}
Andrea Silvestri and Steven~W. Barwick.
\newblock {Constraints on Extragalactic Point Source Flux from Diffuse Neutrino
  Limits}.
\newblock {\em Phys. Rev. D}, 81:023001, 2010.

\bibitem{Ahlers:2014ioa}
Markus Ahlers and Francis Halzen.
\newblock {Pinpointing Extragalactic Neutrino Sources in Light of Recent
  IceCube Observations}.
\newblock {\em Phys. Rev. D}, 90(4):043005, 2014.

\bibitem{Murase:2016gly}
Kohta Murase and Eli Waxman.
\newblock {Constraining High-Energy Cosmic Neutrino Sources: Implications and
  Prospects}.
\newblock {\em Phys. Rev. D}, 94(10):103006, 2016.

\bibitem{Dekker:2018cqu}
Ariane Dekker and Shin'ichiro Ando.
\newblock {Angular power spectrum analysis on current and future high-energy
  neutrino data}.
\newblock {\em JCAP}, 02:002, 2019.

\bibitem{Stecker:1991vm}
F.W. Stecker, C.~Done, M.H. Salamon, and P.~Sommers.
\newblock {High-energy neutrinos from active galactic nuclei}.
\newblock {\em Phys. Rev. Lett.}, 66:2697--2700, 1991.
\newblock [Erratum: Phys.Rev.Lett. 69, 2738 (1992)].

\bibitem{Piran:2004ba}
Tsvi Piran.
\newblock {The physics of gamma-ray bursts}.
\newblock {\em Rev. Mod. Phys.}, 76:1143--1210, 2004.

\bibitem{Wang:2011ip}
Xiang-Yu Wang, Ruo-Yu Liu, Zi-Gao Dai, and K.S. Cheng.
\newblock {Probing the tidal disruption flares of massive black holes with
  high-energy neutrinos}.
\newblock {\em Phys. Rev. D}, 84:081301, 2011.

\bibitem{IceCube:2018cha}
M.G. Aartsen et~al.
\newblock {Neutrino emission from the direction of the blazar TXS 0506+056
  prior to the IceCube-170922A alert}.
\newblock {\em Science}, 361(6398):147--151, 2018.

\bibitem{IceCube:2018dnn}
M.G. Aartsen et~al.
\newblock {Multimessenger observations of a flaring blazar coincident with
  high-energy neutrino IceCube-170922A}.
\newblock {\em Science}, 361(6398):eaat1378, 2018.

\bibitem{Stein:2020xhk}
Robert Stein et~al.
\newblock {A high-energy neutrino coincident with a tidal disruption event}.
\newblock 5 2020.

\bibitem{Gao:2018mnu}
Shan Gao, Anatoli Fedynitch, Walter Winter, and Martin Pohl.
\newblock {Modelling the coincident observation of a high-energy neutrino and a
  bright blazar flare}.
\newblock {\em Nature Astron.}, 3(1):88--92, 2019.

\bibitem{Cerruti:2018tmc}
M.~Cerruti, A.~Zech, C.~Boisson, G.~Emery, S.~Inoue, and J.-P. Lenain.
\newblock {Leptohadronic single-zone models for the electromagnetic and
  neutrino emission of TXS 0506+056}.
\newblock {\em Mon. Not. Roy. Astron. Soc.}, 483(1):L12--L16, 2019.

\bibitem{Keivani:2018rnh}
A.~Keivani et~al.
\newblock {A Multimessenger Picture of the Flaring Blazar TXS 0506+056:
  implications for High-Energy Neutrino Emission and Cosmic Ray Acceleration}.
\newblock {\em Astrophys. J.}, 864(1):84, 2018.

\bibitem{Rodrigues:2018tku}
Xavier Rodrigues, Shan Gao, Anatoli Fedynitch, Andrea Palladino, and Walter
  Winter.
\newblock {Leptohadronic Blazar Models Applied to the 2014--2015 Flare of TXS
  0506+056}.
\newblock {\em Astrophys. J. Lett.}, 874(2):L29, 2019.

\bibitem{Winter:2020ptf}
Walter Winter and Cecilia Lunardini.
\newblock {A concordance scenario for the observation of a neutrino from the
  Tidal Disruption Event AT2019dsg}.
\newblock {\em Nat. Astron.}, 2021.

\bibitem{Murase:2020lnu}
Kohta Murase, Shigeo~S. Kimura, B.~Theodore Zhang, Foteini Oikonomou, and Maria
  Petropoulou.
\newblock {High-Energy Neutrino and Gamma-Ray Emission from Tidal Disruption
  Events}.
\newblock {\em Astrophys. J.}, 902(2):108, 2020.

\bibitem{Liu:2020isi}
Ruo-Yu Liu, Shao-Qiang Xi, and Xiang-Yu Wang.
\newblock {Neutrino emission from an off-axis jet driven by the tidal
  disruption event AT2019dsg}.
\newblock {\em Phys. Rev. D}, 102(8):083028, 2020.

\bibitem{Liu:2018utd}
Ruo-Yu Liu, Kai Wang, Rui Xue, Andrew~M. Taylor, Xiang-Yu Wang, Zhuo Li, and
  Huirong Yan.
\newblock {Hadronuclear interpretation of a high-energy neutrino event
  coincident with a blazar flare}.
\newblock {\em Phys. Rev. D}, 99(6):063008, 2019.

\bibitem{Winter:2012xq}
Walter Winter.
\newblock {Neutrinos from Cosmic Accelerators Including Magnetic Field and
  Flavor Effects}.
\newblock {\em Adv. High Energy Phys.}, 2012:586413, 2012.

\bibitem{Winter:2011jr}
Walter Winter.
\newblock {Interpretation of neutrino flux limits from neutrino telescopes on
  the Hillas plot}.
\newblock {\em Phys. Rev.}, D85:023013, 2012.

\bibitem{Mucke:1999yb}
A.~Mucke, Ralph Engel, J.P. Rachen, R.J. Protheroe, and Todor Stanev.
\newblock {SOPHIA: Monte Carlo simulations of photohadronic processes in
  astrophysics}.
\newblock {\em Comput. Phys. Commun.}, 124:290--314, 2000.

\bibitem{Kashti:2005qa}
Tamar Kashti and Eli Waxman.
\newblock {Flavoring astrophysical neutrinos: Flavor ratios depend on energy}.
\newblock {\em Phys. Rev. Lett.}, 95:181101, 2005.

\bibitem{Lipari:2007su}
Paolo Lipari, Maurizio Lusignoli, and Davide Meloni.
\newblock {Flavor Composition and Energy Spectrum of Astrophysical Neutrinos}.
\newblock {\em Phys. Rev. D}, 75:123005, 2007.

\bibitem{Hummer:2010ai}
S.~Hummer, M.~Maltoni, W.~Winter, and C.~Yaguna.
\newblock {Energy dependent neutrino flavor ratios from cosmic accelerators on
  the Hillas plot}.
\newblock {\em Astropart. Phys.}, 34:205--224, 2010.

\bibitem{Hummer:2010vx}
S.~Hummer, M.~Ruger, F.~Spanier, and W.~Winter.
\newblock {Simplified models for photohadronic interactions in cosmic
  accelerators}.
\newblock {\em Astrophys. J.}, 721:630--652, 2010.

\bibitem{Baerwald:2010fk}
Philipp Baerwald, Svenja Hummer, and Walter Winter.
\newblock {Magnetic Field and Flavor Effects on the Gamma-Ray Burst Neutrino
  Flux}.
\newblock {\em Phys. Rev. D}, 83:067303, 2011.

\bibitem{Baerwald:2011ee}
Philipp Baerwald, Svenja Hummer, and Walter Winter.
\newblock {Systematics in the Interpretation of Aggregated Neutrino Flux Limits
  and Flavor Ratios from Gamma-Ray Bursts}.
\newblock {\em Astropart. Phys.}, 35:508--529, 2012.

\bibitem{Roulet:2020yye}
Esteban Roulet and Francesco Vissani.
\newblock {On the energy of the protons producing the very high-energy
  astrophysical neutrinos}.
\newblock 11 2020.

\bibitem{Rasmussen:2017ert}
Rasmus~W. Rasmussen, Lukas Lechner, Markus Ackermann, Marek Kowalski, and
  Walter Winter.
\newblock {Astrophysical neutrinos flavored with Beyond the Standard Model
  physics}.
\newblock {\em Phys. Rev. D}, 96(8):083018, 2017.

\bibitem{Winter:2013cla}
Walter Winter.
\newblock {Photohadronic Origin of the TeV-PeV Neutrinos Observed in IceCube}.
\newblock {\em Phys. Rev. D}, 88:083007, 2013.

\bibitem{Bustamante:2020bxp}
Mauricio Bustamante and Irene Tamborra.
\newblock {Using High-Energy Neutrinos As Cosmic Magnetometers}.
\newblock 9 2020.

\bibitem{Ambrosone:2020evo}
Antonio Ambrosone, Marco Chianese, Damiano F.~G. Fiorillo, Antonio Marinelli,
  Gennaro Miele, and Ofelia Pisanti.
\newblock {Starburst galaxies strike back: a multi-messenger analysis with
  Fermi-LAT and IceCube data}.
\newblock {\em Mon. Not. Roy. Astron. Soc.}, 503(3):4032, 2021.

\bibitem{Rodrigues:2017fmu}
Xavier Rodrigues, Anatoli Fedynitch, Shan Gao, Denise Boncioli, and Walter
  Winter.
\newblock {Neutrinos and Ultra-High-Energy Cosmic-Ray Nuclei from Blazars}.
\newblock {\em Astrophys. J.}, 854(1):54, 2018.

\bibitem{Guepin:2017dfi}
Claire Gu\'epin and Kumiko Kotera.
\newblock {Can we observe neutrino flares in coincidence with explosive
  transients?}
\newblock {\em Astron. Astrophys.}, 603:A76, 2017.

\bibitem{Murase:2006mm}
Kohta Murase, Kunihito Ioka, Shigehiro Nagataki, and Takashi Nakamura.
\newblock {High Energy Neutrinos and Cosmic-Rays from Low-Luminosity Gamma-Ray
  Bursts?}
\newblock {\em Astrophys. J. Lett.}, 651:L5--L8, 2006.

\bibitem{Murase:2008sp}
Kohta Murase.
\newblock {Prompt High-Energy Neutrinos from Gamma-Ray Bursts in the
  Photospheric and Synchrotron Self-Compton Scenarios}.
\newblock {\em Phys. Rev. D}, 78:101302, 2008.

\bibitem{Kashiyama:2012zn}
Kazumi Kashiyama, Kohta Murase, Shunsaku Horiuchi, Shan Gao, and Peter
  Meszaros.
\newblock {High energy neutrino and gamma ray transients from relativistic
  supernova shock breakouts}.
\newblock {\em Astrophys. J. Lett.}, 769:L6, 2013.

\bibitem{Murase:2013ffa}
Kohta Murase and Kunihito Ioka.
\newblock {TeV\textendash{}PeV Neutrinos from Low-Power Gamma-Ray Burst Jets
  inside Stars}.
\newblock {\em Phys. Rev. Lett.}, 111(12):121102, 2013.

\bibitem{Lunardini:2016xwi}
Cecilia Lunardini and Walter Winter.
\newblock {High Energy Neutrinos from the Tidal Disruption of Stars}.
\newblock {\em Phys. Rev. D}, 95(12):123001, 2017.

\bibitem{Senno:2016bso}
Nicholas Senno, Kohta Murase, and Peter Meszaros.
\newblock {High-energy Neutrino Flares from X-Ray Bright and Dark Tidal
  Disruption Events}.
\newblock {\em Astrophys. J.}, 838(1):3, 2017.

\bibitem{Band:1993eg}
D.~Band et~al.
\newblock {BATSE observations of gamma-ray burst spectra. 1. Spectral
  diversity.}
\newblock {\em Astrophys. J.}, 413:281--292, 1993.

\bibitem{Palladino:2018lov}
Andrea Palladino, Xavier Rodrigues, Shan Gao, and Walter Winter.
\newblock {Interpretation of the diffuse astrophysical neutrino flux in terms
  of the blazar sequence}.
\newblock {\em Astrophys. J.}, 871(1):41, 2019.

\bibitem{Rodrigues:2020pli}
Xavier Rodrigues, Jonas Heinze, Andrea Palladino, Arjen van Vliet, and Walter
  Winter.
\newblock {AGN jets as the origin of UHECRs and perspectives for the detection
  of EeV astrophysical neutrinos}.
\newblock 3 2020.

\bibitem{Hillas:1985is}
A.~M. Hillas.
\newblock {The Origin of Ultrahigh-Energy Cosmic Rays}.
\newblock {\em Ann. Rev. Astron. Astrophys.}, 22:425--444, 1984.

\bibitem{Hummer:2011ms}
Svenja Hummer, Philipp Baerwald, and Walter Winter.
\newblock {Neutrino Emission from Gamma-Ray Burst Fireballs, Revised}.
\newblock {\em Phys. Rev. Lett.}, 108:231101, 2012.

\bibitem{Gao:2016uld}
Shan Gao, Martin Pohl, and Walter Winter.
\newblock {On the direct correlation between gamma-rays and PeV neutrinos from
  blazars}.
\newblock {\em Astrophys. J.}, 843(2):109, 2017.

\bibitem{Song:2020nfh}
Ningqiang Song, Shirley~Weishi Li, Carlos~A. Arg\"uelles, Mauricio Bustamante,
  and Aaron~C. Vincent.
\newblock {The Future of High-Energy Astrophysical Neutrino Flavor
  Measurements}.
\newblock 12 2020.

\bibitem{Feldman:1997qc}
Gary~J. Feldman and Robert~D. Cousins.
\newblock {A Unified approach to the classical statistical analysis of small
  signals}.
\newblock {\em Phys. Rev. D}, 57:3873--3889, 1998.

\bibitem{Abbasi:2009ig}
R.~Abbasi et~al.
\newblock {Search for muon neutrinos from Gamma-Ray Bursts with the IceCube
  neutrino telescope}.
\newblock {\em Astrophys. J.}, 710:346--359, 2010.

\bibitem{Baerwald:2014zga}
Philipp Baerwald, Mauricio Bustamante, and Walter Winter.
\newblock {Are gamma-ray bursts the sources of ultra-high energy cosmic rays?}
\newblock {\em Astropart. Phys.}, 62:66--91, 2015.

\bibitem{Zyla:2020zbs}
P.A. Zyla et~al.
\newblock {Review of Particle Physics}.
\newblock {\em PTEP}, 2020(8):083C01, 2020.

\bibitem{Rodrigues:2020fbu}
Xavier Rodrigues, Simone Garrappa, Shan Gao, Vaidehi~S. Paliya, Anna
  Franckowiak, and Walter Winter.
\newblock {Multi-wavelength and neutrino emission from blazar PKS 1502+106}.
\newblock 9 2020.

\bibitem{Heinze:2020zqb}
Jonas Heinze, Daniel Biehl, Anatoli Fedynitch, Denise Boncioli, Annika Rudolph,
  and Walter Winter.
\newblock {Systematic parameter space study for the UHECR origin from GRBs in
  models with multiple internal shocks}.
\newblock {\em Mon. Not. Roy. Astron. Soc.}, 498(4):5990--6004, 2020.

\bibitem{Oikonomou:2019djc}
Foteini Oikonomou, Kohta Murase, Paolo Padovani, Elisa Resconi, and Peter
  M\'esz\'aros.
\newblock {High energy neutrino flux from individual blazar flares}.
\newblock {\em Mon. Not. Roy. Astron. Soc.}, 489(3):4347--4366, 2019.

\bibitem{Beacom:2003nh}
John~F. Beacom, Nicole~F. Bell, Dan Hooper, Sandip Pakvasa, and Thomas~J.
  Weiler.
\newblock {Measuring flavor ratios of high-energy astrophysical neutrinos}.
\newblock {\em Phys. Rev. D}, 68:093005, 2003.
\newblock [Erratum: Phys.Rev.D 72, 019901 (2005)].

\bibitem{Bustamante:2015waa}
Mauricio Bustamante, John~F. Beacom, and Walter Winter.
\newblock {Theoretically palatable flavor combinations of astrophysical
  neutrinos}.
\newblock {\em Phys. Rev. Lett.}, 115(16):161302, 2015.

\bibitem{Bustamante:2019sdb}
Mauricio Bustamante and Markus Ahlers.
\newblock {Inferring the flavor of high-energy astrophysical neutrinos at their
  sources}.
\newblock {\em Phys. Rev. Lett.}, 122(24):241101, 2019.

\bibitem{Datarelease}
Damiano F.~G. Fiorillo, Arjen Van~Vliet, Stefano Morisi, and Walter Winter.
\newblock Unified-thermal-model.
\newblock \url{https://github.com/damianofiorillo/Unified-thermal-model}.

\bibitem{Abbasi:2020zmr}
R.~Abbasi et~al.
\newblock {Measurement of Astrophysical Tau Neutrinos in IceCube's High-Energy
  Starting Events}.
\newblock 11 2020.

\bibitem{Palladino:2015zua}
A.~Palladino, G.~Pagliaroli, F.~L. Villante, and F.~Vissani.
\newblock {What is the Flavor of the Cosmic Neutrinos Seen by IceCube?}
\newblock {\em Phys. Rev. Lett.}, 114(17):171101, 2015.

\bibitem{Adrian-Martinez:2016fdl}
S.~Adrian-Martinez et~al.
\newblock {Letter of intent for KM3NeT 2.0}.
\newblock {\em J. Phys.}, G43(8):084001, 2016.

\bibitem{Glashow:1960zz}
Sheldon~L. Glashow.
\newblock {Resonant Scattering of Antineutrinos}.
\newblock {\em Phys. Rev.}, 118:316--317, 1960.

\bibitem{Biehl:2016psj}
Daniel Biehl, Anatoli Fedynitch, Andrea Palladino, Tom~J. Weiler, and Walter
  Winter.
\newblock {Astrophysical Neutrino Production Diagnostics with the Glashow
  Resonance}.
\newblock {\em JCAP}, 01:033, 2017.

\bibitem{IceCubeGlashow2021}
M.~G. Aartsen et~al.
\newblock Detection of a particle shower at the glashow resonance with icecube.
\newblock {\em Nature}, 591(7849):220--224, 2021.

\bibitem{Dermer:2012rg}
Charles~D. Dermer, Kohta Murase, and Hajime Takami.
\newblock {Variable Gamma-ray Emission Induced by Ultra-High Energy Neutral
  Beams: Application to 4C +21.35}.
\newblock {\em Astrophys. J.}, 755:147, 2012.

\bibitem{Murase:2014foa}
Kohta Murase, Yoshiyuki Inoue, and Charles~D. Dermer.
\newblock {Diffuse Neutrino Intensity from the Inner Jets of Active Galactic
  Nuclei: Impacts of External Photon Fields and the Blazar Sequence}.
\newblock {\em Phys. Rev. D}, 90(2):023007, 2014.

\bibitem{Pakvasa:2010jj}
Sandip Pakvasa.
\newblock {Neutrino Flavor Detection at Neutrino Telescopes and Its Uses}.
\newblock 4 2010.

\bibitem{Aartsen:2016zhm}
M.G. Aartsen et~al.
\newblock {Search for annihilating dark matter in the Sun with 3 years of
  IceCube data}.
\newblock {\em Eur. Phys. J. C}, 77(3):146, 2017.
\newblock [Erratum: Eur.Phys.J.C 79, 214 (2019)].

\bibitem{Backstrom:2018plo}
Anton Bäckström, Riccardo Catena, and Carlos Pérez de~los Heros.
\newblock {Assessing the sensitivity of PINGU to effective dark matter-nucleon
  interactions}.
\newblock {\em JCAP}, 05:023, 2019.

\bibitem{Honda:2015fha}
M.~Honda, M.~Sajjad~Athar, T.~Kajita, K.~Kasahara, and S.~Midorikawa.
\newblock {Atmospheric neutrino flux calculation using the NRLMSISE-00
  atmospheric model}.
\newblock {\em Phys.\ Rev.\ D}, 92(2):023004, 2015.

\bibitem{Kierans:2021kpg}
Carolyn~A. Kierans.
\newblock {AMEGO: Exploring the Extreme Multimessenger Universe}.
\newblock {\em Proc. SPIE Int. Soc. Opt. Eng.}, 11444:1144431, 2020.

\end{thebibliography}

\end{document}